\documentclass[3p]{elsarticle}

\usepackage{amssymb}

\newtheorem{thm}{Theorem}[section]
\newtheorem{lem}[thm]{Lemma}
\newtheorem{cor}[thm]{Corollary}
\newtheorem{prop}[thm]{Proposition}

\newdefinition{rema}[thm]{Remark}
\newdefinition{defi}[thm]{Definition}
\newdefinition{nota}[thm]{Notation}
\newdefinition{exam}[thm]{Example}
\newproof{prf}{Proof}
\long\def\proof#1\qed{%
 \begin{prf}#1\qed\end{prf}}
\makeatletter
\def\shrinktopsep{%
 \setlength{\@sep}{.2\baselineskip plus.1\baselineskip minus.1\baselineskip}}
\makeatother

\def\vspec#1{\special{ps:#1}}
\edef\zerocapsule#1{\vspec{%
 /scaleDvips {
 8.3 8.3 scale .5 setlinewidth currentpoint
 translate [1 0 0 -1 0 0] concat }def
 gsave 1 1 scale scaleDvips #1 grestore}}

\newdimen\latticeUnit
\newdimen\boxMargin \newdimen\layerMargin

\newbox\tbox \newbox\ttbox
\chardef\tc=8 \chardef\tcc=9
\newdimen\tdimen \newdimen\ttdimen
\newdimen\sdimen \newdimen\ssdimen
\newtoks\ttoks \newtoks\tttoks
\newcount\tcount \newcount\ttcount
\newcount\tttcount \newcount\ttttcount

\newbox\diagbox
\newdimen\rightlim \newdimen\leftlim
\newdimen\upperlim \newdimen\lowerlim
\newdimen\dimA \newdimen\dimB
\newdimen\dimC \newdimen\dimD
\newdimen\dimE \newdimen\dimF \newdimen\dimG

\def\clist{\\}

\def\rightappend#1\to#2{\ttoks={#1\\}\tttoks=\expandafter{#2}%
 \edef#2{\the\tttoks\the\ttoks}}

{\catcode`p=12 \catcode`t=12 \gdef\gazonc#1pt{#1}}
\let\getfactor=\gazonc
\def\dimensionToNumber#1{\expandafter\getfactor\the#1}
\let\ptless=\dimensionToNumber

\def\nodebox#1{%
 \futurelet\com\nnodeboxx#1\with}

\def\nnodeboxx{%
 \ifx\com\invisible
  \let\next=\invisibleNodeBox
 \else
  \let\next=\visibleNodeBox
 \fi\next}

\def\visibleNodeBox#1\with#2#3{%
 \def\bm{#3}
 \setbox#2=\hbox{\kern\bm\vbox{\kern\bm\hbox{$\displaystyle{#1}$%
  }\kern\bm}\kern\bm}}

\def\invisible{}

\def\invisibleNodeBox#1\with#2#3{%
 \def\bm{#3}
 \setbox#2=\hbox{\kern\bm\vbox{\kern\bm\hbox{$\displaystyle{#1}$%
  }\kern\bm}\kern\bm}
 \setbox#2=\hbox{\vrule width0pt height\ht#2 depth\dp#2%
  \vbox to0pt{\hrule height0pt depth0pt width\wd#2}}}

\gdef\object(#1,#2)=#3{%
 \nodebox{#3}\tbox\boxMargin
 \tdimen=.5\wd\tbox \ttdimen=.5\ht\tbox \advance\ttdimen by.5\dp\tbox
 \edef\bbbox{(#1,#2)(\ptless{\tdimen}pt,\ptless{\ttdimen}pt)}%
 \expandafter\rightappend\bbbox\to\clist
 \advance\tdimen by-#1\latticeUnit \advance\ttdimen by-#2\latticeUnit
 \tdimen=-\tdimen \ttdimen=-\ttdimen
 \sdimen=\tdimen
  \ifdim\sdimen<\leftlim \global\leftlim=\sdimen \fi
  \advance\sdimen by\wd\tbox
  \ifdim\sdimen>\rightlim \global\rightlim=\sdimen \fi
 \sdimen=\ttdimen
  \ifdim\sdimen<\lowerlim \global\lowerlim=\sdimen \fi
  \advance\sdimen by\ht\tbox
  \ifdim\sdimen>\upperlim \global\upperlim=\sdimen \fi
 \put(\ptless{\tdimen},\ptless{\ttdimen}){\unhbox\tbox}
}

\def\refer(#1,#2)\to(#3,#4){\def\a{#1}\def\b{#2}%
 \tcount=#3 \ttcount=#4
 \expandafter\rreferr\clist\empty}
\def\rreferr\\#1{\ifx#1\empty \let\next=\relax
 \else \let\next=\rrreferrr \fi \next}
\def\rrreferrr#1,#2)(#3,#4){\def\aa{#1}\def\bb{#2}%
 \ifdim\a pt=\aa pt \ifdim\b pt=\bb pt
  \dimen\tcount=#3 \dimen\ttcount=#4
 \fi\fi \rreferr}

\gdef\edge{\let\lcommand=\line
 \morphismSwitch}

\gdef\morphism{\let\lcommand=\vector
 \morphismSwitch}

\def\morphismSwitch(#1,#2)to(#3,#4){%
 \def\mNext{\ifx\farg[\morphismBody(#1,#2)(#3,#4)%
  \else \slantMorphism(#1,#2)(#3,#4)\fi}
 \futurelet\farg\mNext}

\def\morphismBody(#1,#2)(#3,#4){%
 \slantMorphism(#1,#2)(#3,#4)%
 \expandafter\attachSwitch}

\def\abs#1{\ifdim#1<0pt-#1\else#1\fi}

\def\attachSwitch#1#2]{
 \nodebox{{\scriptstyle #2}}\tbox\boxMargin
 \rightsidefalse
 \diagonalfalse \horizontalfalse \verticalfalse
 \def\tNext{\ifx\targ[\expandafter\attachOption
  \else \attachBody\fi}
 \futurelet\targ\tNext}

\newif\ifrightside
\newif\ifhorizontal
\newif\ifvertical
\newif\ifdiagonal

\def\attachOption#1{%
 \attachOptionLoop}
\def\attachOptionLoop#1{%
 \ifx#1]
  \let\next\attachBody
 \else
  \ifx#1R \rightsidetrue \fi
  \ifx#1H \horizontaltrue \fi
  \ifx#1V \verticaltrue \fi
  \ifx#1D \diagonaltrue \fi
  \let\next\attachOptionLoop
 \fi
 \next
}

\newtoks\labelPosition
\labelPosition={.5}

\def\attachBody{%
 \dimC=\xC \advance\dimC by-\xB
 \dimD=\yC \advance\dimD by-\yB
 \dimA=\xB \advance\dimA by\the\labelPosition\dimC
 \dimB=\yB \advance\dimB by\the\labelPosition\dimD
 \ifhorizontal \ifdim0pt<\abs\dimD \attachBodyHorizontal \fi
 \else \ifvertical \ifdim0pt<\abs\dimC \attachBodyVertical \fi
 \else \ifdiagonal \attachBodyDiagonal
 \else
  \tdimen.2\dimD \tdimen=\abs\tdimen
  \ifdim\abs\dimC<\tdimen
   \attachBodyHorizontal
  \else \tdimen.2\dimC \tdimen=\abs\tdimen
  \ifdim\abs\dimD<\tdimen
   \attachBodyVertical
  \else
   \attachBodyDiagonal
  \fi\fi
 \fi\fi\fi}

\def\attachBodyHorizontal{%
 \dimG=.5\ht\tbox \advance\dimG by.5\dp\tbox
 \ifrightside \dimE=\dimD \else \dimE=-\dimD \fi
 \dimF=\dimC
 \ifdim\dimD<0.0pt
  \ifrightside \advance\dimA by-.5\wd\tbox
  \else \advance\dimA by.5\wd\tbox \fi
 \else
  \ifrightside \advance\dimA by.5\wd\tbox
  \else \advance\dimA by-.5\wd\tbox \fi
 \fi
 \displaceLabel(\ptless{\dimA},\ptless{\dimB})(\ptless{\dimE},\ptless{\dimF},\ptless{\dimG},1)
}

\def\attachBodyVertical{%
 \dimG=.5\wd\tbox
 \ifrightside \dimE=\dimC \else \dimE=-\dimC \fi
 \dimF=\dimD
 \ifdim\dimC<0.0pt
  \ifrightside \advance\dimB by.5\ht\tbox \advance\dimB by.5\dp\tbox
  \else \advance\dimB by-.5\ht\tbox \advance\dimB by-.5\dp\tbox \fi
 \else
  \ifrightside \advance\dimB by-.5\ht\tbox \advance\dimB by-.5\dp\tbox
  \else \advance\dimB by.5\ht\tbox \advance\dimB by.5\dp\tbox \fi
 \fi
 \displaceLabel(\ptless{\dimA},\ptless{\dimB})(\ptless{\dimE},\ptless{\dimF},\ptless{\dimG},-1)
}

\def\attachBodyDiagonal{%
 \ifdim\dimD<0.0pt
  \ifrightside
   \advance\dimA by-.5\wd\tbox \advance\dimA by.5\boxMargin
  \else
   \advance\dimA by.5\wd\tbox \advance\dimA by-.5\boxMargin
  \fi
 \else
  \ifrightside
   \advance\dimA by.5\wd\tbox \advance\dimA by-.5\boxMargin
  \else
   \advance\dimA by-.5\wd\tbox \advance\dimA by.5\boxMargin
  \fi
 \fi
 \ifdim\dimC<0.0pt
  \ifrightside
   \advance\dimB by.5\ht\tbox \advance\dimB by.5\dp\tbox
   \advance\dimB by-.5\boxMargin
  \else
   \advance\dimB by-.5\ht\tbox \advance\dimB by-.5\dp\tbox
   \advance\dimB by.5\boxMargin
  \fi
 \else
  \ifrightside
   \advance\dimB by-.5\ht\tbox \advance\dimB by-.5\dp\tbox
   \advance\dimB by.5\boxMargin
  \else
   \advance\dimB by.5\ht\tbox \advance\dimB by.5\dp\tbox
   \advance\dimB by-.5\boxMargin
  \fi
 \fi
 \displaceLabel(\ptless{\dimA},\ptless{\dimB})(1.0,0.0,0.0,0)
}

\def\displaceLabel(#1,#2)(#3,#4,#5,#6){%
 \put(#1,#2){%
  \vspec{gsave
   currentpoint currentpoint translate
   /adif {#5 #4 abs mul #3 div 8.3 mul} def
   /bdif {0.0} def
   #6 0 gt {adif bdif translate} {bdif adif translate} ifelse
   neg exch neg exch translate}%
  \hbox to0pt{\hss \vbox to0pt{\vss
   \hbox{\copy\tbox}%
   \vss}\hss}%
  \vspec{currentpoint grestore moveto}%
 }
}

\newdimen\hd \newdimen\vd \newdimen\cd \newdimen\md
\newdimen\xB \newdimen\yB
\newdimen\xC \newdimen\yC

\def\slantMorphism(#1,#2)(#3,#4){%
 \hd=#3\latticeUnit \advance\hd by-#1\latticeUnit
 \vd=#4\latticeUnit \advance\vd by-#2\latticeUnit
 \vectorPosition(#1,#2)(#3,#4)%
 \edef\arg{{\ptless{\xB}}{\ptless{\yB}}{\ptless{\xC}}{\ptless{\yC}}}
  \ifx\lcommand\vector
   \expandafter\drawLine\arg
   \expandafter\putTip\arg
  \else
   \expandafter\drawLine\arg
  \fi
}

\font\tenline = line10
\def\tip{\vbox to0pt{\hbox to0pt{{\tenline\char"36}\hss}\vss}}

\def\putTip#1#2#3#4{%
 \put(#3,#4){%
 \vspec{gsave
  currentpoint currentpoint translate
  /tanrot {50 10 sub 30 20 sub atan rotate} def
  #4 #2 eq not {#3 #1 sub #4 #2 sub atan rotate}
   {#1 #3 lt {90 rotate} {270 rotate} ifelse} ifelse
  neg exch neg exch translate}%
 \tip%
 \vspec{currentpoint grestore moveto}}}

\newif\iflayer
\layerfalse

\def\drawLine#1#2#3#4{%
 \iflayer
  \put(0,0){\zerocapsule{2.5 setlinewidth 1 setgray
   #1 #2 moveto #3 #4 lineto stroke
   .5 setlinewidth 0 setgray
   #1 #2 moveto #3 #4 lineto stroke}}%
 \else
  \put(0,0){\zerocapsule{#1 #2 moveto #3 #4 lineto stroke}}%
 \fi
}

{\catcode`p=12 \catcode`t=12 \gdef\hazonc#1.#2pt{#1}}
\def\toJnt#1{\expandafter\hazonc\the#1}
\let\getInt=\toJnt

\newdimen\xd \newdimen\yd \newdimen\xe \newdimen\ye

\def\vectorPosition(#1,#2)(#3,#4){%
 \ifdim \hd>0.0pt
  \ifdim \vd>0.0pt
   \obtainDelta(#1,#2)(#3,#4)
  \else
   \vd=-\vd
   \obtainDelta(#1,#2)(#3,#4)
   \vd=-\vd \yd=-\yd \ye=-\ye
  \fi
 \else
  \hd=-\hd
  \ifdim \vd>0.0pt
   \obtainDelta(#1,#2)(#3,#4)
   \xd=-\xd \xe=-\xe
  \else
   \vd=-\vd
   \obtainDelta(#1,#2)(#3,#4)
   \vd=-\vd \xd=-\xd \xe=-\xe \yd=-\yd \ye=-\ye
  \fi
  \hd=-\hd
 \fi
 \xB=#1\latticeUnit \advance\xB by\xd
 \yB=#2\latticeUnit \advance\yB by\yd
 \xC=#3\latticeUnit \advance\xC by\xe
 \yC=#4\latticeUnit \advance\yC by\ye
}

\def\obtainDelta(#1,#2)(#3,#4){%
 \refer(#1,#2)\to(\tc,\tcc)%
 \ifdim\hd<\vd 
  \dimA=\dimen\tcc
  \multiply\dimA by\getInt\hd
  \divide\dimA by\getInt\vd    
  \dimB=\dimen\tc
  \ifdim \dimB<\dimA
   \dimA=\dimen\tc
   \multiply\dimA by\getInt\vd
   \divide\dimA by\getInt\hd   
   \xd=\dimen\tc \yd=\dimA
  \else
   \xd=\dimA \yd=\dimen\tcc
  \fi
 \else 
  \dimA=\dimen\tc
  \multiply\dimA by\getInt\vd
  \divide\dimA by\getInt\hd    
  \dimB=\dimen\tcc
  \ifdim \dimB<\dimA
   \dimA=\dimen\tcc
   \multiply\dimA by\getInt\hd
   \divide\dimA by\getInt\vd   
   \xd=\dimA \yd=\dimen\tcc
  \else
   \xd=\dimen\tc \yd=\dimA
  \fi
 \fi
 \refer(#3,#4)\to(\tc,\tcc)%
 \ifdim\hd<\vd 
  \dimA=\dimen\tcc
  \multiply\dimA by\getInt\hd
  \divide\dimA by\getInt\vd    
  \dimB=\dimen\tc
  \ifdim \dimB<\dimA
   \dimA=\dimen\tc
   \multiply\dimA by\getInt\vd
   \divide\dimA by\getInt\hd   
   \xe=-\dimen\tc \ye=-\dimA
  \else
   \xe=-\dimA \ye=-\dimen\tcc
  \fi
 \else 
  \dimA=\dimen\tc
  \multiply\dimA by\getInt\vd
  \divide\dimA by\getInt\hd    
  \dimB=\dimen\tcc
  \ifdim \dimB<\dimA
   \dimA=\dimen\tcc
   \multiply\dimA by\getInt\hd
   \divide\dimA by\getInt\vd   
   \xe=-\dimA \ye=-\dimen\tcc
  \else
   \xe=-\dimen\tc \ye=-\dimA
  \fi
 \fi
}

\def\ptToCoord#1{%
 \tcount=\expandafter\toInt\number\ptless\latticeUnit
 \divide#1 by\tcount}

\newenvironment{diagramme}{%
  \latticeUnit=100pt \boxMargin=3pt \layerMargin=3pt
  \rightlim=0pt \leftlim=0pt \upperlim=0pt \lowerlim=0pt
  \setbox\diagbox=\hbox\bgroup
  \begin{picture}(0,0)}%
 {\end{picture}\egroup
  \tdimen=\rightlim \advance\tdimen by-\leftlim
  \ttdimen=\upperlim \advance\ttdimen by-\lowerlim
  \begin{picture}(\ptless{\tdimen},\ptless{\ttdimen})%
                 (\ptless{\leftlim},\ptless{\lowerlim})
   \put(0,0){\unhbox\diagbox}
  \end{picture}%
}

\def\mor#1{\mathrel{\mathop{\longrightarrow}\limits^{\vbox
 to0pt{\vss \hbox{$\scriptstyle #1$}\kern-2pt}}}}
\def\leftmor#1{\mathrel{\mathop{\longleftarrow}\limits^{\vbox
 to0pt{\vss \hbox{$\scriptstyle #1$}\kern-2pt}}}}
\def\twomor#1{\mathrel{\mathop{\Longrightarrow}\limits^{\vbox
 to0pt{\vss \hbox{$\scriptstyle #1$}\kern-1pt}}}}
\def\shortmor#1{\mathrel{\mathop{\rightarrow}\limits^{\vbox
 to0pt{\vss \hbox{$\scriptstyle #1$}\kern-2pt}}}}
\def\shortleftmor#1{\mathrel{\mathop{\leftarrow}\limits^{\vbox
 to0pt{\vss \hbox{$\scriptstyle #1$}\kern-2pt}}}}
\def\fitarr#1{\vcenter{\offinterlineskip
 \halign{\hfil ##\hfil &$##$\hfil \crcr
 \vbox to0pt{\vss \hbox{\kern2pt$\scriptstyle #1$\kern3pt}}\crcr
 \noalign{\vskip1pt}
 \leaders\hrule height.7pt depth-.3pt\hfill $\mkern2mu$
  &\vbox to0pt{\vss \llap{$\rightarrow$}\vss} \crcr}}}

\def\sqda#1#2#3#4#5#6#7#8{%
 \let\ss=\scriptstyle
 \latticeUnit=1pt \boxMargin=3pt
 \object(0,\vert)={#1}
 \object(\hori,\vert)={#2}
 \object(0,0)={#3}
 \object(\hori,0)={#4}
 \morphism(0,\vert)to(\hori,\vert)[\ss{#5}]
 \morphism(0,\vert)to(0,0)[\ss{#6}][R]
 \morphism(\hori,\vert)to(\hori,0)[\ss{#7}]
 \morphism(0,0)to(\hori,0)[\ss{#8}][R]
}

\def\squarediagabs#1#2{%
 \def\hori{#1}\def\vert{#2}%
 \argsqda}

\def\argsqda#1#2#3#4#5#6#7#8{%
 \hbox{\begin{diagramme}
  \sqda{#1}{#2}{#3}{#4}{#5}{#6}{#7}{#8}
 \end{diagramme}}}

\newdimen\rotdimen
\def\vspec#1{\special{ps:#1}}
\def\rotstart#1{\vspec{gsave currentpoint currentpoint translate
   #1 neg exch neg exch translate}}
\def\rotfinish{\vspec{currentpoint grestore moveto}}

\newdimen\dimSt \newdimen\dimCt
\newdimen\dimSin \newdimen\dimCos
\newcount\numX

\def\sincos#1{
 \numX=5
 \dimSin=.087156pt\dimCos=.996195pt
 \loop\ifnum \numX<#1%
 \dimCt=\dimCos \dimSt=\dimSin
 \dimCos=.996195\dimCt \advance\dimCos by-.087156\dimSt
 \dimSin=.087156\dimCt \advance\dimSin by.996195\dimSt
 \advance \numX by5\repeat}

{\catcode`p=12 \catcode`t=12 \gdef\elimPT#1pt{#1}}
\let\getfactor=\elimPT
\def\dimtonum#1{\expandafter\getfactor\the#1}

 \newcount\rct\newbox\rbx
\def\rot#1#2{%
 \ifnum #2<91%
  \prot{#1}{#2}\else
 \ifnum #2<181%
  \rct=#2 \advance\rct by-90%
  \setbox\rbx\hbox{\prot{#1}{\the\rct}}%
  \prot{\rbx}{90}\else
 \ifnum #2<271%
  \rct=#2 \advance\rct by-180%
  \setbox\rbx\hbox{\prot{#1}{\the\rct}}%
  \setbox\rbx\hbox{\prot\rbx{90}}%
  \prot{\rbx}{90}\else
 \ifnum #2<361%
  \rct=#2 \advance\rct by-270%
  \setbox\rbx\hbox{\prot{#1}{\the\rct}}%
  \setbox\rbx\hbox{\prot\rbx{90}}%
  \setbox\rbx\hbox{\prot\rbx{90}}%
  \prot{\rbx}{90}
 \fi\fi\fi\fi}

 \newdimen\hep\newdimen\gap
\def\prot#1#2{\sincos{#2}%
 \hep=\ht#1\advance\hep by\dp#1%
 \hbox{%
  \expandafter\gap\dimtonum\dimSin\hep
  \kern\gap
  \vbox{%
   \rotstart{#2 rotate}\copy#1
   \expandafter\gap\dimtonum\dimCos\hep
   \advance\gap by-\hep
   \expandafter\hep\dimtonum\dimSin\wd#1%
   \advance\hep by\gap
   \kern\hep
   }%
  \expandafter\gap\dimtonum\dimCos\wd#1%
  \advance\gap by-\wd#1%
  \kern\gap
  }\rotfinish}

\def\llapem#1#2{\llap{\hbox to#1em{\rm #2\hss}}}

\def\uplet#1#2#3{#1\mathbin\upharpoonright #2\mathbin{:=}#3}%

\def\lc{(\mskip-4mu|}%
\def\rc{|\mskip-4mu)}%

\font\teneufm=eufm10
\def\eufm#1{\hbox{\teneufm #1}}

\newbox\tcd\setbox\tcd=\hbox{$\omega$}
\def\agemo{\mathord{\hbox{\rot\tcd{180}}}}

\def\dbot{\bot\,\llap{$\bot$}}

\def\scirc{\mathbin{\vcenter{\hbox{\scriptsize $\circ$}}}}

\def\lbra#1{%
 \vtop{\hbox{\strut}%
  \kern-\baselineskip\kern-\lineskip
  \hbox{$\left\{\vcenter
   {\halign{\strut $\>##$\qquad \hfil &##\hfil \crcr
 #1}}\right.$}}}

\def\exraise#1{\mathop{{\sf raise}\,#1}}%

\begin{document}

\title{Complete Call-by-Value Calculi of Control Operators, I}

\author{Ryu Hasegawa}	
\address{Graduate School
 of Mathematical Sciences, The University of Tokyo, Komaba 3-8-1,
 Meguro-ku, Tokyo 153-8914, Japan}	
\ead{ryu@ms.u-tokyo.ac.jp}  


\begin{abstract}
  \noindent
We give new call-by-value calculi of control operators that
 are complete for the continuation-passing style semantics.
Various anticipated computational properties are induced from
 the completeness.
In the first part of a series of papers, we give the characterization
 of termination properties using the continuation-passing style semantics
 as well as the union-intersection type discipline.
\end{abstract}

\maketitle

\section*{Introduction}

After the presentation of the
 first formal system embodying control
 operators by Felleisen et al.~\cite{krj20},
 and after the striking exposition
 of its relevance to the classical logic
 by Griffin \cite{ucb27},
 the studies of first-class continuation exploded.
Various systems have been proposed
 by now.
If we restrict the search range to call-by-value calculi, we still
 find an abundance; \cite{krj20}\cite{kwx35}\cite{zfk00}
 to name a few.
In this paper, we add new systems further to the end of the long list.
Our contribution is to present new call-by-value calculi that
 are complete and still satisfy good properties,
 as explained shortly in turn.

The semantics of call-by-value
 calculi (as well as call-by-name
 calculi) is traditionally
 given by the continuation-passing style (CPS) translation.
From its start, it has been used in connection with the implementation
 of programming languages \cite{aaj61}.
Flanagan et al.\ suggests relationship to optimization \cite{wmx35}.
With regard to mathematical logic, it corresponds
 to the Kuroda translation that
 is a variant of G\"odel's double negation translation
 \cite{tuj96}\cite{sjt10}.
Now the CPS translation is commonly accepted as the standard semantics.

Given the semantics, completeness is one of the most desirable properties.
The completeness asserts that every equality in the semantics is reflected
 in the calculus.
If a system is not complete, it suffers a loss of a part of the wealth
 that is inherited from the target of the semantics.
The target of the CPS semantics is simply the lambda calculus, the
 $\beta\eta$-reduction of which
 is the condensation of the essence
 of computation.
If not complete, thus failing
 to inherit the essence, the concerned system
 definitely misses certain quality that it should
 intrinsically possess.

The widespread systems are not complete, however.
None of the calculi listed in the first paragraph are complete.
Therefore the quest for complete call-by-value systems
 with control operators has been pursued
 over many years.
After the first system presented by
 Sabry, a number of systems are proposed in the literature
 \cite{yrs06}\cite{vyh68}\cite{lss57}\cite{iba58}\cite{fck86}\cite{gnj37}%
 \cite{iab45}\cite{vtp88}\cite{mhx11}.

Though many complete call-by-value
 systems are proposed so far,
 the computational properties
 are hardly explored.
Consider the confluence property or termination of evaluation,
 for example.
Almost all of the previous systems are presented by equational axioms,
 having no notion of computation, thus hopeless even of
 dictating such statements.
It seems difficult to design a complete but well-behaved calculus.
When we say calculus, we intend a system equipped with computation.

One of the reasons why we lack computational properties
 is that we must add a number of exotic rules to a core system
 to restore completeness.
For instance, several systems have a rule
 $(\lambda x.\,E[x])M=E[M]$ where $E[\cdot]$ is a call-by-value
 evaluation context.
It deviates from the general pattern of reduction rules, which
 is concerned only with a combination of a few basic constructors.
Moreover, if the number of rules increases, the intricacy
 of verifying properties is liable to accelerate.
In order to obtain anticipated results, therefore,
 we must try to (i) restrict the added rules to manageable ones and
 (ii) find a proof method that overcomes the difficulty caused
 by the increasing number of rules.

Our main contribution is to show that a
 simple (indeed, very simple) modification
 in the call-by-value lambda calculus leads us
 to complete calculi.
We add a little ``twist'' to its syntax.
If we view the calculus as a two-story
 structure of syntax (terms) and reductions,
 we modify only the level of syntax.
As a consequence, we have only standard reduction rules found
 in the literature, achieving the goal (i) above.
This contrasts with previous approaches where
 the syntax is largely maintained and the reductions are modified.

Making use of our syntax, we verify a sharpened completeness
 result (Thm.~\ref{gbu84}).
It is a kind of completeness that takes reduction sequences into account,
 not only equational relations.
Similar results are given for different calculi
 \cite{ndy68}\cite{eiq93}\cite{pjd29}.
For call-by-value calculi
 with control operators, this type of completeness is
 proven first (as we comment above, most previous systems have no notion
 of reduction from the beginning).

The sharpened completeness is a key result.
It provides us with machinery to verify
 computational properties.
If the target calculus satisfies a certain property, the sharpened
 completeness informs us how to reinstall it in the source language.
Definitely, this type of argument is not our original idea.
Fujita especially uses
 similar technique extensively \cite{eiq93}\cite{pjd29}.
We can find a similar approach in de~Groote \cite{owr18}, Sabry and
 Wadler \cite{ndy68}, and Crolard \cite{ltb70}
 as well.
We systematically apply the method to prove various properties.
This way we achieve the goal (ii) above.

We dub this method proof by parasitism.
It is the image of depriving its host of nourishment at low
 cost.
The source language controls the target calculus,
 and let it work hard for him.
The properties of the target calculus are
 sucked through a channel prepared by completeness.

We apply the method of proof by parasitism
 to verify desirable syntactic properties.
Let $[\![M]\!]$ denote the CPS translation of a term $M$.
We first consider an untyped system.
We prove the following.

\begingroup
\let\tempar\par \def\par{{\tempar}}%
	\vskip0ex
        \hangafter0\hangindent1.8em
        \noindent
\llapem{1.8}{(1)}%
Our calculus enjoys the Church-Rosser property (Thm.~\ref{pje26}).

	\vskip0ex
        \noindent
\llapem{1.8}{(2)}%
$M$ is normalizable if and only if $[\![M]\!]$ is normalizable (Thm.~\ref{okm14}).

	\vskip0ex
        \noindent
\llapem{1.8}{(3)}%
The call-by-value evaluation of $M$ terminates
 if and only if $[\![M]\!]$ is solvable (Thm.~\ref{pin99}).

	\vskip0ex
\endgroup
\noindent
Next, we turn to type theory.
We give a union-intersection type discipline for
 the calculus.
The main results are
 the following type-theoretic characterization of computational
 properties.
\begingroup
\let\tempar\par \def\par{{\tempar}}%
	\vskip0ex
        \hangafter0\hangindent1.8em
        \noindent
\llapem{1.8}{(4)}%
The type system satisfies the subject reduction/expansion property
 (Thm.~\ref{ehw91}).

	\vskip0ex
        \noindent
\llapem{1.8}{(5)}%
A term terminates with respect
 to the call-by-value evaluation if and only if the term is
 typeable (Thm.~\ref{tno58}, (1)).
 
	\vskip0ex
        \noindent
\llapem{1.8}{(6)}%
A term is weakly normalizing
 if and only if
 it has a type judgment that contains neither empty intersection
 nor empty union (Thm.~\ref{tno58}, (2)).

	\vskip.5ex
\endgroup
\noindent
We repeat that complete call-by-value
 systems with control operators in the literature
 scarcely have the notion of computation (except Church-Rosser in \cite{mhx11}).
Hence our results are new at the outset.
For the call-by-name $\lambda\mu$-calculus, van~Bakel developed
 a type system that incorporated union and intersection \cite{fxo45}.
Subject reduction/expansion are verified, though
 computational properties are not explicitly characterized.
As far as the author knows, there has been
 no attempt to prove the type-theoretic characterization
 for the call-by-value $\lambda\mu$-calculus.

We have similar results characterizing
 the strong termination property
 by the CPS semantics as well as by types.
However, to avoid the paper's being too long, we decide to discuss
 strong termination in the second part of the series
 of papers \cite{ycq31}.

Finally we disclose the tricks to obtain our complete calculi.
Landin introduced the let-construct ${\sf let}\ x=M\ {\sf in}\ N$
 in \cite[p.~91]{ljk61}.
This notation is inherited by Reynolds \cite{aaj61}, and is widely
 accepted through the spread of programming
 language ML.
In the context of call-by-value lambda calculus, the let notation
 is conveniently used.
However we notice that, in an earlier influential
 paper \cite{rsw67}, Landin first used the where-construct
 $N\ {\sf where}\ x=M$.
It has the same effect as the let-construct, but the body is
 written first.
Later he turned to preferring the let-construct, saying
 ``The only consideration in choosing {\bf let} and {\bf where} will be
 relative convenience of writing an auxiliary definition before or after the
 expression it qualifies'' \cite{ljk61}.
We propose in this paper to return to Landin's original choice
 to write the let-binding to the right of its body.

Another leap is audacious identification between terms.
It means that we deviate from the standard tradition
 that terms are built in the form of trees.
We have an analogy in arithmetic.
For instance, an expression $2+4+6$
 allows two tree forms, $(2+4)+6$ and $2+(4+6)$.
However, we, humans, can understand the expression
 simultaneously as whichever of these trees, or even as the addition of
 three numbers, with intended ambiguity.
We appeal to this power of humans, who can identify the
 entities that look similar.
While this ability is not shared by computers,
 there is no harm, for the behavior of identified terms is
 indistinguishable observationally.
The main philosophy in our design is to build a system that
 is usable for humans.
We believe that the complexity caused by the ambiguity
 is paid off by a number of good properties, and is mitigated
 by the verification method using parasitism.

These modifications affect only
 the structure of the call-by-value lambda calculus.
It is orthogonal to control operators in a sense.
So we incorporate control operators into our call-by-value lambda calculus
 in two different ways.
The first is the $\lambda\mu$-calculus
 and the second the catch/throw calculus after Crolard \cite{ltb70}.
We focus mainly on the former, giving only a short discussion for the
 latter in \S\ref{weg04}.

\section{The CCV $\lambda\mu$-calculus}\label{xvn96}

In this section, we propose a new type-free system of
 call-by-value $\lambda\mu$-calculus.
We call it the CCV $\lambda\mu$-calculus (CCV is the acronym of
 ``complete call-by-value'').
In fact, the calculus is complete for the standard
 CPS transformation \ref{rcr59}.
We verify sharpened completeness involving reduction, not
 only equality, in Thm.~\ref{gbu84}.

\subsection{Syntax}

The let-construct after Landin is frequently used in the syntax
 of languages.
Traditionally it is written in the form ${\sf let}\ x=M\ {\sf in}\ N$.
Some authors use it as a syntax sugar of $(\lambda x.\,N)M$, while
 others use it as a primitive.
For example, Moggi's $\lambda_c$-calculus is an example of the latter
 \cite{dvp96}.
The let-flat rule of the $\lambda_c$-calculus is

	\vskip2ex
        \noindent\kern5em
${\sf let}\ x=({\sf let}\ y=N\ {\sf in}\ M)\ {\sf in}\ L\quad
 \rightarrow\quad
 {\sf let}\ y=N\ {\sf in}\ ({\sf let}\ x=M\ {\sf in}\ L)$

	\vskip2ex
        \noindent
 where we assume that $y$ is not free in $L$.

We also take the let-construct as a primitive distinct from the
 lambda-construct.
However, we write the let-binding to the right of its body.
We adopt the following syntax:

	\vskip2ex
        \noindent\kern5em
$\uplet LxM$

	\vskip2ex
        \noindent
 instead of ${\sf let}\ x=M\ {\sf in}\ L$.
Note that the order of $L$ and $M$ is reversed.
The two terms of the let-flat rule above correspond to

	\vskip2ex
        \noindent\kern5em
$\uplet Lx{(\uplet MyN)}$\quad and\quad
 $\uplet{(\uplet LxM)}yN$

	\vskip2ex
        \noindent
 in the new form.
Let us observe that these two terms differ only in associativity of
 brackets.

The new let-form works in harmony with the $\mu$-binding
 of the $\lambda\mu$-calculus \cite{ygh05}\cite{lhr24}.
The call-by-value variant of the $\lambda\mu$-calculus is
 given in \cite{zfk00}\cite{nty36}.
In the rest of this subsection, we give the definition of
 the CCV $\lambda\mu$-calculus based on this new form
 of the let-construct.

Variables are split into two categories: ordinary variables $x,y,\ldots$
 and continuation variables $k,l,\ldots$.

	\vskip0ex

\begin{defi}\label{adx54}
The {\it terms} $M$ and {\it jumps} $J$ of the CCV $\lambda
 \mu$-calculus are defined by the following syntax:

	\vskip2ex

\halign{\kern5em $#$\hfil &${}\ \mathrel{::=}\ #$\hfil\cr
 M & x\ \ \ |\ \ \ \lambda x.\,M\ \ \ |
       \ \ \ MM\ \ \ |
       \ \ \ \uplet MxM\ \ \ |\ \ \ \mu k.\,J\cr
 J & [k]M\ \ \ |\ \ \ \uplet JxM\cr
}

	\vskip2ex
        \noindent
 where $x$ ranges over ordinary variables and $k$ over continuation
 variables.
We call the prefix $[k]$ a {\it jumper}.
\end{defi}

	\vskip0ex

\noindent
The terms and jumps are syntactically separated.
For example, $\lambda x.\,[k]M$ is disallowed.
The lambda-abstraction and the mu-abstraction bind
 the abstracted variables as usual.
Moreover, the let-sentences $\uplet LxM$ and $\uplet JxM$ bind
 the variable $x$ the scopes of which are $L$ and $J$.
Definition of the set ${\it FV}(M)$ or ${\it FV}(J)$ of free
 (ordinary and continuation) variables is standard.
For example, ${\it FV}(\uplet MxN)$ is given by
 $({\it FV}(M)-\{x\})\cup {\it FV}(N)$.
The terms and the jumps obey ordinary $\alpha$-conversion
 rules.
For example, $\uplet LxM$ equals $\uplet {L\{y/x\}}yM$ if
 $y$ is not free in $L$.

	\vskip0ex

\begin{nota}\label{eua38}\hfill
\shrinktopsep
\begin{enumerate}[(1)]
\item
We use braces for substitution to avoid too much of the
 overloading of the square brackets.
Namely $\{M/x\}$ denotes to substitute $M$ for $x$.
If $M$ is too long, we also use $\{x\mapsto M\}$
 for readability.

\item
We write $x\in M$ in place of $x\in {\rm FV}(M)$ for simplicity.
\end{enumerate}
\end{nota}

	\vskip0ex

\noindent
As usual, we regard $\alpha$-convertible terms to be equal.
In addition to this,
 we introduce new identification rules between terms involving
 the let notation.

	\vskip3ex

\begin{defi}\label{yoq54}
We assume the following equality axioms:

	\vskip2ex

\halign{\kern5em $#$\hfil &${}\ =\ #$\qquad\hfil&#\hfil \cr
 \uplet Lx{(\uplet MyN)} & \uplet{(\uplet LxM)}yN &if $y\not\in L$\cr
 \uplet{(\mu k.\,J)}xM & \mu k.\,(\uplet JxM) &if $k\not\in M$\cr
 [k](\uplet LxM) & \uplet{([k]L)}xM.\cr
}

	\vskip2ex
        \noindent
The first rule is the associativity of let-constructs, the
 second the commutativity between $\mu$ and let, and
 the third the commutativity between jumper and let.
To be precise, the first equality remains valid when $y\in L$ but $x=y$.
However we may assume $x\neq y$ by applying $\alpha$-conversion if needed.
\end{defi}

	\vskip0ex

\noindent
Let us consider the first rule of Def.~\ref{yoq54}.
In rewriting from left to right, the side condition
 $y\not \in L$ does not actually matter.
The violation of the condition disappears
 if we $\alpha$-convert the bound variable $y$ with fresh $z$
 as in $\uplet Lx{(\uplet {M\{z/y\}}zN)}$.
For the rewriting from right to left, in contrast, the side
 condition must be definitely fulfilled.
For example, $\uplet{(\uplet{xy}xM)}yN$ never equals
 $\uplet{xy}x{(\uplet MyN)}$, since $y$ spills out of
 its scope.
Similarly, the second equality of the definition is always applicable
 from left to right, up to the $\alpha$-conversion
 of $k$ if needed.
Hence, if we follow the convention that we do $\alpha$-renaming
 whenever needed, we may assume that the brackets are always
 set as in the right-hand sides of Def.~\ref{yoq54}.

	\vskip0ex

\begin{rema}\label{bmv07}
We often omit brackets as

	\vskip2ex

\halign{\kern5em $#$\hfil\cr
 \uplet Lx{\uplet MyN}\cr
 \uplet{\mu k.\,J}xM.\cr
}

	\vskip2ex
        \noindent
If the side conditions of Def.~\ref{yoq54} fail, these
 are forcibly understood to be the right-hand sides of the definition.
If the conditions hold, we can regard the bracketing to be in
 either way.
If we must select one for some reason, we adopt
 the convention that the right-hand sides are canonical.
We omit also brackets between jump and let-binding as in

	\vskip2ex
        \noindent\kern5em
$[k]\uplet LxM$.

	\vskip2ex
        \noindent
We do not fix which bracketing is canonical for this pattern.
\end{rema}

	\vskip0ex

\noindent
The introduction of equality
 means that we work on equivalence classes of syntactic trees.
There is no mathematical reason to reverse the order
 of the body and binding in let expressions.
However, it would be unbearable to work with equivalences such
 as $[k]({\sf let}\ x=M\ {\sf in}\ L)\sim
 {\sf let}\ x=M\ {\sf in}\ [k]L$ throughout.
Our syntax alleviates the burden to manipulate
 equivalence classes.
It also simplifies the verification of properties.

The let-flat rule mentioned above turns out to
 be one of the equality rules.
A crucial point is that also the reverse of the let-flat
 is implicitly allowed by the equality.
It is indispensable to assure
 the properties discussed in this paper.
Likewise, we have a rule flipping the order of
 $\mu$ and let, which is not explicit
 in other systems.
For example,
 the reader may challenge a non-trivial exercise to prove
 the corresponding equality
 $(\lambda x\mu k.\,J)M=\mu k.\,(\lambda x.\,J)M$ from the
 axioms in Tab.~10 of \cite[p.~246]{fck86}.

	\vskip0ex

\begin{defi}\label{zhl05}
A term is a {\it value} if it is either a variable $x$ or a
 lambda abstraction $\lambda x.\,M$.
Otherwise, it is called a {\it non-value}.
Letter $V$ is preferably used to denote values, often with no mention.
\end{defi}

\begin{defi}\label{ofe40}
The following are the reduction rules of the CCV
 $\lambda\mu$-calculus, where the leftmost column gives the names of rules for
 future reference:

	\vskip0ex

\halign{\rlap{\footnotesize $(#)$}&\kern5em
  $#$\hfil &${}\ \rightarrow\ #$\kern2em\hfil &#\hfil \cr
 {\it ad}_1 & NM & \uplet{zM}zN &($N$ is a non-value)\cr
 {\it ad}_2 & VN & \uplet{Vz}zN &($V$ is a value; $N$ is a non-value)\cr
 \beta_\lambda & (\lambda x.\,M)V & \uplet MxV &($V$ is a value)\cr
 \beta_{\it let} & \uplet MxV & M\{V/x\} &($V$ is a value)\cr
 \beta_\mu & \uplet Mx\mu k.\,J & \rlap{$\mu k.\,J\{[k]\square\mapsto
  [k]\uplet Mx\square\}$}\cr
 \beta_{\it jmp} & [l]\mu k.\,J & J\{l/k\}\cr
 \eta_\lambda & \lambda x.\,Vx & V &($V$ is a value and $x\not\in V$)\cr
 \eta_{\it let} & \uplet xxM & M\cr
 \eta_\mu & \mu k.\,[k]M & M &($k\not\in M$).\cr
}
\end{defi}

	\vskip1ex

\noindent
The first two rules are called {\it administrative rules}.
In them, $z$ is a fresh variable.
The notation $J\{[k]\square\mapsto [k]\uplet Mx\square\}$ is
 a standard context substitution in the $\lambda\mu$-calculus.
Here $\square$ signifies a hole.
Each occurrence of $[k]Q$ in $J$ is,
 the hole being regarded to be filled
 with $Q$, substituted with $[k]\uplet MxQ$,
This process is done recursively.
Namely, as $Q$ contains further occurrences of
 $[k]R$ in general,
 the substitution is applied repeatedly.
For example, $[k]L(\mu l.\,[k]N)\{[k]\square\mapsto [k]\uplet Mx\square\}$
 becomes $[k]\uplet MxL(\mu l.\,[k]\uplet MxN)$, provided
 that $k\not\in L,N$.
The substitution $J\{l/k\}$ is a shorthand of
 $J\{[k]\square\mapsto [l]\square\}$.

As usual, substitutions $\{V/x\}$ and $\{[k]\square
 \mapsto[k]\uplet Mx\square\}$ invoke $\alpha$-renaming
 to avoid accidental collisions of bound variables.
We recall that the let-construct binds a variable
 as well as $\lambda$ and $\mu$.
Hence, for example,

	\vskip2ex
        \noindent\kern5em
$\uplet{\uplet{M}yN}x{y}\ \ \rightarrow
 \ \ \uplet{M\{z/y\}\{y/x\}}z{N\{y/x\}}$,

	\vskip2ex
        \noindent
 where we need the renaming of
 bound $y$ with fresh $z$ to avoid a capture
 by the substitution $\{y/x\}$.

	\vskip0ex

\begin{rema}\rm\label{yxu20}\hfill
\shrinktopsep
\begin{enumerate}[(1)]
\advance\parindent by15pt
\item
By virtue of $\alpha$-renaming, substitution operations
 respect equalities between terms.
This warrants the safety of omission of brackets in Rem.~\ref{bmv07}.

\item
There is ambiguity of the contexts captured by $\mu$-reduction,
 caused by the equality rules given in Def.~\ref{yoq54}.
For example, let us consider

	\vskip2ex
        \noindent\kern5em
$\uplet Lx{\uplet My{\mu k.\,J}}\kern3em (y\not\in L)$.

	\vskip2ex
        \noindent
We have two ways of bracketing.
If we read it left-associative,
 $\mu$-reduction captures the context $\uplet Lx{\uplet My\square}$ as

	\vskip2ex
        \noindent\kern5em
$\mu k.\,J\{[k]\square\mapsto [k]\uplet Lx{\uplet My\square}\}$.

	\vskip2ex
        \noindent
However, if we
 regard the let-bindings to be right-associative,
 $\mu$-reduction captures a partial context $\uplet My\square$ as

	\vskip2ex
        \noindent\kern5em
$\uplet Lx{\mu k.\,J\{[k]\square\mapsto \uplet My\square\}}$.

	\vskip2ex
        \noindent
This ambiguity is intended.
Both reductions are allowed in our system.
We note that the latter converges to the former by one more application
 of rule $(\beta_\mu)$.

Similarly two ways of bracketing in
 $\mu k.\,[l]\uplet Mx\mu h.\,J$ with $k\not\in J$ admit
 different reductions:

	\vskip2ex
        \noindent\kern5em
$\mu k.\,[l]\mu h.\,J\{[h]\square\mapsto [h]\uplet Mx\square\}$

	\vskip2ex
        \noindent
 and

	\vskip2ex
        \noindent\kern5em
$\mu h.\,J\{[h]\square\mapsto [h]\mu k.\,[l]\uplet Mx\square\}$.

	\vskip2ex
        \noindent
These terms converge to a common term if we apply
 rule $(\beta_{\it jmp})$ to them.

\item
As derived from the reduction rules, we have

	\vskip2ex
        \noindent\kern5em
$\uplet{J_0}x{\mu k.\,J}\quad \mor*\quad J\{[k]\square
 \mapsto \uplet{J_0}x\square\}$

	\vskip2ex
        \noindent
 where $J_0$ and $J$ are jumps.
In fact, if $J_0$ has the shape $[l]M$, the left-hand side contracts
 by rule $(\beta_\mu)$ to $[l]\mu k.\,J\{[k]\square
 \mapsto [k]\uplet Mx\square\}$, from which
 we obtain the right hand by rule $(\beta_{\it jmp})$.
If $J_0$ has the shape $\uplet{J_1}yM$, we transfer it
 to the shape $[l]M'$ by
 the equality rules of \ref{yoq54}.
We do not need the second rule exchanging precedence between
 $\mu$ and let, though.
(The last miscellaneous comment is returned later in \ref{xkd18}.)

The derived rule above corresponds to
 the second case of rule $(\mu\hbox{-}\zeta_V)$
 in \cite[Fig.~5, p.~347]{iba58}.
They split cases by types.
Although our system is type-free, the behaviors are different
 between terms and jumps.
The latter correspond to the case of type $\bot$.
\end{enumerate}
\end{rema}

\begin{rema}\rm\label{lkh08}
All rules in Def.~\ref{ofe40} are totally standard,
 save our particular syntax of the let-construct.
Two administrative rules are introduced in \cite{dvp96}.
The combination of $(\beta_\lambda)$ and $(\beta_{\it let})$ splits the ordinary
 $\beta_v$-rule after Plotkin \cite{pag41} into two steps.
The combination is found
 in the core scheme of Flanagan et al.~\cite{wmx35}.
Also the call-by-value $\lambda$-calculus by Curien and Herbelin \cite{qoq47}
 have similar rules, though the syntax is quite different.
Rule $(\beta_\mu)$ and $(\beta_{\it jmp})$ are standard in
 the $\lambda\mu$-calculus.
Rule $(\eta_\lambda)$ is also standard in the call-by-value lambda
 calculus.
Rule $(\eta_{\it let})$ is found in \cite{dvp96} as well as
 in \cite{wmx35} in the context of the optimization transforming
 to tail recursion.
Rule $(\eta_\mu)$ exists in \cite{zfk00}.
These data are by no means exhaustive.
\end{rema}

\begin{defi}\label{hio53}
The smallest congruence relation containing the reduction rules
 in Def.~\ref{ofe40} is denoted by $=_{\it ccv}$, and
 referred as the {\it CCV equality}.
\end{defi}

\begin{exam}\rm\label{isn97}
Many complete systems in the literature
 \cite{vyh68}\cite{yrs06}\cite{iba58}\cite{gnj37}\cite{vtp88}\cite{mhx11}
 contain the axiom

	\vskip2ex
        \noindent\kern5em
$(\lambda x.\,E[x])M\ =\ E[M]$

	\vskip2ex
        \noindent
 where $E$ is the evaluation context of the call-by-value calculus
 (see Def.~\ref{xuc89}).
Understanding the equality to be $=_{\it ccv}$, we can derive this axiom.
If $E$ is the void context $\square$, the equality
 is derivable by reduction
 only, as $(\lambda x.\,x)M\mor{{\it ad}_2}\uplet{(\lambda x.\,x)z}zM
 \mor{\beta_{\lambda}}\uplet{\uplet xxz}zM\mor{\beta_{\it let}}
 \uplet zzM\mor{\eta_{\it let}}M$, where we assume that $M$ is a non-value
 (it is easy if $M$ is a value).
The general case is derived from $(\uplet{E[x]}xM)=_{\it ccv}E[M]$ that is
 verified by induction on the construction of $E$.
\end{exam}

	\vskip0ex

\noindent
Let us consider two typical examples in the following.
We use them as milestones to check the behavior of the calculus
 throughout the paper.

	\vskip3ex

\begin{exam}\rm\label{ozp13}
As the fixed-point combinator suitable for the call-by-value
 calculus, we suggest to using

	\vskip2ex
        \noindent\kern5em
$Y\ =_{\it def}\ \lambda fz.\,D_fD_fz,\qquad D_f\ =\ \lambda xw.\,
 f(\lambda v.\,xxv)w$.

	\vskip2ex
        \noindent
The difference from the ordinary fixed-point combinator \`a la Curry
 is that our $Y$ is fully $\eta$-expanded.
Hasegawa and Kakutani propose three axioms that the
 fixed-point combinator in the call-by-value calculus should satisfy,
 in the simply typed setting \cite{dhx18}.
Among the three, the first two are equational axioms, while
 the last axiom has a denotational feature, stated in a conditional phrase.
So here we consider the first two axioms: the fixed-point axiom
 $YF=\lambda x.\,F(YF)x$, and the stability axiom
 $YF=Y(\lambda yx.\,Fyx)$, in both of which $F$ is supposed to be a value.
We do not mind types here.
The encoded $Y$ above satisfies the fixed-point axiom,
 as $YF=\lambda z.\,D_FD_Fz=\lambda z.\,F(\lambda v.\,D_FD_Fv)z
 =\lambda z.\,F(YF)z$,
 where the equality symbol is understood to mean $=_{\it ccv}$.
Also the stability axiom is satisfied since $D_F=_{\it ccv}D_G$
 holds for $G=\lambda yx.\,Fyx$.
The $\eta$-expansion by $z$ is needed for the fixed-point axiom,
 the expansion by $w$ for the stability axiom, and the expansion
 by $v$ for both.
We comment that Plotkin \cite{pag41} suggests the version where only $xx$
 is $\eta$-expanded.
We return to this example in \ref{iiu81} and \ref{udf62}.
\end{exam}

\begin{exam}\rm\label{fci53}
As an example that uses control operators, we take
 cooperative multitasking.
We consider the case where two agents work in
 a cooperative manner.
The coding below is essentially a simplification
 of the implementation of coroutine
 in \cite[\S17.1]{pea10}.

The following computation demonstrates
 that Felleisen's $\mathcal{C}$-operator
 \cite{krj20} plays the role of the yield command switching live processes.
Let $\hat\tau$ be a fixed continuation variable, representing
 the topmost continuation point.
Let us define

	\vskip2ex
        \noindent\kern5em
$M\mathrel{\vartriangleright^q_r}N\ \ =_{\it def}\ \ \uplet M
 q{\lambda r\mu\delta.\,[\hat\tau]N}$.

	\vskip2ex
        \noindent
We understand $M\mathrel{\vartriangleright^q_r}N$
 to mean that the current agent runs
 $M$ while the other agent is idle waiting for a chance to run $N$
 that is encapsulated in a closure.
Let us say that $M$ is active.

Operator $\mathcal{C}$ is used to insert a break-point
 to switch the active agent.
In the $\lambda\mu$-calculus, we can encode the operator
 as $\mathcal{C}M=\mu k.\,[\hat\tau]M(\lambda x\mu\delta.\,[k]x)$
 where $\delta$ is a dummy continuation variable.
We have the following reduction sequence:

	\vskip2ex
        \noindent
\halign{\kern5em $#$\hfil &${}\ \ \mor*\ \ #$\hfil\cr
 [\hat\tau]\uplet Mq{\mathcal{C}q}\mathrel{\vartriangleright^q_r}N
  & [\hat\tau]\uplet Mq{\mu k.\,[\hat\tau](\lambda r\mu\delta.\,[\hat\tau]N)
     (\lambda q\mu\delta.\,[k]q)}\cr   
  & [\hat\tau](\lambda r\mu\delta.\,[\hat\tau]N)
     (\lambda q\mu\delta.\,[\hat\tau]M)\cr
  & [\hat\tau]N\mathrel{\vartriangleright^r_q}M\cr}

	\vskip2ex
        \noindent
Note that $N$ becomes active.
Two occurrences of $q$ in $q\mathrel{:=}\mathcal{C}q$
 refer distinct variables.
So we may write $\uplet{M\{q'/q\}}{q'}{\mathcal{C}q}$ instead.
But we use the same variable name deliberately.
Intuitively the variable $q$
 is understood to be the channel that passes
 information from one agent to the other.
The variable $r$ is the channel in the reverse direction.

We can interpret the use of $\uplet{M}q{\mathcal{C}q}$
 as inserting a break-point in a process.
Namely $\mathcal{C}q$ plays the role of the yield command
 in the terminology of multitasking.
At the break-point, the current agent pauses
 and the rest of computation $M$ is preserved in the closure
 $\lambda q\mu\delta.\,[\hat\tau]M$.
The right of execution is conceded to the other agent who starts
 $N$.
If a break-point $\uplet{N_0}r{\mathcal{C}r}$
 has been built in $N$, it may be executed eventually.
Then $[\hat\tau]\uplet{N_0}r{\mathcal{C}r}\mathrel{\vartriangleright^r_q}M\mor*
 [\hat\tau]M\mathrel{\vartriangleright^q_r}N_0$.
Namely, the right of execution is returned to the first agent who
 restarts the stopped computation from $M$.

For example,
 ${\sf wall}=Y(\lambda fq.\,(\uplet{fq}q{\mathcal{C}q}))$ is
 a program doing nothing
 whenever invoked and pauses immediately conceding
 the right of execution to the other agent.
Namely $[\hat\tau]\mathop{\sf wall}q\vartriangleright^q_rN=_{\it ccv}
 [\hat\tau]N\vartriangleright^r_q\mathop{\sf wall}q$.

The implementation of multitasking
 is well-behaved under the condition that
 there is at most one occurrence of free $q$ in $M$
 of $M\mathrel{\vartriangleright^q_r}N$.
It is the programmer's task to fulfill this constraint.
In certain cases, it will be useful to use multiple occurrences of $q$.
For example, if we implement a board game where players can
 retract their moves, we may reinstall the same continuation more than once.

Since our calculus has fewer features than the SML programming language,
 we adopt several modifications and simplifications from the original.
Since we consider only two agents, we do not need
 a queue to record the processes waiting for restart.
Our calculus does not have mutable stores.
So we must pass continuation $q$ explicitly in the yield command
 $\mathcal{C}q$.
We return to this example in Rem.~\ref{sdp73} to discuss
 the call-by-value evaluation using evaluation contexts
 and in Rem.~\ref{gia05} for typeability.
\end{exam}

\subsection{CPS translation}

In this subsection, we define the continuation-passing style
 translation of the CCV $\lambda\mu$-calculus and verify its soundness.
It follows the idea of the colon translation to save
 redundant $\beta$-redices \cite{pag41}\cite{ndy68}.
For readability, we use the notation $\lc M\rc[K]$ instead
 of $M:K$.
At this stage, we understand $K$ to be an arbitrary lambda term.
However, as we do later, we can restrict the range of $K$ to the terms of
 particular forms.
Namely, it is a term of sort $K$ given in Def.~\ref{eob58}.
The translation is standard except that it is adjusted
 to the syntax of our calculus.

	\vskip3ex

\begin{defi}\label{vcy79}
We define the CPS translation $[\![M]\!]$.
It depends on an auxiliary translations $\lc M\rc [K]$,
 $\lc J\rc $, and $V^*$.

	\vskip2ex

\halign{\kern5em $#$\hfil &${}\ \mathrel{:=}\ #$\hfil\cr
 \lc V\rc[K] & KV^*\cr
 \lc V_1V_2\rc[K] & V_1^*V_2^*K\cr
 \lc VN\rc [K] & \lc N\rc [\lambda y.\,V^*yK]\cr
 \lc NV\rc [K] & \lc N\rc [\lambda x.\,xV^*K]\cr
 \lc N_1N_2\rc [K] & \lc N_1\rc [\lambda x.\,\lc N_2\rc [
  \lambda y.\,xyK]]\cr
 \lc \uplet LxM\rc [K] & \lc M\rc [\lambda x.\,\lc L\rc [K]]\cr
 \lc \mu k.\,J\rc [K] & (\lambda k.\,\lc J\rc )K\cr
 \lc [k]M\rc  & \lc M\rc [k]\cr
 \lc \uplet JxM\rc  & \lc M\rc [\lambda x.\,\lc J\rc ]\cr
 \noalign{\vskip2ex}
 x^* & x\cr
 (\lambda x.\,M)^* & \lambda xk.\,\lc M\rc [k]\cr
 \noalign{\vskip2ex}
 [\![M]\!] & \lambda k.\,\lc M\rc [k]\cr
}

	\vskip2ex
	\noindent
Herein $V$ is a value and $N$ a non-value.
The ordinary variables $x$ and $y$ for three cases of application
 are fresh, and the continuation variables $k$ in $(\lambda x.\,M)^*$
 and $[\![M]\!]$ are fresh.
Other occurrences of variables $x$ and $k$ share common symbols
 in the source language and the target language.
For example, $[\![(\uplet{xy}x{\mu h.\,[h]y})y]\!]=
\lambda k.\,(\lambda h.\,hy)(\lambda x.\,xy(\lambda z.\,zyk))$ where
 $z$ is a fresh variable chosen to translate the outermost application.

In case $\lc\uplet LxM\rc[K]$, if $K$ contains $x$ as a free variable,
 we must perform $\alpha$-conversion as
 $\lc M\rc[\lambda z.\,\lc L\{z/
	\penalty-3000
x\}\rc [K]]$ using fresh $z$.
This is a slippery point.
The careful reader will find this observation used in
 the proof of Prop.~\ref{fgv86}.
\end{defi}

	\vskip0ex

\noindent
Now the setup is finished.
We begin with verifying the
 soundness of the CPS translation with respect to
 the CCV equality in the source and the $\beta\eta$-equality
 in the target.

	\vskip0ex

\begin{lem}\label{ghv97}\hfill
\shrinktopsep
\begin{enumerate}[(1)]
\item
We have $\lc M\{V/x\}\rc [K]=\lc M\rc[K]\{V^*/x\}$ provided
 $x\not\in K$.
\item
We have $\lc J\{[k]\square\mapsto [k]\uplet Mx\square\}\rc=
 \lc J\rc\{k\mapsto \lambda x.\,\lc M\rc [k]\}$.
\end{enumerate}
\end{lem}

\proof\hskip.5em
Easy by compositionality of the colon translation, Def.~\ref{vcy79}
\qed

	\vskip0ex

\noindent
The following is the soundness of the CPS translation,
 first verified for the lambda calculus by Plotkin~\cite[Thm.~3, p.~148]{pag41}.

	\vskip0ex

\begin{prop}\label{fgv86}
If $L=_{\it ccv}M$, then $[\![L]\!]=[\![M]\!]$ holds with respect
 to $\beta\eta$-equality.
\end{prop}

\proof\hskip.5em
We verify $\lc L\rc [K]=\lc M\rc [K]$.
Among the axioms in Def.~\ref{yoq54}, both sides yield the
 same lambda terms, save the second axiom.
For this axiom, the translations of both sides contract
 to a common term
 by $\beta$-reduction (no need of $\eta$).

Next we consider reduction rules $L\rightarrow M$ in Def.~\ref{ofe40}.
For two administrative rules, both sides translate into the same terms.
For rules $\beta_{\it let}$, $\beta_{\it jmp}$, $\eta_{\mu}$,
 $\eta_\lambda$, and $\eta_{\it let}$, we have
 $\lc L\rc [K]\shortmor* \lc M\rc [K]$ by $\beta$-reduction or $\eta$-reduction.
For the remaining two $\beta_{\lambda}$ and $\beta_{\mu}$,
 we need $\beta$-expansion as well as
 $\beta$-reduction.
For example, the left-hand side of $\beta_{\mu}$ is
 translated into
 $(\lambda k.\,\lc J\rc )(\lambda x.\,\lc M\rc [K])$ while
 the right hand side into $(\lambda k.\,\lc J\rc \{k
 \mapsto \lambda x.\,\lc M\rc [k]\})K$ by Lem.~\ref{ghv97}, (2).
The former $\beta$-equals to the latter by reduction followed by expansion.
\qed

\begin{rema}\label{bqb30}\hfill
\shrinktopsep
\begin{enumerate}[(1)]
\advance\parindent by15pt
\item
A subtle problem is that the equality determined by Def.~\ref{yoq54}
 is not strictly respected.
As mentioned in the proof of Prop.~\ref{fgv86}, $[\![\uplet{(\mu k.\,J)}xM]\!]$
 and $[\![\mu k.\,(\uplet JxM)]\!]$ are not exactly identical,
 only $\beta$-equal.
Hence $[\![M]\!]$ is determined only up to $\beta$-equality.
It may cause serious trouble in certain cases.
Fortunately, the ambiguity does not injure
 the proof of the main theorem~\ref{gbu84}
 and its applications presented in this paper.
It is because we are involved only in
 the properties of the target calculus that are $\beta$-invariant.
In the second of our series of papers \cite{ycq31}, we deal with
 strong termination, for which the equality becomes hazardous
 since the strong termination is not closed
 under the $\beta$-equality.

We stress that only the second rule
 is problematic.
The remaining two rules are strictly respected by the CPS translation.

\item
Suppose $L\shortmor*M$.
By inspection of the proof of \ref{fgv86}, we see that
 $\beta\eta$-reduction and $\beta$-expansion are needed
 to obtain $\lc M\rc [K]$ from $\lc L\rc [K]$.
We emphasize that $\eta$-expansion is not needed.
This observation is used in Thm.~\ref{ehw91}
 later.
\end{enumerate}
\end{rema}

\subsection{Inverse translation}

Following the idea of Sabry and Felleisen \cite{yrs06}\cite{vyh68}, we define the
 inverse of the CPS translation.
Up to equational theories, the inverse translation actually gives
 the inverse.
We prove a sharpened result taking into account the orientation
 of reduction in Thm.~\ref{gbu84}.

Until now, the target calculus of the CPS translation is
 the ordinary untyped lambda calculus.
To define the inverse translation, however, we regard the target
 as a sorted lambda calculus as in \cite{vyh68}\cite{yrs06}.

We prepare four sorts $T,Q,W$, and $K$.
They are called terms, jumps, values, and continuations, respectively.
Moreover, the variables are split into
 the variables $x$ of sort $W$ and
 the variables $k$ of sort $K$.
The former are called ordinary variables and the latter continuation variables.
We let $T$ itself represent the lambda terms of sort $T$,
 and likewise for other sorts.

	\vskip0ex

\begin{defi}\label{eob58}
The syntax of the target calculus is defined as follows:

	\vskip2ex

\halign{\kern5em #\kern3em\hfil &$#$\hfil &${}\ \mathrel{::=}\ #$\hfil\cr
 Term& T & \lambda k.\,Q\ \ |\ \ WW\cr
 Jump& Q & KW\ \ |\ \ TK\cr
 Value& W & x\ \ |\ \ \lambda x.\,T\cr
 Continuation& K & k\ \ |\ \ \lambda x.\,Q\cr}

	\vskip2ex
        \noindent
 where $x$ ranges over ordinary variables and $k$ over continuation variables.
We take ordinary $\beta\eta$-reduction as reduction rules.
We comment that each sort is closed under $\beta\eta$-reduction as is
 straightforward by definition.
\end{defi}

	\vskip0ex

\noindent
It is easy to see that, according to the translations of
 Def.~\ref{vcy79}, $[\![M]\!]$ belongs to sort $T$, both
 $\lc M\rc[K]$ and $\lc J\rc$ to sort $Q$, and $V^*$ to sort $W$.

	\vskip0ex

\begin{defi}\label{dfn37}
The {\it inverse translation} $(\hbox{-})^{-1}$ is
 a mapping from the target calculus
 to the CCV $\lambda\mu$-calculus.
The definition is given by induction:

	\vskip2ex

\halign{\kern5em $#$\kern3em\hfil &$#^{-1}$\hfil &${}\ \mathrel{:=}\ #$\hfil\cr
 T^{-1}: & (\lambda k.\,Q) & \mu k.\,Q^{-1}\cr
 & (W_1W_2) & W_1^{-1}W_2^{-1}\cr
 \noalign{\vskip1ex}
 Q^{-1}: & (KW) & K^{-1}[W^{-1}]\cr
 & (TK) & K^{-1}[T^{-1}]\cr
 \noalign{\vskip1ex}
 W^{-1}: & x & x\cr
 & (\lambda x.\,T) & \lambda x.\,T^{-1}\cr
 \noalign{\vskip1ex}
 K^{-1}: & k & [k]\square\cr
 & (\lambda x.\,Q) & \uplet{Q^{-1}}x\square.\cr
}

	\vskip2ex
        \noindent
$T^{-1}$ and $W^{-1}$ are terms of the CCV
 $\lambda\mu$-calculus, $Q^{-1}$ jumps, and $K^{-1}$ jumps
 with a single hole $\square$.
The notation $K^{-1}[M]$ signifies to fill the hole $\square$ of $K^{-1}$
 with term $M$.
\end{defi}

\begin{rema}\label{gyv73}
In view of Def.~\ref{dfn37}, it is clear that
 application appearing in consequence of the translation
 is restricted to the shape $VV'$
 between values.
Hence we cannot apply administrative reductions to
 $T^{-1}$ etc.
Moreover, once the application is restricted so, reduction steps
 preserve the shape since they send values to values (note
 that variables are able to be substituted only with values).
So administrative rules can never be applied later on. 
\end{rema}

\begin{defi}\label{bme70}
Reduction by a non-administrative rule is
 called {\it practical}.
For technical reasons, we call reduction by rule
 $(\eta_\mu):\mu k.\,[k]M\rightarrow M$ {\it vertical}.
\end{defi}

\begin{lem}\label{uai79}
Vertical reductions are strongly terminating and confluent.
\end{lem}

\proof\hskip.5em
Easy.
\qed

\begin{lem}\label{mus37}
Suppose $L'\leftmor{V}L\mor{P}M$ holds where $V$ denotes
 a one-step vertical reduction
 and $P$ a one-step practical reduction.
There is a term $M'$ such that

	\vskip2ex
        \noindent\kern5em
\squarediagabs{50}{40}%
 LM{L'}{M'}{P}{V}{V^*}{P?}

	\vskip2ex
        \noindent
 where the bottom $P?$ denotes a one-step practical
 reduction or an equality
 and the right $V^*$ a finite
 number of vertical reductions (different from the usage in
 Def.~\ref{vcy79}).
\end{lem}

\proof\kern.5em
There are only three patterns of overlapping redices yielding critical
 pairs: $\uplet Lx{\mu k.\,[k]M}$
 and $[l]\mu k.\,[k]M$ where $k\not\in M$ in either,
 and $\mu k.\,[k]\mu l.\,J$ where $k\not\in \mu l.\,J$.
Proof is easy.
\qed

	\vskip0ex

\noindent
Note that Lem.~\ref{mus37} excludes administrative reductions.
Indeed it fails, since vertical reductions may create values.

	\vskip0ex

\begin{cor}\label{tvn51}
Suppose $L'\leftmor{V^*}L\mor{P^*}M$ where $V^*$ denotes a finite
 number of vertical reductions and $P^*$ a finite
 sequence of practical reductions.
There is a term $M'$ such that

	\vskip2ex
        \noindent\kern5em
\squarediagabs{50}{40}%
 LM{L'}{M'}{P^*}{V^*}{V^*}{P^*}.

	\vskip2ex
        \noindent
\end{cor}

\proof\hskip.5em
We stack the square diagram of Lem.~\ref{mus37} vertically to
 replace $V$ with $V^*$, then horizontally to replace $P$ with
 $P^*$.
(Our specific syntactic design
 notably simplifies the proof here.
We do not have
 to be bothered by transformations between equivalent terms,
 as we can identify terms only by ignoring brackets.
It enables
 verification by simply laying square bricks.)
\qed

	\vskip0ex

\noindent
Corollary \ref{tvn51} is used later in Thm.~\ref{gbu84}.
Next, let us prove (a half of) that
 the inverse translation is actually the
 inverse of the CPS.

	\vskip0ex

\begin{lem}\label{wcg04}
Provided $x\not\in K$,
 we have $K^{-1}[\uplet LxM]=\uplet{K^{-1}[L]}xM$ by
 the equality rules of Def.~\ref{yoq54} (but with
 no use of the second rule interchanging $\mu$ and let; see \ref{xkd18}).
\end{lem}

\proof\hskip.5em
Easy.
\qed

\begin{prop}\label{ffa70}
Let $M$ be a term of the CCV $\lambda\mu$-calculus.
There is a term $M^\dagger$ of the calculus such that
 $M\mor{A^*}M^\dagger\leftmor{V^*}[\![M]\!]^{-1}$ (notice the direction).
Here $A^*$ denotes a finite number of administrative reductions
 and $V^*$ a finite number of vertical reductions.
As $[\![M]\!]$ is determined only
 up to $\beta$-equality (Rem.~\ref{bqb30}),
 we choose an arbitrary one among those obtained by bracketing between $\mu$
 and let.
\end{prop}

\proof\kern.5em
For terms of the form of application define we $M^\dagger$ as follows:

	\vskip2ex

\halign{\kern5em $(#)^\dagger$\hfil &${}\ \mathrel{:=}\ #$\hfil\cr
 V_1V_2 & V_1^\dagger V_2^\dagger\cr
 VN & \uplet{V^\dagger y}yN^\dagger\cr
 NV & \uplet{xV^\dagger}xN^\dagger\cr
 N_1N_2 & \uplet{xy}y{\uplet{\smash{N_2^\dagger}}x{\smash{N_1^\dagger}}}\cr
}

	\vskip2ex
        \noindent
 where $V$ denotes a value and $N$ a non-value.
For terms of other forms and for jumps, definition of $M^\dagger$ (and
 $J^\dagger$) is simply homomorphic.
For example, $(\lambda x.\,M)^\dagger\mathrel{:=}\lambda x.\,M^\dagger$.
By definition, $M\mor{A^*}M^\dagger$ is evident.
To verify $[\![M]\!]^{-1}\mor{V^*}M^\dagger$, we show
 $(\lc M\rc [K])^{-1}\mor{V^*}K^{-1}[M^\dagger]$.
Proof is straightforward.
To manipulate the case $M=N_1N_2$, we need Lem.~\ref{wcg04}.
Since the lemma does not depend on the equality
 interchanging $\mu$ and let-binding, $[\![M]\!]^{-1}
 \mor{V^*}M^\dagger$ does not change the chosen bracketing structure.
This is why we can choose arbitrary bracketing for $[\![M]\!]$.
\qed

\begin{cor}\label{iyy02}
$[\![M]\!]^{-1}=_{\it ccv}M$ holds for every CCV $\lambda\mu$-term $M$.
\qed
\end{cor}

\begin{rema}\label{vba20}\hfill
\shrinktopsep
\begin{enumerate}[(1)]
\item
The term $[\![M]\!]^{-1}$ is free from administrative redices by
 Rem.~\ref{gyv73}.
Namely, it is fully let-expanded.
This property
 corresponds to the main result related to A-normal forms
 in Flanagan et al. \cite[\S4]{wmx35}.
In that paper, $(kW)^{-1}$ is defined as $W^{-1}$, the
 variable $k$ discarded.
In our definition, $(kW)^{-1}=[k]W^{-1}$.
Hence we need vertical reductions to collapse $\mu k.\,[k]$.

\item
Closely related (and often better) results are proved for
 different systems in previous works.
For example, Sabry and Felleisen
 verify $[\![M]\!]^{-1}\mor*M$ for the call-by-value calculus with no
 control operators \cite[Thm.~10, p.~310]{yrs06}.
Fujita verifies $[\![M]\!]^{-1}
 \equiv M$ for the call-by-name second-order calculus
 \cite[Thm.~32, p.~204]{eiq93}\cite[Cor.~17, p.~329]{pjd29}.
\end{enumerate}
\end{rema}

\subsection{Completeness}

By virtue of the inverse translation, we can verify the completeness
 of the CPS translation.
The completeness found in the previous works
 is concerned with equalities.
Namely, two terms are equal in the source language
 if their translations are equal in the target.
This type of completeness is given in Cor.~\ref{rcr59}.
However, we verify a stronger result (Thm.~\ref{gbu84})
 respecting reduction, not only equality.
The result in this subsection is main vehicles
 in the rest of this paper.
The $\beta$-redices of the target calculus are manipulated
 in Lem.~\ref{qca10} and \ref{aom70}, and the $\eta$-redices
 in Lem.~\ref{cai90}.

	\vskip0ex

\begin{lem}\label{qca10}
We have $((\lambda k.\,Q)K)^{-1}\mor+
 (Q\{K/k\})^{-1}$.
\end{lem}

\proof\hskip.5em
First we observe that
 $(Q\{\lambda x.\,Q_0/k\})^{-1}$
 is syntactically equal to $Q^{-1}\{[k]\square
 \mapsto \uplet{Q_0^{-1}}x\square\}$.
Proof of the lemma is by case splitting.
The case of $K=k$ is easy.
If $K=\lambda x.\,Q_0$, the left-hand side equals
 $\uplet{Q_0^{-1}}x{\mu k.\,Q^{-1}}$, which contracts to
 $Q^{-1}\{[k]\square\mapsto \uplet{Q_0^{-1}}x\square\}$
 as noted in Rem.~\ref{yxu20}, (3).
Now we use the observation above.
In this proof, we use
 rule $(\beta_\mu)$ and rule $(\beta_{\it jmp})$
 of Def~\ref{ofe40}.
\qed

\begin{lem}\label{aom70}\hfill
\shrinktopsep
\begin{enumerate}[(1)]
\item
We have $((\lambda x.\,T)W)^{-1}\mor+(T\{W/x\})^{-1}$.

\item
We have $((\lambda x.\,Q)W)^{-1}\mor+(Q\{W/x\})^{-1}$.
\end{enumerate}
\end{lem}

\proof\hskip.5em
We show (1).
The left- hand side equals $(\lambda x.\,T^{-1})W^{-1}$, which
 contracts to $\uplet{T^{-1}}x{W^{-1}}$, then to $T^{-1}\{W^{-1}/x\}$
 since $W^{-1}$ is a value.
It is easy to show that $T^{-1}\{W^{-1}/x\}$ is identical
 to $(T\{W/x\})^{-1}$.
(2) is similar.
In this proof, we use rule $(\beta_{\it let})$ and rule
 $(\beta_\lambda)$ of Def.~\ref{ofe40}.
\qed

\begin{lem}\label{cai90}\hfill
\shrinktopsep
\begin{enumerate}[(1)]
\item
We have $(\lambda k.\,Tk)^{-1}\mor+T^{-1}$.

\item
We have $(\lambda x.\,Wx)^{-1}\mor+W^{-1}$.

\item
We have $(\lambda x.\,Kx)^{-1}\mor+K^{-1}$.
\end{enumerate}
\end{lem}

\proof\hskip.5em
(1) is an immediate consequence of rule ($\eta_\mu$) in
 Def.~\ref{ofe40}, and (2) of rule ($\eta_\lambda$).
We show (3).
We split cases by the shape of $K$.
The case of $K=k$ is easy.
If $K=\lambda y.\,Q$, the left side equals $\uplet{\uplet{Q^{-1}}yx}x\square$,
 which contracts to $\uplet{Q^{-1}}y\square$, i.e., $(\lambda y.\,Q)^{-1}$.
In both cases, we use rule ($\eta_{\it let}$) of Def.~\ref{ofe40}.
We note that the equality rules of Def.~\ref{yoq54} are
 implicitly used in the proof of (3).
However, the interchange law of $\mu$ and let is not used.
See \ref{xkd18}.
\qed

\begin{prop}\label{lzv28}
We have $P^{-1}\mor{P^+}Q^{-1}$ in the CCV $\lambda\mu$-calculus,
 whenever $P\rightarrow Q$ holds with respect to $\beta\eta$-reduction in the
 target calculus.
Here $P$ and $Q$ represent lambda terms of all sorts
 in the target calculus, and $\mor{P^+}$ denotes one or more
 practical reductions.
\end{prop}

\proof\hskip.5em
It is a consequence of Lem.~\ref{qca10}, \ref{aom70}, and \ref{cai90}.
No administrative reductions are used, as pointed out in the proofs
 of the lemmata.
\qed

	\vskip0ex

\noindent
Similar results are proved by Sabry and Felleisen
 \cite[Thm.~13, p.~312]{yrs06}\cite[Thm.~3.11, p.~45]{vyh68}
 and by Fujita \cite[Prop.~27, p.~203]{eiq93}\cite[Lem.~18,
 p.~330]{pjd29} for different systems
 with no control operators.

	\vskip0ex

\begin{cor}\label{rcr59}
If $[\![M_1]\!]=[\![M_2]\!]$ holds up to $\beta\eta$-equality in
 the target calculus, then $M_1=_{\it ccv}M_2$ holds.
\end{cor}

\proof\hskip.5em
Applying Prop.~\ref{lzv28} to a zigzag of $\beta\eta$-reductions, we
 have $[\![M_1]\!]^{-1}=_{\it ccv}[\![M_2]\!]^{-1}$.
We have also $M_i=_{\it ccv}[\![M_i]\!]^{-1}$ by Cor.~\ref{iyy02}.
\qed

	\vskip0ex

\noindent
Corollary \ref{rcr59} refers only to equality.
We show sharpened completeness with regard to reductions
 in Thm.~\ref{gbu84}.

	\vskip0ex

\begin{defi}\label{ier42}
We let $M^\downarrow$ denote the normal form with respect to vertical
 reductions for each term $M$ of the CCV $\lambda\mu$-calculus.
The existence of the normal form comes from Lem.~\ref{uai79}.
\end{defi}

	\vskip0ex

\noindent
Before stating the theorem, we recall ambiguity in the definition
 of the CPS translation $[\![M]\!]$ discussed in Rem.~\ref{bqb30}, (i).
If we choose different bracketing between $\mu$ and let, we
 obtain different $[\![M]\!]$.

	\vskip0ex

\begin{thm}\label{gbu84}
Let $M$ be a term of the CCV $\lambda\mu$-calculus.
If $[\![M]\!]\shortmor*N$ with respect to $\beta\eta$-reduction
 in the target calculus, $M\shortmor*(N^{-1})^\downarrow$ holds
 in the CCV $\lambda\mu$-calculus.
Here we understand $[\![M]\!]$ to mean arbitrarily chosen one
 among those obtained by bracketing between $\mu$ and let.
\end{thm}

\proof\hskip.5em
By Prop.~\ref{lzv28} and \ref{tvn51}, we have $M'$ satisfying

	\vskip2ex
        \noindent\kern5em
\squarediagabs{60}{40}
 {[\![M]\!]^{-1}}{N^{-1}}{M^\dagger}{M'}%
 {P^*}{V^*}{V^*}{P^*}

	\vskip2ex
        \noindent
 where $M^\dagger$ is introduced in Prop.~\ref{ffa70}
 that is valid irrelevant of the choice of bracketing.
We can replace $M'$ with $(N^{-1})^\downarrow$, performing
 further vertical reductions if needed.
Thence we have $M\mor{*}M^\dagger\mor{*}
 (N^{-1})^\downarrow$.
\qed

	\vskip0ex

\noindent
Similar results are verified by Sabry and Wadler
 for Moggi's $\lambda_c$-calculus \cite[Thm~8.2, p.~130]{ndy68}, and
 by Fujita for the second-order call-by-name
 $\lambda\mu$-calculus \cite[Thm.~1, p.~330]{pjd29}.

	\vskip0ex

\begin{rema}\label{xkd18}
The equality axioms in Def.~\ref{yoq54}
 are the key for Thm.~\ref{gbu84}.
So let us analyze where the axioms are used throughout
 the proof.
They are used in Prop.~\ref{ffa70} (through Lem.~\ref{wcg04}),
 Lem.~\ref{qca10} (through Rem.~\ref{yxu20}, (3)), and
 Lem.~\ref{cai90}.
As commented in these parts,
 only the first and the third axioms are used
 among the three axioms of \ref{yoq54}.
The second axiom interchanging $\mu$ and let is never used.

This is not strange, since the second axiom is derivable from others
 if the equality is our concern, the orientation of the reduction
 being ignored.
Let us assume that the interchange axiom between
 jumper and let is valid.
Then we derive the second axiom
 as $\mu k.\,(\uplet JzM)=\mu k\,(\uplet{([k](\mu k.\,J))}zM)
 =\mu k.\,([k](\uplet{(\mu k.\,J)}zM))=\uplet{(\mu k.\,J)}zM$
 where three equalities are the reverse of $\beta_{\it jmp}$, the
 interchange law of jumper and let, and $\eta_\mu$, respectively.

As an alternative choice, therefore, we may take
 the reduction rule

	\vskip2ex
        \noindent\kern5em
$\uplet{(\mu k.\,J)}xM
 \quad\rightarrow\quad\mu k.\,(\uplet JxM)$

	\vskip2ex
        \noindent
 in place of the equality axiom.
Here we assume that $k$ is not free in $M$.
We do $\alpha$-renaming otherwise.
In this paper, we stick to the equality form,
 because it is difficult for humans to
 distinguish the things that look similar by
 tracking brackets carefully.
\end{rema}

\section{Proof by Parasitism}

The rest of the paper is mostly devoted to the applications of the sharpened
 completeness theorem \ref{gbu84}.
It succeeds in dealing with reduction, not only with equality.

The target calculus of the CPS translation is just the ordinary
 lambda calculus except that it is equipped with sorts.
We have a stock of the results on the lambda calculus
 through a long history of research.
By the sharpened completeness, we can reflect the properties of the lambda
 calculus to the CCV $\lambda\mu$-calculus.

We coin the term ``proof by parasitism'' to describe
 the technique to obtain syntactic properties through
 completeness.
A parasite (the source calculus) infests a host (the target calculus)
 to suck the nurture from the latter.
The nurture in this case is a desired syntactic property.
Through completeness, the source calculus can take over
 the target calculus and make it work hard to provide for him.
Fruits are collected by the host and the free-loader robs
 them at the lowest cost.

In this section, we demonstrate
 several samples of the results that use the proof by parasitism.
We do not claim that this type of argument is original at all.
For instance, de~Groote \cite[Prop.~3.2,Lem.~4.2]{owr18},
 Sabry and Wadler \cite[Prop.~7.1.2, p.~126]{ndy68}
 Crolard \cite[Thm.~3.5.1]{ltb70},
 and Fujita \cite[Cor.~34,35]{eiq93}\cite[Cor.~19]{pjd29}
 use the same approaches to
 prove some of their results.
Here we deploy the outfit systematically.

\subsection{Proof by parasitism, I: Church-Rosser property}

We start with the Church-Rosser property.
Its proof is a typical example of an advantage of
 the technique.

	\vskip0ex

\begin{thm}\label{pje26}
The CCV $\lambda\mu$-calculus enjoys the Church-Rosser property.
\end{thm}

\proof\hskip.5em
Suppose $M_1=_{\it ccv}M_2$.
By soundness~\ref{fgv86} concerning equality, we have $[\![M_1]\!]
=[\![M_2]\!]$ up to $\beta\eta$-equality.
Since the lambda calculus fulfills Church-Rosser
 with respect to $\beta\eta$-reduction (thus
 the target calculus does) \cite[Thm.~3.3.9]{zxr70},
 there is a term $N$ such that $[\![M_1]\!]\mor*N\leftmor*[\![M_2]\!]$
 in the target calculus.
Now Thm.~\ref{gbu84} yields $M_1\mor*(N^{-1})^\downarrow\leftmor*M_2$.
\qed

\begin{rema}\label{kgy11}
The fragment of our system with no $\mu$-operators is
 essentially the same as Moggi's $\lambda_c$-calculus \cite{dvp96}.
If we identify $\uplet MxN$ with ${\sf let}\ x=N\ {\sf in}\ M$,
 the reduction rules are quite similar, except that the
 $\lambda_c$-calculus has the let-flat rule for associativity of
 the let-construct, while our calculus has it as one of the
 equality rules \ref{yoq54}.
This type of relation of the complete call-by-value calculus
 to the $\lambda_c$-calculus is first pointed out in \cite{vyh68}\cite{yrs06}.

With respect to equational theories,
 the CCV $\lambda\mu$-calculus is a conservative extension
 of the $\lambda_c$-calculus, up to the identification of the let-constructs.
It is a consequence of the Church-Rosser property \ref{pje26}.
Although the $\lambda_c$-calculus is defined as a typed system,
 let us ignore the typing here.
Let $=_{\lambda_c}$ denote the equational theory generated from
 the reduction rules in Def.~6.2 of \cite{dvp96}.
We show that $L=_{\lambda_c}M$ is equivalent to $L=_{\it ccv}M$.
By the Church-Rosser, $L=_{\it ccv}M$
 is derivable only by reduction from $L$ and $M$.
Since $L$ and $M$ are supposed to have no $\mu$ or jumpers,
 the reduction paths use the rules for $\lambda$, applications,
 and let only.
Therefore $L=_{\lambda_c}M$ holds.
The converse is trivial.
\end{rema}

\subsection{Proof by parasitism, II: Weak normalization}

In this subsection, we verify that $M$ is normalizable iff $[\![M]\!]$
 is normalizable.
First, we give a partial characterization of the normal forms
 in the CCV $\lambda\mu$-calculus.
It is said to be partial,
 since it does not exclude the redices of the $\eta$-type in
 Def.~\ref{ofe40}, that is, $\lambda x.\,Vx$, $\uplet xxM$,
 or $\mu k.\,[k]M$.

	\vskip0ex

\begin{defi}\label{siw66}
A term or a jump in the CCV $\lambda\mu$-calculus is {\it quasi-normal}
 if it contains none of the first six redices
 in Def.~\ref{ofe40} (viz., only $\eta$-type redices
 are allowed).
If a term or a jump contracts to a quasi-normal form with no use
 of the $\eta$-type rules, we say that it is {\it quasi-normalizable}.
\end{defi}

	\vskip0ex

\noindent
A quasi-normal form may contain a redex
 $\lambda x.\,Vx$ of rule $\eta_\lambda$.
However, if the value $V$ is a lambda abstraction
 $\lambda y.\,L$, the subterm $Vx=(\lambda y.\,L)x$ is
 a forbidden $\beta$-type redex.
Hence only the form $\lambda x.\,yx$ is allowed.

	\vskip0ex

\begin{defi}\label{ayb92}
We define four classes $\eufm M,\eufm B,\eufm N$, and
 $\eufm V$ of CCV $\lambda\mu$-terms, and a class $\eufm H$
 of jumps by the following syntax:

	\vskip2ex

\halign{\kern5em $#$\hfil &${}\ \ \mathrel{::=}\ \ #$\hfil\cr
 \eufm M & \eufm B\ \ \ |\ \ \ \mu k.\,\eufm H\cr
 \eufm H & [k]\eufm B\cr
 \eufm B & \eufm V\ \ \ |\ \ \ \eufm N\ \ \ |\ \ \ \uplet{\eufm B}x{\eufm N}\cr
 \eufm N & x\eufm V\cr
 \eufm V & x\ \ \ |\ \ \ \lambda x.\,\eufm M\cr
}

	\vskip2ex
        \noindent
 where $x$ ranges over ordinary variables and $k$ continuation variables.
Here we assume the bracketing convention in Rem.~\ref{bmv07}.
\end{defi}

\begin{rema}\label{fyl89}\hfill
\shrinktopsep
\begin{enumerate}[(1)]
\advance\parindent by15pt
\item
It is easy to see that the terms and jumps defined by the syntax
 of \ref{ayb92} are quasi-normal.

\item
Conversely, quasi-normal terms are correctly
 characterized by Def.~\ref{ayb92}.
To this end, we adopt the canonical bracketing rules in Rem.~\ref{bmv07}.
Between jumper and let, we suppose that $[k](\uplet LxM)$ is canonical.

If a quasi-normal term has the form of application $M_1M_2$, both
 $M_i$ must be values so that it does not form a redex of
 the administrative rules.
Moreover, value $M_1$ must be a variable for otherwise it is
 a lambda-abstraction, forming a $\beta_\lambda$-redex $(\lambda x.\,M)V$.
Hence the term must be of the form $\eufm N=x\eufm V$ where $\eufm V$ is
 a quasi-normal value. 
If a quasi-normal term is $\uplet LxM$, then $M$ is
 neither a value nor a $\mu$-term.
Hence $M$ must be an application $\eufm N$.
There is no restriction to the form of $L$.
However, by canonical bracketing rules, we can assume that $L$ is
 not a $\mu$-term.
Hence the term must be of the form $\uplet{\eufm B}x{\eufm N}$.
For a quasi-normal jump $[k]M$, the term $M$ does not start with
 $\mu$.
Hence it is of the form $[k]\eufm B$.
The other cases are similar.
\end{enumerate}
\end{rema}

\begin{lem}\label{hqc54}
A term or a jump of the CCV $\lambda\mu$-calculus is normalizable
 if it is quasi-normalizable.
\end{lem}

\proof\hskip.5em
It suffices to show that quasi-normal terms reduce to quasi-normal forms.
Straightforward.
\qed

\begin{defi}\label{bsy53}
The $\beta$-normal forms in the target calculus
 are characterized by the following:

	\vskip2ex

\halign{\kern5em $#$\hfil &${}\ \mathrel{::=}\ #$\hfil\cr
 T_{\it NF} & \lambda k.\,Q_{\it NF}\ \ |\ \ xW_{\it NF}\cr
 Q_{\it NF} & xW_{\it NF}K_{\it NF}\ \ |\ \ kW_{\it NF}\cr
 W_{\it NF} & x\ \ |\ \ \lambda x.\,T_{\it NF}\cr
 K_{\it NF} & k\ \ |\ \ \lambda x.\,Q_{\it NF}.\cr
}

	\vskip2ex
        \noindent
We remark that $\beta$-normalizability in the lambda calculus (thus
 in the target calculus) is equivalent to $\beta\eta$-normalizability
 \cite[Cor.~15.1.5]{zxr70}.
\end{defi}

	\vskip0ex

\noindent
Now we verify that $M$ is normalizable iff $[\![M]\!]$ is normalizable.
First, we prove the if-part, and then the only-if part.
Verification is done by a simple comparison of quasi-normal forms and
 $\beta$-normal forms.

Following the convention of the formal language
 theory, we use the name of a class also as a meta-symbol
 signifying an element of the class.
We also make rough use of equality.
For example, $T_{\it NF}^{-1}=\eufm M$ in the proof of
 the next lemma means that, for every
 term $T$ in the class $T_{\it NF}$, its inverse $T^{-1}$ is
 equal to a term in $\eufm M$.

	\vskip0ex

\begin{lem}\label{jct82}
If $N$ is a $\beta$-normal term in the target calculus, $N^{-1}$
 is quasi-normal.
\end{lem}

\proof\hskip.5em
We verify that $T_{\it NF}^{-1}=\eufm M$, $Q_{\it NF}^{-1}=\eufm H$,
 $W_{\it NF}^{-1}=\eufm V$, and $K_{\it NF}^{-1}[\eufm N]=\eufm H$
 by induction on the construction of the grammar in Def.~\ref{bsy53}.
If $T_{\it NF}=xW_{\it NF}$, then $(xW_{\it NF})^{-1}=xW_{\it NF}^{-1}=
 x\eufm V$ by induction hypothesis.
Thence $(xW_{\it NF})^{-1}=x\eufm V=\eufm N=\eufm B=\eufm M$.
The other cases are similar.
\qed

\begin{prop}\label{jno24}
Let $M$ be a term of the CCV $\lambda\mu$-calculus.
If $[\![M]\!]$ is normalizable, $M$ is normalizable.
\end{prop}

\proof\hskip.5em
Let us take a reduction sequence $[\![M]\!]\shortmor*N$ where $N$
 is $\beta$-normal in the target calculus.
Theorem~\ref{gbu84} implies $M\shortmor*(N^{-1})^\downarrow$ in
 the source calculus.
By Lem.~\ref{jct82}, $N^{-1}$ is quasi-normal.
Thus, by Lem.~\ref{hqc54}, $(N^{-1})^\downarrow$ is normal.
\qed

\begin{prop}\label{wdw96}
If $M$ is a normalizable CCV $\lambda\mu$-term,
 $[\![M]\!]$ is normalizable
\end{prop}

\proof\hskip.5em
If $M_1=_{\it ccv}M_2$ then $[\![M_1]\!]=[\![M_2]\!]$ by the equational
 soundness~\ref{fgv86}.
Since the lambda calculus satisfies the Church-Rosser property,
 normalizability of $[\![M_i]\!]$ implies that of the other.
Hence, from the beginning, we may assume that $M$ is normal.
We show
 that $\lambda k.\,\lc M\rc [k]$ is normalizable in case $M=\eufm M$.
To this end, we verify $\lc \eufm M\rc [k]$, $\lc \eufm H\rc $,
 $\lc \eufm B\rc [k]$, and $\lc \eufm N\rc [K_{\it NF}]$
 contract to $Q_{\it NF}$ by $\beta$-reductions.
Moreover, we show $\eufm V^*$ contracts to $W_{\it NF}$.
The proof is by simultaneous induction on the construction
 of Def.~\ref{ayb92} (up to the renaming of continuation
 variables).
If $\eufm M=\mu l.\,\eufm H$, then $\lc \mu l.\eufm H\rc [k]
 =(\lambda l.\,\lc \eufm H\rc )k\rightarrow
 \lc \eufm H\{k/l\}\rc $ by $\beta$-reduction.
This is the only part we need $\beta$-reduction essentially.
The rest are easy.
We take the case of $\eufm B$.
If $\eufm B=\eufm V$, then $\lc \eufm V\rc [k]=k\eufm V^*=kW_{\it NF}
 =Q_{\it NF}$.
If $\eufm B=(\uplet{\eufm B}x{\eufm N})$, we have
 $\lc \uplet{\eufm B}x{\eufm N}\rc [k]=\lc \eufm N\rc [\lambda x.\,
 \lc \eufm B\rc [k]]
 =\lc \eufm N\rc [\lambda x.\,Q_{\it NF}]
 =\lc \eufm N\rc [K_{\it NF}]$ that contracts to $Q_{\it NF}$
 by induction hypothesis.
\qed

\begin{thm}\label{okm14}
Let $M$ be a term of the CCV $\lambda\mu$-calculus.
Then $M$ is normalizable if and only if $[\![M]\!]$ is normalizable.
\end{thm}

\proof\hskip.5em
By Prop.~\ref{jno24} and \ref{wdw96}.
\qed

\begin{rema}\label{boy03}
The converse of Lem.~\ref{hqc54} is true, but it is
 not trivial, contrary to the appearance.
In the ordinary lambda calculus,
 $\beta\eta$-normalizability implies $\beta$-normalizability
 (it is not trivial) \cite[Cor.~15.1.5]{zxr70}\cite[\S3]{lrs98}.
We use this fact.
\end{rema}

\begin{rema}\label{btd46}
Conceptually,
 Thm.~\ref{okm14} has an idea common to the reduction-free
 normalization by Berger and Schwichtenberg \cite{yhq71}.
Both methods extract a normal form by the inverse of the
 interpretation.
They use a full model on lambda terms while we use the CPS semantics.
The reduction-free normalization interweaves
 the process to find a normal form in the construction
 of reification-reflection pairs defined by induction on types.
In our case, we consign the task to the target calculus.
If a normal form of $[\![M]\!]$ is found by chance,
 a normal form of $M$ is obtained from its inverse translation.
If we combine the inverse translation with the method
 by Berger and Schwichtenberg, we can give
 a reduction-free normalization for the CCV $\lambda\mu$-calculus.
This idea will be pursued in a forthcoming paper.
\end{rema}

\subsection{Proof by parasitism, III: Termination of the call-by-value
 evaluation}\label{dbo22}

In this subsection, we give the operational semantics
 using evaluation contexts for the CCV $\lambda\mu$-calculus.
We prove that the call-by-value evaluation of $M$ terminates
 iff $[\![M]\!]$ is solvable.
Recall that a lambda term is solvable iff it has
 a head normal form~\cite[Thm.~8.3.14]{zxr70}.

The operational semantics using evaluation contexts
 are introduced by Plotkin~\cite{pag41} for the lambda calculus
 to argue the execution on an abstract machine.
Thereafter the evaluation contexts are used as a standard technique to give
 the handy operational semantics of calculi.
For example, Felleisen et al.~\cite{krj20} gives the operational
 semantics of the calculus with control operators $\mathcal{C}$
 and $\mathcal{A}$.
Here we apply the technique to the CCV $\lambda\mu$-calculus.

The operational semantics determines a canonical order of
 evaluation.
The evaluation context is used to indicate which is the next
 redex to be contracted.
We should be sensitive to the scope of each subterm.
Hence, when we speak of the operational semantics,
 we stop the ellipsis of brackets in
 \ref{bmv07} and follow the rule in Rem.~\ref{jiv96} below.

This switch of convention reflects the difference of focused targets.
The calculus in this paper is designed so as to be usable for
 humans.
The omission of brackets is introduced because
 tracking brackets is cumbersome for us.
On the other hand, computers cannot work on
 equivalence classes.
If we talk by analogy to arithmetic, $(3+4)+5$
 and $3+(4+5)$ have different order of evaluation.
When we are concerned with implementation,
 we must distinguish these expressions.
Meanwhile, $(3+4)+5=3+(4+5)$ should not be violated by
 the difference of evaluation.
No matter which expressions are chosen,
 the results of computation must be operationally
 equal.
Our stance to the equalities in Def.~\ref{yoq54}
 is similar.
To discuss the implementation,
 we temporarily stop obeying the rules.
Afterward, we argue that the evaluation is consistent
 with the equalities.
See Rem.~\ref{nhi19}.

	\vskip0ex

\begin{rema}\label{jiv96}
When we speak of the operational semantics, we
 take the following convention.
We distinguish $\uplet{(\mu k.\,J)}xM$
 from $\mu k.\,(\uplet JxM)$, as well as
 $\uplet{(\uplet LxM)}yN$ from $\uplet Lx{(\uplet MyN)}$.
For jumper and let, we keep on omitting brackets.
We understand $[k]\uplet MxN$ to mean $[k](\uplet MxN)$.
\end{rema}

\begin{defi}\label{xuc89}
The {\it evaluation contexts} $E$ in the CCV $\lambda\mu$-calculus
 are defined by the following syntax:

	\vskip2ex
        \noindent\kern5em
$E\ \ \mathrel{::=}\ \ \square\ \ \ |\ \ \ E[V\square]
 \ \ \ |\ \ \ E[\square M]\ \ \ |\ \ \ E[\uplet Mx\square]$

	\vskip2ex
        \noindent
 where, for example, $E[V\square]$ is understood to fill
 a unique hole $\square$ in $E$ by $V\square$.
\end{defi}

\begin{rema}\label{ffv12}\hfill
\shrinktopsep
\begin{enumerate}[(1)]
\item
We may define the evaluation contexts as $E\mathrel{::=}\square\ |\ VE
 \ |\ EM\ |\ \uplet MxE$.
This style is more usual.
Since the evaluation contexts are compositional,
 either definition works.
Indeed, we may define them even as

	\vskip2ex
        \noindent\kern3.2em
$E\ \ \mathrel{::=}\ \ \square\ \ \ |\ \ \ E[VE]\ \ \ |
 \ \ \ E[ME]\ \ \ |\ \ \ E[\uplet MxE]$

	\vskip2ex
        \noindent
 combining both styles.
We use this remark in Lem.~\ref{zfg99}.

\item
Each term is decomposed into the form of either $E[V]$ or $E[\mu k.\,J]$.
Indeed, if a subterm is neither a value nor $\mu$-abstraction,
 we can dismantle it further by a clause of Def.~\ref{xuc89}.
\end{enumerate}
\end{rema}

	\vskip0ex

\noindent
We define the call-by-value evaluation semantics as rewriting rules.
It is called $E$-rewriting.
Beforehand we introduce the $E_0$-rewriting rules.
This relation rewrites terms of the form
 either $E[V]$ or $E[\mu k.\,J]$ with non-void evaluation
 context $E$.

	\vskip0ex

\begin{defi}\label{yyl85}
The rewriting relation $\mor{E_0}$ is defined by the following five rules:

	\vskip2ex

\halign{\kern5em$#$\hfil &${}\ \mor{E_0}\ #$\hfil\cr
 E[(\mu k.\,J)M] & E[\uplet{zM}z{\mu k.\,J}]\cr
 E[V(\mu k.\,J)] & E[\uplet {Vz}z{\mu k.\,J}]\cr
 E[(\lambda x.\,M)V] & E[\uplet MxV]\cr
 E[\uplet MxV] & E[M\{V/x\}]\cr
 E[\uplet Mx{\mu k.\,J}] & E[\mu k.\,J\{[k]\square\mapsto
  [k]\uplet Mx\square\}]\cr
}

	\vskip2ex
        \noindent
In the first two rules, $z$ is a fresh variable.
The first two rules are called {\it administrative}.
\end{defi}

\begin{rema}\label{avp79}\hfill
\shrinktopsep
\begin{enumerate}[(1)]
\item
Each term is decomposed into a unique one of the
 left-hand sides of Def.~\ref{yyl85} if any.
Hence the computation path by the $E_0$-rewriting is
 uniquely determined.

\item
A term $E[xV]$ has the shape of $E'[V]$ with
 non-void evaluation context $E'=E[x\square]$.
However, there is no rewriting from $E[xV]$.
In this case the computation stalls.
It is the only situation where the evaluation procedure
 deadlocks.

\item
A value $V$ does not have a shape of $E[V]$ or
 $E[\mu k.\,J]$ with non-void $E$.
So the $E_0$-computation terminates if a value is encountered.
Contrary to the case above, we regard this as successful termination.
Likewise, the $E_0$-computation terminates if $\mu k.\,J$ is reached.
However, the latter may be computed more by the $E$-rewriting
 introduced below.
See (1) of Rem.~\ref{zbj43}.

\item
No terms of the left-hand sides of Def.~\ref{yyl85} start with $\mu$.
However, for example, $\uplet{(\mu k.\,J)}xV$ may occur.
Recall that we are sensitive to bracketing in speaking of
 the operational semantics.
\end{enumerate}
\end{rema}

\begin{defi}\label{hhv52}
The rewriting relation $\mor E$ is defined by the
 following three derivation rules:

	\vskip2ex
        \noindent\kern5em
\begingroup\small
\def\strut{%
 \newdimen\dimA\dimA=\ht\strutbox\advance\dimA by1.5pt
 \newdimen\dimB\dimB=\dp\strutbox
 \hbox{\vrule width0pt height\dimA depth\dimB}}%
\vbox{\offinterlineskip
 \halign{\strut $#$\hfil\cr
  \noalign{\hrule}
  \mu k.[l]\mu m.\,J\mor E\mu k.\,J\{l/m\}\cr
}}\kern4em
\vbox{\offinterlineskip
 \halign{\strut\hfil $#$\hfil\cr
  L\mor{E_0}M\cr
  \noalign{\hrule}
  L\mor{E}M\cr
}}\kern4em
\vbox{\offinterlineskip
 \halign{\strut\hfil $#$\hfil\cr
  L\mor{E_0}M\cr
  \noalign{\hrule}
  \mu k.\,[l]L\mor{E}\mu k.\,[l]M\cr
}}
\endgroup

	\vskip2ex
        \noindent
We comment that it is uniquely
 determined which rule is applicable, if any,
 by (4) of Rem.~\ref{avp79}.
\end{defi}

\begin{rema}\label{zbj43}\hfill
\shrinktopsep
\begin{enumerate}[(1)]
\advance\parindent by15pt
\item
If a term starts with $\mu k_1.\,[l_1]\mu k_2.\,[l_2]\cdots \mu k_n.\,[l_n]$,
 the $E$-computation first collapses
 the pairs of a jump and a $\mu$-binder, until we reach
 $\mu k.\,[l]M$ where $M$ does not start with $\mu$.
Then $E_0$-rewriting is applied to $M$.
Afterward, if the resulting term starts again with
 a sequence of the two or more
 pairs of $\mu k_i$ and $[l_i]$, we collapse them.
Therefore the computation path by the $E$-rewriting
 is unique, by (1) of Rem.~\ref{avp79}.

This is for cleanup only.
The essence of evaluation lies in the $E_0$-rewriting.
The results in this section
 do not change if the
 $E$-rewriting is reasonably modified.
For example,
 we may skip the collapsing process, leaving
 the head $\mu k_1.\,[l_1]\mu k_2.\ldots$ intact.
Or else, we may rewrite $\mu k.\,[k]M$ to $M$
 if $k\not\in M$ in analogy to a topmost rule
 ${\cal A}M\vartriangleright M$ in \cite{krj20}.
These modifications add or reduce only a finite
 number of rewriting steps, thus not affecting
 computational properties.
We chose the rewriting above so that extraneous arguments
 are minimized.

\item
We see that
 the $E$-computation is classified into
 three cases, viewing (2) and (3) of Rem.~\ref{avp79}. 
First, it normally terminates returning
 either a value $V$ or $\mu k.\,[l]V$.
Second, the computation stalls with $E[xV]$ or $\mu k.\,[l]E[xV]$
 where $x$ is free.
Third, it may never terminate.

\item
If a term $M$ is closed, the $E$-computation from $M$
 either normally terminates or runs for good.
It never stalls.
If the system contains constants $c$, deadlock $E[cV]$ may happen.
\end{enumerate}
\end{rema}

\begin{defi}\label{agd76}\hfill
\shrinktopsep
\begin{enumerate}[(1)]
\item
An {\it $E$-normal form} is a term
 admitting no $E$-rewriting.
By remark (2) of \ref{zbj43}, an $E$-normal form
 is one of $V$, $\mu k.\,[l]V$,
 $E[xV]$, and $\mu k.\,[l]E[xV]$.
We note that the definition does not exclude the stalling cases.

\item
The {\it call-by-value evaluation} from $M$ terminates
 if there is an $E$-rewriting sequence $M\mor{E^*}N$
 such that $N$ is $E$-normal.
\end{enumerate}
\end{defi}

\begin{exam}\label{iiu81}\hfill
\shrinktopsep
\begin{enumerate}[(1)]
\item
Let us recall the definition of the fixed-point combinator
 $Y$ in \ref{ozp13}.
For variables $f$ and $z$, it is
 easy to see $Yfz\mor{E^*}f(\lambda v.\,D_fD_fv)z$, which
 is an $E$-normal form.
The call-by-value evaluation from $Yfz$ terminates.

\item
In contrast, if we set $D_f=\lambda x.\,f(xx)$ in definition
 of $Y=\lambda fz.\,D_fD_fz$, we have an infinite
 rewriting sequence $Yfz\mor{E^*}f(D_fD_f)z\mor{E^*}
 f(f(D_fD_f))z \mor{E^*}\cdots$.
\end{enumerate}
\end{exam}

	\vskip0ex

\noindent
First, we relate (1) of Def.~\ref{agd76} to its CPS semantics.
It is an easy task.
The characterization of (2) is given later.

	\vskip0ex

\begin{defi}\label{jgv10}
The head normal form in the target calculus
 is defined by the following syntax:

	\vskip2ex

\halign{\kern5em $#$\hfil &${}\ \mathrel{::=}\ #$\hfil\cr
 T_{\it HN} & \lambda k.\,Q_{\it HN}\ \ |\ \ xW\cr
 Q_{\it HN} & kW\ \ |\ \ xWK\cr
}

	\vskip2ex
        \noindent
 where $K$ and $W$ are as in Def.~\ref{eob58}.
\end{defi}

\begin{lem}\label{qmr77}
If $N$ is in the head normal form in the target calculus,
 $N^{-1}$ is $E$-normal.
\end{lem}

\proof\hskip.5em
Let us see $Q_{\it HN}^{-1}$.
We have $(lW)^{-1}=[l]W^{-1}$.
On the other hand, $(xWK)^{-1}$ is equal to $[l]xW^{-1}$
 or $\uplet{Q^{-1}}y{xW^{-1}}$ according to $K$'s shape,
 $l$ or $\lambda y.\,Q$.
Therefore $T_{\it HN}^{-1}$ is one of $xW^{-1}$, $\mu k.\,[l]W^{-1}$,
 $\mu k.\,[l]xW^{-1}$, and $\mu k.\,\uplet{Q^{-1}}y{xW^{-1}}$.
Since $W^{-1}$ is a value,
 these are all $E$-normal by (2) of Rem.~\ref{zbj43}.
The last case may need more comments.
For example, if $Q=(\lambda m.\,Q_0)l$, it is equal to $\mu k.\,
 ([l]\uplet{(\mu m.\,Q_0^{-1})}y{xW^{-1}})$.
The $E$-computation does not collapse $[l]\mu m$,
 contrary to the appearance.
By the convention in Rem.~\ref{jiv96}, the term should read
 $\mu k.\,[l](\uplet{(\mu m.\,Q_0^{-1})}y{xW^{-1}})$.
Hence $[l]$ and $\mu m$ do not face each other.
\qed

\begin{lem}\label{vxj29}
If a term $M$ is $E$-normal in the CCV $\lambda\mu$-calculus,
 $[\![M]\!]$ is solvable.
\end{lem}

\proof\hskip.5em
$[\![M]\!]$ is almost in head normal form.
We need a single step of head reduction in certain cases.

First we claim that, if $N$ is a non-value, $\lc E[N]\rc [K]
 =\lc N\rc [\mathcal{K}_E]$ holds for suitably defined
 continuation $\mathcal{K}_E$.
It is defined by induction on the construction of $E$ in
 Def.~\ref{xuc89}.
(i) $\mathcal{K}_\square=K$.
(ii) $\mathcal{K}_{E[V\square]}=\lambda y.\,V^*y\mathcal{K}_E$.
(iii) $\mathcal{K}_{E[\square M]}$ equals
 $\lambda x.\,xV^*\mathcal{K}_E$ if $M$ is a value $V$.
If $M$ is a non-value, $\mathcal{K}_{E[\square M]}$
 equals $\lambda x.\,\lc M\rc [\lambda y.\,xy\mathcal{K}_E]$.
(iv) $\mathcal{K}_{E[\uplet Mx\square]}=\lambda x.
 \,\lc M\rc [\mathcal{K}_E]$.
Proof of the claim is immediate from the definition of the colon
 translation~\ref{vcy79}.
Since $\mathcal{K}_E$ depends on given continuation $K$,
 we write $\mathcal{K}_{E,K}$ below.

We prove that $[\![M]\!]=\lambda k.\,\lc M\rc [k]$ belongs to $T_{\it HN}$
 in Def.~\ref{jgv10} up to head reduction.
We split cases by the shape of $E$-normal $M$.
If $M$ equals $E[xV]$, the claim above yields
 $\lc M\rc[k]=\lc xV\rc [\mathcal{K}_{E,k}]=xV^*
 \mathcal{K}_{E,k}=Q_{\it HN}$.
If $M$ equals $\mu k.\,[l]E[xV]$, then we have $(\lambda k.\,
 \lc E[xV]\rc [l])k$, that is, $(\lambda k.\,x
 V^*\mathcal{K}_{E,l})k$.
This contracts to $xV^*\mathcal{K}_{E,l}=Q_{\it HN}$ by
 one-step head reduction.
In either case, $[\![M]\!]=_{\it \beta}\lambda k.\,Q_{\it HN}=T_{\it HN}$.
If $M=V$, then $[\![V]\!]=\lambda k.\,kV^*=T_{\it HN}$.
If $M=\mu k.\,[l]V$, one-step head-reduction yields
 $\lambda k.\,lV^*=T_{\it HN}$.
\qed

	\vskip0ex

\noindent
From Lem.~\ref{vxj29} and soundness~\ref{fgv86},
 it is immediate to see that if
 the call-by-value evaluation of $M$ terminates then $[\![M]\!]$ is
 solvable.
The converse is true as is verified below.

	\vskip0ex

\begin{lem}\label{daf26}
If $T_0\rightarrow T_1$ is a head reduction of sort $T$ terms
 in the target calculus,
 $T_0^{-1}\mor{E^*}T_1^{-1}$ holds where $E^*$ denotes a finite sequence
 of $E$-rewriting with no use of the administrative rules in Def.~\ref{yyl85}
\end{lem}

\proof\hskip.5em
We return to Prop.~\ref{lzv28}.
As remarked in Rem.~\ref{gyv73}, the administrative rules
 are never applicable to the term obtained by the inverse translation.
There are only four patterns for the head redices
 of sort $T$ terms:
 $(\lambda x.\,T)W$, $\lambda k.\,(\lambda x.\,Q)W$,
 $\lambda k.\,(\lambda m.\,Q)K$, and $\lambda k.\,(\lambda x.\,T)WK$.
We verify the lemma for the third pattern.
If $K=l$, the inverse
 is $\mu k.\,[l]\mu m.\,Q^{-1}$, which is by \ref{hhv52} rewritten to
 $\mu k.\,Q^{-1}\{l/m\}$, viz., $(\lambda k.\,Q\{l/m\})^{-1}$.
If $K=\lambda y.\,Q_0$, the inverse is $\mu k.\,
 (\uplet{Q_0^{-1}}y{\mu m.\,Q^{-1}})$.
Since $Q_0^{-1}$ has the form of $[l]M_0$, we have $\mu k.\,
 [l](\uplet {M_0}y{\mu m.\,Q^{-1}})$.
(We need re-bracketing as commented in the proof of
 Lem.~\ref{qmr77}.)
By the $E$-rewriting it is rewritten to $\mu k.\,[l]\mu m.Q^{-1}\{[m]\square
 \mapsto [m]\uplet{M_0}y\square\}$, then to
 $\mu k.\,Q^{-1}
 \{[m]\square\mapsto [l]\uplet{M_0}y\square\}$.
The last equals to the inverse of $\lambda k.\,Q\{m\mapsto
 \lambda y.Q_0\}$.
The remaining patterns are similar.
\qed

\begin{cor}\label{eei59}
There is a rewriting sequence $[\![M]\!]^{-1}\mor{E^*}N$ to
 some $E$-normal $N$, whenever $[\![M]\!]$ is solvable.
Here $E^*$ denotes a finite number of $E$-rewriting with
 no use of the administrative rules in Def.~\ref{yyl85}.
\end{cor}

\proof\hskip.5em
By solvability, there is a head reduction sequence $[\![M]\!]
 \shortmor*T$ where $T$ is in head normal form.
Apply Lem.~\ref{daf26} and \ref{qmr77}.
\qed

	\vskip0ex

\noindent
Hence we know that the call-by-value evaluation of $[\![M]\!]^{-1}$
 terminates if $[\![M]\!]$ is solvable.
We want to transfer the evaluation of $[\![M]\!]^{-1}$ to
 the evaluation of $M$.
To this end, we recall
 that we have $M\mor{A^*}M^\dagger\leftmor{V^*}[\![M]\!]^{-1}$
 by Prop.~\ref{ffa70}.
Let us analyze the interaction of $E$-rewriting with administrative
 reductions $A^*$ and vertical reductions $V^*$.

In Prop.~\ref{ffa70} the arrow labeled with $V^*$ is
 in actuality a mixture of true vertical reductions
 and applications of equality axioms
 save the second axiom (see Lem.~\ref{wcg04}).
As the $E$-rewriting is sensitive to bracketing, we must take care
 of this fact.
In the following lemmata, we should understand $\mor{V^*}$ to
 mean such a mixture.

	\vskip0ex

\begin{lem}\label{zfg99}
In the following, $V^*$ and $A^*$ denote finite sequences of
 vertical reduction and administrative reduction respectively.

\shrinktopsep
\begin{enumerate}[(1)]
\item
If $N\mor{V^*}N'$ and $N$ is $E$-normal, then $N'$ is $E$-normal.

\item
Suppose $N\mor{A^*}N'$.
Then $N$ is $E$-normal if $N'$ is $E$-normal.
\end{enumerate}
\end{lem}

\proof\hskip.5em
(1)
Vertical reduction $\mu k.\,[k]M\rightarrow M$
 preserves the property of being a value or $E[xV]$.
Moreover, the first and the third equality axioms of Def.~\ref{yoq54}
 respect $E$-normal forms.
(2)
It suffices to manipulate one-step administrative reduction.
We consider ${\it ad}_1$-reduction $C[NM]\mor AC[\uplet{zM}zN]$.
The case of ${\it ad}_2$-reduction is similar.
We suppose that $C[\uplet{zM}zN]$ is $E$-normal.
Recall (2) of Rem.~\ref{zbj43} for the shapes of $E$-normal forms.
Interested cases are $E[xV]$ and $\mu k.\,[l]E[xV]$.
We consider the former.
The essential case is $E=E'[\uplet{zM}z{E''}]$ so
 that $C[\uplet{zM}zN]$ is equal to $E'[\uplet{zM}z{E''[xV]}]$.
Then $C[NM]$ is equal to $E'[E''[xV]M]$ that is $E$-normal.
\qed

	\vskip0ex

\noindent
We first show that the $E$-rewriting commutes with vertical reductions.
Second, we verify that administrative reductions can be postponed.

	\vskip0ex

\begin{lem}\label{ncp00}\hfill
If $M'\leftmor{V^*}M\mor{E^*}N$ holds (pay attention
 to the directions of arrows) and $N$ is $E$-normal,
 there is an $E$-normal term $N'$ such that

	\vskip2ex

\halign{\kern5em #\hfil\cr
\squarediagabs{50}{40}%
 MN{M'}{N'}{E^*}{V^*}{V^*}{E^*}
\cr}

	\vskip2ex
        \noindent
 where
 $E^*$ a finite sequence of $E$-rewriting with no
 use of administrative rules in Def.~\ref{yyl85}.
For $V^*$, see the comment immediately before Lem.~\ref{zfg99}.
\end{lem}

\proof\hskip.5em
$N'$ is $E$-normal by Lem.~\ref{zfg99}, (1).
Although $\mor{V^*}$ is a mixture,
 we can shuffle the order so that
 first true vertical reductions are applied, then equality axioms.
So we first prove the lemma for pure vertical reductions.
It is easy.
(However, we comment that the lemma fails if administrative $E$-rewriting
 is included, since vertical reduction creates a value
 in case $\mu k.\,[k]V\rightarrow V$.)
Second we prove that if $M_1=M_2$ save the second axiom and
 if $M_1\mor{E^*}N_1$
 to an $E$-normal form $N_1$, then $M_2\mor{E^*}N_2$ with $N_1=N_2$.
A crucial case is $E[\uplet{(\uplet LxM)}y{\mu k.\,J}]=
 E[\uplet Lx{(\uplet My{\mu k.\,J})}]$.
The one-step $E$-rewriting from the former splits into two steps
 from the latter.
\qed

\begin{lem}\label{clo14}
If $M\mor{A^*}M_1\mor{E^*}N_1$ holds,
 (compare the directions of arrows with those in \ref{ncp00}), then

	\vskip2ex

\halign{\kern5em #\hfil\cr
\squarediagabs{50}{40}%
 M{M_1}N{N_1}{A^*}{E^*}{E^*}{A^*}
\cr}

	\vskip2ex
        \noindent
 for some $N$.
Here $A^*$ denotes a finite sequence of administrative
 reduction of Def.~\ref{ofe40}, and $E^*$ a finite sequence
 of $E$-rewriting.
We stress that administrative rules
 of Def.~\ref{yyl85} are allowed in $E^*$ here.
\end{lem}

\proof\hskip.5em
We follow the standard parallel reduction argument.
We define parallel reduction $\twomor{A}$ allowing simultaneous
 administrative reduction.
We verify the following: if $M\twomor{A}M_1\mor{E^*}N_1$ holds,
 there is a term
 $N$ such that $M\mor{E^*}N\twomor{A}N_1$.
Then we use equivalence of $L\twomor{A^*}L_1$
 and $L\mor{A^*}L_1$
\qed

\begin{thm}\label{pin99}
Let $M$ be a term of the CCV $\lambda\mu$-calculus.
The call-by-value evaluation of $M$ terminates if and only if
 $[\![M]\!]$ is solvable. 
\end{thm}

\proof\hskip.5em
We verify the if-part.
By Cor.~\ref{eei59}, we have $[\![M]\!]^{-1}\mor{E^*}N$ for $E$-normal
 $N$ where $E^*$ does not use the administrative rules.
By Prop.~\ref{ffa70}, we have $M\mor{A^*}M^\dagger\leftmor{V^*}
 [\![M]\!]^{-1}$.
Hence Lem.~\ref{ncp00}, (1), gives $E$-normal $N'$ satisfying

	\vskip2ex
        \noindent\kern5em
\hbox{\begin{diagramme}
 \latticeUnit=1pt
 \boxMargin=3pt
 \object(0,0)={M}
 \object(50,0)={M^\dagger}
 \object(50,40)={[\![M]\!]^{-1}}
 \object(100,0)={N'}
 \object(100,40)={N}
 \morphism(50,40)to(100,40)[E^*]
 \morphism(50,40)to(50,0)[V^*][R]
 \morphism(100,40)to(100,0)[V^*]
 \morphism(0,0)to(50,0)[A^*][R]
 \morphism(50,0)to(100,0)[E^*][R]
\end{diagramme}}.

	\vskip2ex
        \noindent
Deferring the administrative reductions $M\mor{A^*}M^\dagger$
 by Cor.~\ref{clo14}, we finally obtain
 $M\mor{E^*}N''\mor{A^*}N'$.
By (2) of Lem.~\ref{zfg99}, $N''$ is $E$-normal.
\qed

\begin{cor}\label{lly89}
If the call-by-value evaluation of $M$ terminates returning $N$,
 and if $M=_{\it ccv}M'$, the call-by-value evaluation of
 $M'$ terminates returning some $N'$.
Moreover the $E$-normal forms satisfy $N=_{\it ccv}N'$.
\end{cor}

\proof\hskip.5em
By soundness \ref{fgv86}, $[\![M]\!]=[\![M']\!]$ holds.
If the evaluation of $M$ terminates, $[\![M]\!]$ is
 solvable by Thm.~\ref{pin99}.
Since solvability is closed under
 $\beta\eta$-equality \cite[Thm.~15.1.7]{zxr70},
$[\![M']\!]$ is solvable.
Hence the evaluation of $M'$ terminates by the
 theorem.
The last part of the corollary is immediate since $M=_{\it ccv}N$
 and $M'=_{\it ccv}N'$.
\qed

	\vskip0ex

\noindent
From Cor.~\ref{lly89}, we know that the operational meaning of a
 term does not change if we transfer it to another term by
 CCV-equality.
This corresponds to \cite[Thm.~5]{pag41} and
 \cite[Thm.~4.11, (i)]{krj20}.
If we extend the calculus with constants, however,
 the $\eta_\lambda$ equality must be considered carefully.
For a constant $c$, whether $c$ should be operationally
 equal to $\lambda x.\,cx$ depends on the design and
 implementation of the calculus.
This problem is discussed in \cite{yrs06}.

	\vskip0ex

\begin{exam}\label{sdp73}
We return to the implementation of the cooperative multitasking in \ref{fci53}.
The call-by-value evaluation of $[\hat\tau](\uplet Mq{\mathcal{C}q})
 \mathrel{\vartriangleright^q_r}N$ proceeds to
 $[\hat\tau]\uplet{(\mu\delta.\,[\hat\tau]N)}r{\lambda q.\,(
 \mu\delta.\,[\hat\tau](\uplet Mqq))}$, which is not accurately
 equal $[\hat\tau]N\mathrel{\vartriangleright^r_q}M$.
However, Cor.~\ref{lly89} ensures that, if the call-by-value
 evaluation of either halts, the other halts,
 and they yield the $E$-normal forms equal up to $=_{\it ccv}$.
So it is safe to regard $[\hat\tau](\uplet Mq{\mathcal{C}q})
 \mathrel{\vartriangleright^q_r}N$ to have the effect of
 switching a live process
 by invoking $[\hat\tau]N\mathrel{\vartriangleright^r_q}M$.
\end{exam}

\begin{rema}\label{nhi19}
We can dynamically
 change the scopes of let and mu bindings during execution.
Suppose $L\mor{E^*}M$ and $M=M'$ where the latter is obtained by
 re-bracketing of Def.~\ref{yoq54}.
By Cor.~\ref{lly89}, then, $L$ terminates iff $M'$
 terminates, and they return equal results.
Repeatedly doing so does not damage the terminating
 situation (by K\"onig's lemma).
This ensures the safety of optimization by changing scopes, such as
 let-flattening \cite{nie60}.
We note that our equality enables optimization by oppositely
 narrowing the scope of let-bindings.

\end{rema}

\begin{rema}\label{qjd67}
Theorem~\ref{pin99} is an extension of
 \cite[p.~148, Thm.~2]{pag41}.
Results closely related to Cor.~\ref{lly89} are proved
 in \cite[p.~142, Cor.~1]{pag41} for the call-by-value lambda calculus
 and in \cite[p.~231, Cor.~4.9]{krj20}
 for the calculus with control operators $\mathcal{C}$ and $\mathcal{A}$.
Since completeness fails in these calculi, the proof strategy
 is different.
They  appeal to the standardization theorem to shuffle the order
 of computation.
We use completeness.
We need only the restricted shuffle patterns in Lem.~\ref{clo14}.
\end{rema}

\section{Type-Theoretic Characterization}\label{opg15}

In the theory of the ordinary lambda calculus,
 various computational properties are characterized by
 the intersection type discipline.
A well-known theorem by Barendregt, Coppo, and Dezani-Ciancaglini
 exhibits that normalizability and solvability
 are reduced to certain typeability problems~\cite{jix62}.
In the previous section, we have verified (Thm.~\ref{okm14},
 \ref{pin99}) that these two properties of the target calculus
 respectively correspond to
 normalizability and the termination of call-by-value evaluation
 of the source language.
It is, therefore, natural to think of extending the type-theoretic
 characterization to the CCV $\lambda\mu$-calculus.

In this section, we develop the union-intersection type discipline
 for the CCV $\lambda\mu$-calculus.
We verify the following two theorems in \ref{tno58}:
(i) the call-by-value evaluation
 of a CCV $\lambda\mu$-term $M$ terminates
 if and only if $M$ is typeable;
(ii)
 $M$ is normalizable if and only if $M$ is typeable
 and the typing judgment of $M$ contains neither empty intersection
 nor empty union.
The results are verified by reflecting the intersection type discipline
 of the target calculus.

\subsection{Union-intersection type discipline}

We introduce the union-intersection type discipline for
 the CCV $\lambda\mu$-calculus.
There are plenty of works
 discussing type systems having both union and intersection, e.g.,
 \cite{opo51}\cite{zdl11}\cite{yah42}.
Later we give a comparison with some of the type systems, limited
 to call-by-value calculi, in \S\ref{qjf05}.
For call-by-name $\lambda\mu$-calculus, van~Bakel gives
 a type system equipped with union and intersection, verifying
 subject reduction/expansion \cite{fxo45}.
Indeed, our type system is inspired by this.

	\vskip0ex

\begin{defi}\label{cuq73}
We define raw types $R$, subsidiary types $S$, and types $T$ by
 the following syntax:

	\vskip2ex

\halign{\kern5em $#$\hfil &${}\ \mathrel{::=}\ #$\hfil\cr
 R & \alpha\ \ |\ \ S\rightarrow T\cr
 S & \bigcap R\cr
 T & \bigcup S\cr
}

	\vskip2ex
	\noindent
 where $\alpha$ ranges over atomic types.
$\bigcap R$ denotes a finite formal intersection of raw types, $R_1\cap
 R_2\cap\cdots\cap R_n$.
If $n=0$, we write $\omega$.
Likewise, $\bigcup S$ is a finite formal union $S_1\cup S_2\cup
 \cdots\cup S_n$.
If $n=0$, we write $\agemo$ (agemo?).
We ignore differences by associativity and commutativity of
 $\cap$ and of $\cup$.
\end{defi}

\begin{rema}\label{rdx30}
In Def.~\ref{cuq73}, case $n=1$ is allowed in $S=R_1\cap R_2\cap\cdots\cap R_n$.
Hence each raw type is a subsidiary type.
Likewise each subsidiary type is a type.
\end{rema}

\begin{defi}\label{mmb04}
The subtype relation $\leq$ is defined by the following derivation rules:

	\vskip2ex
\begingroup\small
        \noindent\kern5em
\vbox{\offinterlineskip
 \halign{\strut $#$\hfil\cr
  \noalign{\hrule}
  \alpha\leq \alpha\cr
}}\kern3em
\vbox{\offinterlineskip
 \halign{\strut $#$\hfil\cr
  S'\leq S\qquad T\leq T'\cr
  \noalign{\hrule}
  S\rightarrow T\leq S'\rightarrow T'\cr
}}

	\vskip2ex
        \noindent\kern5em
\vbox{\offinterlineskip
 \halign{\strut $#$\hfil\cr
  S\leq S'\cr
  \noalign{\hrule}
  S\cap S''\leq S'\cr
}}\kern3em
\vbox{\offinterlineskip
 \halign{\strut $#$\hfil\cr
  [\ S\leq S_i\ ]_i\cr
  \noalign{\hrule}
  S\leq \bigcap_iS_i\cr
}}

	\vskip2ex
        \noindent\kern5em
\vbox{\offinterlineskip
 \halign{\strut $#$\hfil\cr
  T\leq T'\cr
  \noalign{\hrule}
  T\leq T''\cup T'\cr
}}\kern3em
\vbox{\offinterlineskip
 \halign{\strut $#$\hfil\cr
  [\ T_i\leq T\ ]_i\cr
  \noalign{\hrule}
  \bigcup_iT_i\leq T\cr
}}.

	\vskip2ex
\endgroup
        \noindent
The notation $[\>S\leq S_i\>]_i$ should be understood to mean
 a sequence of derivations where $i$ ranges over a finite index set.
$[\>T_i\leq T\>]_i$ is similar.
Taking an empty index set, we have $\agemo\leq T$ and $S\leq \omega$.
From the latter, we further obtain $T\leq \omega$ by use of the
 last inference rule.
\end{defi}

	\vskip0ex

\noindent
The subtype relation is a preorder.
For example, $T$ and $T\cup T$ are comparable to each other, but not equal.

We deliberately omit the reflexivity law and the transitivity law:

	\vskip2ex
\begingroup\small
        \noindent\kern5em
\vbox{\offinterlineskip
 \halign{\strut $#$\hfil\cr
  \noalign{\hrule}
  T\leq T\cr
}}\kern4em
\vbox{\offinterlineskip
 \halign{\strut $#$\hfil\cr
  T\leq T'\qquad T'\leq T''\cr
  \noalign{\hrule}
  T\leq T''\cr
}}

	\vskip2ex
\endgroup
        \noindent
 (both for types and subsidiary types) from the table
 in Def.~\ref{mmb04}.
As is verified in Lem~\ref{jsc26}, these rules are redundant.
We can add them with no change of power.
A similar result is asserted in Dunfield and Pfenning
 \cite[Lem.~1, \S5]{mji15} in a more general setting.

	\vskip0ex

\begin{lem}\label{jsc26}
If $T\leq T'$ is derived in the system having the inference rules for
 reflexivity and transitivity in addition to those of Def.~\ref{mmb04},
 then $T\leq T'$ is derived in the pure system of Def.~\ref{mmb04}.
\end{lem}

\proof\hskip.5em
For distinction, let us write $T\leq_1T'$ if we use the system including
 reflexivity and transitivity.
We define the third relation $\leq_2$.
First $\bigcup_{i\in I}S_i\leq_2\bigcup_{j\in J}S'_j$ if
 there is a mapping $I\mor fJ$ such that $S_i\leq_2S'_{f(i)}$
 for every $i$.
Second $\bigcap_{i\in I}R_i\leq_2\bigcap_{j\in J}R'_j$ if
 there is a mapping $J\mor fI$ such that $R_{f(j)}\leq_2R'_j$
 for every $j$.
Finally $R\leq_2R'$ if both $R$ and $R'$ are the same atomic type
 or $R=S\rightarrow T$ and $R'=S'\rightarrow T'$ with
 $S'\leq_2S$ and $T\leq_2T'$.
Now it is easy to verify that $T\leq_1T'$ implies $T\leq_2T'$ by
 induction on the construction of the derivation trees.
It is also easy to show that $T\leq_2T'$ implies $T\leq T'$.
(As a consequence, all of $\leq,\leq_1$ and $\leq_2$ turn out to
 be the same.)
\qed

\begin{defi}\label{seh62}
We give the inference rules of typing judgments.
A {\it typing judgment} has the form $\Gamma\;\vdash\;M\mathbin:T\;\mathbin|
 \;\Delta$.
Here the typing environment $\Gamma$ is
 a finite sequence of $x_i\mathbin:S_i$, and the typing environment $\Delta$
 is a finite sequence of $k_j\mathbin:T_j$.
As usual, we assume that all variables in the typing environments are
 distinct from each other.
We stress that only subsidiary types are assigned to ordinary variables
 while continuation variables have no limitation.
We assume a special type $\dbot$ for typing jumps.
The following are inference rules:

	\vskip2ex
\begingroup\small
        \noindent\kern5em
\vbox{\offinterlineskip
 \halign{\strut $#$\hfil\cr
  \noalign{\hrule}
  \Gamma,\>x\mathbin:S\ \vdash\ x\mathbin:S\ \mathrel|\ \Delta\cr
}}

	\vskip2ex
        \noindent\kern5em
\vbox{\offinterlineskip
 \halign{\strut $#$\hfil\cr
  [\ \Gamma,\>x\mathbin:S_i\ \vdash\ M\mathbin:T_i
  \ \mathrel|\ \Delta\ ]_{i}\cr
  \noalign{\hrule}
  \Gamma\ \vdash\ \lambda x.\,M\mathbin:\bigcap_{i}(S_i\rightarrow T_i)
   \ \mathrel|\ \Delta\cr
}}

	\vskip2ex
        \noindent\kern5em
\vbox{\offinterlineskip
 \halign{\strut $#$\hfil\cr
  \Gamma\ \vdash\ M\mathbin:\bigcup_{i}\bigcap_{j}(S_{ij}\rightarrow T)
   \ \mathrel|\ \Delta\qquad
   [\ \Gamma\ \vdash\ N\mathbin:\bigcup_{j}S_{ij}
    \ \mathrel|\ \Delta\ ]_{i}\cr
  \noalign{\hrule}
  \Gamma\ \vdash\ MN\mathbin:T\ \mathrel|\ \Delta\cr
}}

	\vskip2ex
        \noindent\kern5em
\vbox{\offinterlineskip
 \halign{\strut $#$\hfil\cr
  [\ \Gamma,\>x\mathbin:S_i\ \vdash\ M\mathbin:T
   \ \mathrel|\ \Delta\ ]_{i}\qquad
   \Gamma\ \vdash\ N\mathbin:\bigcup_{i}S_i\ \mathrel|\ \Delta\cr
  \noalign{\hrule}
  \Gamma\ \vdash\ \uplet MxN\mathbin:T\ \mathrel|\ \Delta\cr
}}

	\vskip2ex
        \noindent\kern5em
\vbox{\offinterlineskip
 \halign{\strut $#$\hfil\cr
  \Gamma\ \vdash\ J\mathbin:\dbot\ \mathrel|\ \Delta,\>k\mathbin:T\cr
  \noalign{\hrule}
  \Gamma\ \vdash\ \mu k.\,J\mathbin:T\ \mathrel|\ \Delta\cr
}}

	\vskip2ex
        \noindent\kern5em
\vbox{\offinterlineskip
 \halign{\strut $#$\hfil\cr
  \Gamma\ \vdash\ M\mathbin:T\ 
   \mathrel|\ \Delta,\>k\mathbin:T\cr
  \noalign{\hrule}
  \Gamma\ \vdash\ [k]M\mathbin:\dbot\ \mathrel|\ \Delta,\>
   k\mathbin:T\cr
}}

	\vskip2ex
        \noindent\kern5em
\vbox{\offinterlineskip
 \halign{\strut $#$\hfil\cr
  [\ \Gamma,\>x\mathbin:S_i\ \vdash\ J\mathbin:\dbot
   \ \mathrel|\ \Delta\ ]_{i}\qquad
   \Gamma\ \vdash\ N\mathbin:\bigcup_{i}S_i\ \mathrel|\ \Delta\cr
  \noalign{\hrule}
  \Gamma\ \vdash\ \uplet JxN\mathbin:\dbot\ \mathrel|\ \Delta\cr
}}

	\vskip2ex
        \noindent\kern5em
\vbox{\offinterlineskip
 \halign{\strut $#$\hfil\cr
  \Gamma\ \vdash\ M\mathbin:T\ \mathrel|\ \Delta\qquad T\leq T'\cr
  \noalign{\hrule}
  \Gamma\ \vdash\ M\mathbin:T'\ \mathrel|\ \Delta\cr
}}

	\vskip2ex
\endgroup
        \noindent
The last inference rule is called the {\it subsumption} rule.
Each suffix ($i$ or $j$) ranges over a finite set.
For instance, the type
 $\bigcup_i\bigcap_j(S_{ij}\rightarrow T)$ is understood
 to be $\smash{\bigcup_{i\in I}\bigcap_{j\in J(i)}}(S_{ij}\rightarrow T)$.
 where $I$ and each $J(i)$ are finite sets.
The notation $[\cdots]_i$ denotes a finite sequence of judgments.
We understand that the parts without suffix $i$ are
 common to all the judgments therein.
In each rule, the corresponding suffixes range over the same set.
For example, in the right premise $[\ \Gamma\ \vdash\ N\mathbin:\bigcup_{j}S_{ij}
 \ \mathrel|\ \Delta\ ]_{i}$ of the third rule,
 the suffix $i$ ranges over
 $I$ and the suffix $j$ in the $i$-th judgment ranges over $J(i)$
 as they do so in $\bigcup_i\bigcap_j(S_{ij}\rightarrow T)$.
Moreover, $\Gamma,N$, and $\Delta$ are common to all judgments in
 the sequence.
\end{defi}

	\vskip0ex

\noindent
The {\it simply typed} CCV $\lambda\mu$-calculus is defined
 by limiting all intersections and unions to range over singleton sets.

	\vskip0ex

\begin{rema}\label{qnd21}\hfill
\shrinktopsep
\begin{enumerate}[(1)]
\item
Note that the values are typed by subsidiary types in
 the first two inference rules of Def.~\ref{seh62}.
Hence, if a value $V$ has type $\bigcup S_i$, it must be
 obtained by an application of the subsumption rule.
So $V$ has type $S_i$ for some $i$, viewing
 the characterization of the subtype relation by $\leq_2$ in
 the proof of \ref{jsc26}.

\item
Not all terms have type $\omega$.
However, every value can have type $\omega$.
For variable $x$, we use the subsumption rule since $S\leq\omega$.
For lambda abstraction $\lambda x.\,M$, let $i$ range over
 an empty set.
In general, a term can have type $\omega$ if and only if
 it is typeable.

\item
Since indices may range over an empty set, the sequences $[\cdots]_i$
 are able to be void.
Hence a term may be typed even if not all of free variables
 occur in the typing environments.
In such a case, we may implicitly
 regard $x\mathbin:\omega$ and $k\mathbin:\agemo$.
\end{enumerate}
\end{rema}

\begin{exam}\label{qcx71}
The typing of the call/cc operator is suggestive.
The operator is encoded as
$\mathop{\sf call/cc}M=\mu k.\,[k]M(\lambda x\mu\delta.\,[k]x)$.
Provided that $M$ has type $(\bigcap_j(S_j\rightarrow\agemo))\rightarrow T$,
 we have

        \nopagebreak
	\vskip2ex
        \nopagebreak
        \noindent\kern5em
$\mathop{\sf call/cc}M\>\mathbin:\>T\cup\bigcup_jS_j$.

	\vskip2ex
        \noindent
Here $\agemo$ can be replaced with any other type.
The type of $\mathop{\sf call/cc}M$ is naturally understood in
 the aspect of side-effects:
 $T$ is the type of a normal exit
 and $\bigcup S_i$ of non-local exits.
The union over $i$ corresponds to possible multiple locations
 where non-local exits are raised.
We can generalize the type of $M$ to $\bigcup_i
 ((\bigcap_j(S_{ij}\rightarrow\agemo))\rightarrow T)$ with
 $\mathop{\sf call/cc}M\mathbin:T\cup\bigcup_{ij}S_{ij}$,
 if it is meaningful.
\end{exam}

\begin{exam}\label{udf62}
Let us recall definition of the call-by-value
 fixed-point combinator $Y$ in Def.\ref{ozp13}.
We show that $f\mathbin:\omega\rightarrow\agemo,\>z\mathbin:S\;
 \vdash\;Yfz\mathbin:T$ where $S$ is an arbitrary subsidiary type
 and $T$ is an arbitrary type.
The key is that $D_f=\lambda xw.\,f(\lambda v.\,xxv)w$ has type
 $\omega\rightarrow S\rightarrow T$ and type $\omega$.
The latter is obvious by Rem.\ref{qnd21}, (2).
For the former, $\lambda v.\,xxv$ has type $\omega$ by the same remark.
Thus $f(\lambda v.\,xxv)$ has type $\agemo$.
Since $\agemo$ is an empty union, $f(\lambda v.\,xxv)z$ can
 have arbitrary type $T$, irrelevant of the type of $z$,
 by letting $i$ range over an empty set in the inference
 rule for application of Def.~\ref{seh62}.
From the two types of $D_f$, we have $D_fD_fz\mathbin:T$.
Therefore $Yfz\mathbin:T$.

In contrast, if we change $D_f$ to $\lambda x.\,f(xx)$, then $Yfz$
 is not typeable.
It is a consequence of non-termination discussed
 in Example~\ref{iiu81}, (2), and the theorem
 verified later in \ref{tno58}, (1).
\end{exam}

\begin{rema}\label{gia05}
The implementation of multitasking in \ref{fci53}
 was done in the type-free calculus.
We have trouble if we want to provide types.
Let us work in the simple types.
Let $\alpha$ be the type of the topmost continuation $\hat\tau$.
In $\uplet M{q'}{\mathcal{C}q}$, if $q'$ has type $Q$, then
 the type of $q$ is $(Q\rightarrow\alpha)\rightarrow\alpha$.
Moreover, in $M\mathrel{\vartriangleright^q_r}N$ (viz.,
 $\uplet Mq{\lambda r\mu\delta.\,[\hat\tau]N}$), the type of $r$ is
 $Q\rightarrow\alpha$.
To say simply, we need one nesting of $(\hbox{-})\rightarrow\alpha$,
 proportional to the number of message-passing through channels.
For example, we cannot type ${\sf wall}$ in Example~\ref{fci53} even if
 we have typed fixed-point combinator, since it allows
 an unbounded number of message passing.

An open question is whether this is an essential phenomenon.
Possibly there is no well-typed implementation of multitasking
 that repeats stop-and-go in an unbounded number of times,
 unless mutable stores are allowed.
Incidentally, as far as the author knows, all implementation
 of multitasking and coroutine in the literature uses mutable
 stores \cite{wgv99}\cite{pea10}\cite{bxp29}%
\footnote{%
The implementation of iterators in \cite{pwf11} does not use
 mutable stores.
However, two processes, an iterator and its caller, are not
 symmetric.}.
\end{rema}

\begin{rema}\label{sxt26}\hfill
\shrinktopsep
\begin{enumerate}[(1)]
\advance\parindent by15pt
\item
Nishizaki gives the translation of the call-by-value calculus
 with the call/cc operator into the linear logic \cite{owr60}.
If we take correspondences between modalities and union-intersection
 by $\mathord!\leftrightsquigarrow\cap$ and $\mathord?\leftrightsquigarrow
 \cup$, his translation is coincidental to ours.
For instance, the implication type is translated to $\mathop!R\multimap
 \mathop?\mathop!R'$ and $\bigcap R\rightarrow\bigcup\bigcap R'$,
 respectively.
This is not surprising.
The modalities in the linear logic are
 the reflection of duplication or deletion of variables \cite{sgh91}.
The union-intersection types are transplanted from
 the intersection types of the lambda calculus,
 and the latter also reflects multiple occurrences of
 variables \cite{jix62}\cite{ftj70}.

\item
The $\lambda_c$-calculus by Moggi
 is designed so as to be sound and complete for the
 Kleisli category of strong monads $T$ \cite{dvp96}.
The CPS semantics of the call-by-value lambda calculus
 is simply a special case where $T$ is a continuation
 monad $\neg\neg(\hbox{-})$.
As discussed in Rem.~\ref{kgy11}, the call-by-value lambda calculus
 has the same equational theories as the $\lambda_c$-calculus.
Hence the continuation monad has a special status among strong monads.
This single strong monad represents the calculus that works for all
 strong monads.

An analysis behind this phenomenon is given by Filinski \cite{gss08}.
In a nutshell, every monad embeds into a codensity monad
 \cite[Exer.~7.3]{brs63}, that is defined using right Kan extension.
The codensity monads share common structures with the continuation monad.
In fact, if we assume the second-order lambda calculus satisfying
 the parametricity condition,
 the codensity monad associated with $T$
 is written $\forall \alpha.\,(X\rightarrow T\alpha)\rightarrow T\alpha$
 (see \cite{mmf55} for the Kan extensions via parametricity),
 while the continuation monad is $(X\rightarrow\dbot)\rightarrow\dbot$.
This type of fact is rediscovered in a different context
 \cite{nuj40}.
See also \cite{jxn04}.
\end{enumerate}
\end{rema}

	\vskip0ex

\noindent
We file an elementary property of the type system
 in Def.~\ref{seh62}.
We verify that the subsumption rule can be limited
 to values.

	\vskip0ex

\begin{defi}\label{fmg13}
We introduce $\Gamma\>\vdash'\>M\mathbin:T\>|\>\Delta$.
This judgment is defined as follows.
We substitute all $\vdash$ in Def.~\ref{seh62} with $\vdash'$.
Moreover, we replace the subsumption rule by the following restricted
 one:

	\vskip2ex
\begingroup\small
        \noindent\kern5em
\vbox{\offinterlineskip
 \halign{\strut $#$\hfil\cr
 \Gamma\ \vdash'\ V\mathbin:S\ |\ \Delta\qquad S\leq T\cr
 \noalign{\hrule}
 \Gamma\ \vdash'\ V\mathbin:T\ |\ \Delta\cr
}}

	\vskip2ex
\endgroup
        \noindent
 where $V$ is a value, $S$ a subsidiary type and $T$ a type.
\end{defi}

	\vskip0ex

\noindent
Let us write $\Gamma\leq\Gamma'$ if the typing environments $\Gamma$
 and $\Gamma'$ contains the same set of ordinary variables $x$
 and if $S\leq S'$ holds for
 $x\mathbin:S$ and $x\mathbin:S'$ occurring in $\Gamma$ and
 $\Gamma'$.
Likewise we define $\Delta\leq\Delta'$.

	\vskip0ex

\begin{lem}\label{czd76}
If $\Gamma\>\vdash\>M\mathbin:T\>|\>\Delta$ holds,
 $\Gamma^-\>\vdash'\>M\mathbin:T^+\>|\>\Delta^+$ holds for
 any typing environments and types satisfying
 $\Gamma^-\leq\Gamma$, $\Delta\leq\Delta^+$, and $T\leq T^+$.
(The notation $T^+$ is irrelevant of that in Def.~\ref{zqv10};
 here we use it simply as a meta-symbol representing a type.)
\end{lem}

\proof\hskip.5em
Easy induction on the derivation of
 $\Gamma\;\vdash\;M\mathbin:T\;\mathbin|\;\Delta$.
We need the transitivity of the subtype relation.
It is verified in Lem.~\ref{jsc26}.
\qed

\begin{cor}\label{iib41}
The following holds

	\vskip2ex
        \noindent\kern5em
$\Gamma\ \vdash\ M\mathbin:T\ |\ \Delta
 \qquad\Longleftrightarrow\qquad
 \Gamma\ \vdash'\ M\mathbin:T\ |\ \Delta$.
\qed
\end{cor}

\begin{rema}
Typing respects equality.
That is to say, if $M=N$ holds with
 respect to the equality axioms in Def.~\ref{yoq54}.
 and if $\Gamma\vdash M:T\,|\,\Delta$ holds, then we have
 $\Gamma\vdash N:T\,|\,\Delta$.
This is easily seen if we use $\vdash'$, viewing Cor.~\ref{iib41}.
\end{rema}

\begin{lem}\label{xsd71}
Except for values, the inversion of inference rules holds.
For instance, if $\Gamma\>\vdash\>\uplet MxN\mathbin:T\>|\>\Delta$ holds,
 there is a finite set of subsidiary types $S_i$ such that
 $\Gamma,\,x\mathbin:S_i\>\vdash\>M\mathbin:T\>|\>\Delta$ for every
 $i$ and $\Gamma\>\vdash\>N\mathbin:\bigcup_iS_i\>|\>\Delta$.
\end{lem}

\proof\hskip.5em
Immediate from Cor.~\ref{iib41}.
This lemma is used later in Lem.~\ref{tju43}.
\qed

\subsection{Type system of the target calculus}\label{ehd63}
The type system for the target calculus
 is the classic intersection type discipline.
It is used to give the type-theoretic characterization
 of syntactic properties of the lambda calculus
 \cite{jix62}\cite{ftj70}.
The most significant property of the discipline
 is the subject-expansion.
Namely, if $M\rightarrow N$ by $\beta$-reduction
 and $N$ has type $\tau$, then $M$ has type $\tau$, too.
Later van~Bakel simplified the system introducing
 the strict type assignment \cite{bkw61}.

It is, however, inconvenient to use the type systems in the literature
 as they are, by the following two reasons.
First, the target calculus has $\eta$-reduction.
The system by van~Bakel is not sound with respect to
 the $\eta$-reduction \cite[p.~145]{bkw61}.
Despite this, we want to take the benefit of
 the strict type assignment of that system.
So we revive the subsumption rule that has been
 eliminated in van~Bakel's.
Second, our target calculus has sorts.
The typing induced by the subject expansion
 must follow the constraints imposed by the sorts.
It is not ensured if we use the results in
 the literature.

One of the simplest ways to clear the obstruction is
 to tailor a system from fabric, not by the adjustment
 of ready-made results.
We develop the intersection type system of the target calculus from
 scratch.
We follow the standard argument, but take care of sorts.
A similar approach is taken by van~Bakel et al.~to
 build a filter model for
 the call-by-name $\lambda\mu$-calculus \cite{qws80}.
The sketch of argrument is summarized in Appendix \ref{ada92}.

	\vskip0ex

\begin{nota}\label{ddn74}
As a special atomic type we prepare $\dbot$.
We write $\neg(\hbox{-})$ in place of $(\hbox{-})\rightarrow\dbot$.
To separate strict types from types, we write underbars for
 types.
\end{nota}

	\vskip0ex

\noindent
We define $\tau$-types, $\kappa$-types, and $\sigma$-types
 corresponding to sort $T$, $K$, and $W$ of
 the target calculus \ref{eob58}.
The atomic type $\dbot$ corresponds to sort $Q$.

	\vskip0ex

\begin{defi}\label{rle89}
We define strict types $\tau,\kappa,\sigma$ and types
 $\underline\kappa,\underline\sigma$ by the following
 syntax:

	\vskip2ex

\halign{&\kern5em $#$\hfil &${}\ \mathrel{::=}\ #$\hfil\cr
 \sigma & \alpha\ \ |\ \ \underline\sigma\rightarrow\tau
  & \underline\sigma & \bigcap \sigma\cr
 \kappa & \neg\underline\sigma & \underline\kappa & \bigcap\kappa\cr
 \tau & \neg\underline\kappa\cr
}

	\vskip2ex
        \noindent
 where $\alpha$ represents atomic types.
The notation $\bigcap\kappa$ signifies finite formal intersection
 $\kappa_1\cap\kappa_2\cap\cdots\cap\kappa_n$ where case
 $n=0,1$ is included.
We do not distinguish a strict type of class $\kappa$ with
 a type of $\underline\kappa$ with $n=1$.
Similar remarks applies to $\bigcap\sigma$.
We manipulate associativity and commutativity of intersection
 implicitly.
An empty intersection is written $\omega$.
\end{defi}

	\vskip0ex

\noindent
We note that the right side
 of the arrow type $\underline\sigma
 \rightarrow\tau$ is limited to a strict type $\tau$.
Moreover, in $\neg\underline\sigma=\underline\sigma\rightarrow\dbot$,
 type $\dbot$ is strict.
Hence we follow the restriction of van~Bakel \cite{bkw61}.

	\vskip0ex

\begin{defi}\label{vzy20}
We define the subtype relation $\leq$ for the target calculus.
Each row of the following table
 gives the definition of the relation between $\underline\sigma$,
 $\underline\kappa$, and $\tau$, respectively.

	\vskip2ex
\begingroup\small
\halign{\kern5em #\hfil&&\kern4em #\hfil\cr
\vbox{\offinterlineskip
 \halign{\strut $#$\hfil\cr
  \noalign{\hrule}
  \alpha\leq\alpha\cr
}}&
\vbox{\offinterlineskip
 \halign{\strut $#$\hfil\cr
  \underline\sigma'\leq \underline\sigma\qquad
   \tau\leq\tau'\cr
  \noalign{\hrule}
  \underline\sigma\rightarrow\tau\leq \underline\sigma'\rightarrow\tau'\cr
}}&
\vbox{\offinterlineskip
 \halign{\strut $#$\hfil\cr
  \underline\sigma\leq\underline\sigma'\cr
  \noalign{\hrule}
  \underline\sigma\cap\underline\sigma''\leq\underline\sigma'\cr
}}&
\vbox{\offinterlineskip
 \halign{\strut $#$\hfil\cr
  [\ \underline\sigma\leq \underline\sigma_i\ ]_i\cr
  \noalign{\hrule}
  \underline\sigma\leq \bigcap_i\underline\sigma_i\cr
}}\cr
	\noalign{\vskip2ex}
&
\vbox{\offinterlineskip
 \halign{\strut $#$\hfil\cr
  \underline\sigma'\leq \underline\sigma\cr
  \noalign{\hrule}
  \neg\underline\sigma\leq \neg\underline\sigma'\cr
}}&
\vbox{\offinterlineskip
 \halign{\strut $#$\hfil\cr
  \underline\kappa\leq\underline\kappa'\cr
  \noalign{\hrule}
  \underline\kappa\cap\underline\kappa''\leq\underline\kappa'\cr
}}&
\vbox{\offinterlineskip
 \halign{\strut $#$\hfil\cr
  [\ \underline\kappa\leq \underline\kappa_i\ ]_i\cr
  \noalign{\hrule}
  \underline\kappa\leq \bigcap_i\underline\kappa_i\cr
}}\cr
	\noalign{\vskip2ex}
&
\vbox{\offinterlineskip
 \halign{\strut $#$\hfil\cr
  \underline\kappa'\leq \underline\kappa\cr
  \noalign{\hrule}
  \neg\underline\kappa\leq \neg\underline\kappa'\cr
}}\cr
}

	\vskip2ex
\endgroup
        \noindent
 where the notation $[\cdots]_i$ denote a finite sequence of
 derivations (void sequences inclusive).
\end{defi}

\begin{rema}\label{rfj19}
We have a characterization similar to $\leq_2$ in the
 proof of Lem.~\ref{jsc26}.
For example, we define $\bigcap_{i\in I}\kappa_i\leq_2\bigcap_{j\in J}\kappa'_j$
 iff there is a mapping $J\mor fI$ such that $\kappa_{f(i)}
 \leq_2\kappa'_i$ for every $i$.
In particular, we can add reflexivity
 and transitivity to \ref{vzy20} with no change of power.
\end{rema}

\begin{defi}\label{amg62}
We give the inference rules of the typing judgments of
 the target calculus.
A typing judgment has two typing environments $\Pi$ and $\Theta$.
The former is a finite sequence of $x\mathbin:\underline\sigma$
 and the latter of $k\mathbin:\underline\kappa$.
The variables occurring in the typing environments are distinct from
 each other.
The following is the table of the inference rules:

	\vskip2ex
\begingroup\small
        \noindent\kern5em
\vbox{\offinterlineskip
 \halign{\strut $#$\hfil\cr
  \Pi,\>\Theta,\>k\mathbin:\underline\kappa\ \vdash_s
   \ Q\mathbin:\dbot\cr
  \noalign{\hrule}
  \Pi,\>\Theta\ \vdash_s\ \lambda k.\,Q\mathbin:\neg\underline\kappa\cr
}}

	\vskip2ex
        \noindent\kern5em
\vbox{\offinterlineskip
 \halign{\strut $#$\hfil\cr
  \Pi,\>\Theta\ \vdash_s\ W_1\mathbin:\underline\sigma\rightarrow\tau\qquad
   [\ \Pi,\>\Theta\ \vdash_s\ W_2\mathbin:\sigma_i\ ]_i\cr
  \noalign{\hrule}
  \Pi,\>\Theta\ \vdash_s\ W_1W_2\mathbin:\tau\cr
}}

	\vskip2ex
        \noindent\kern5em
\vbox{\offinterlineskip
 \halign{\strut $#$\hfil\cr
  \Pi,\>\Theta\ \vdash_s\ K\mathbin:\neg\underline\sigma\qquad
   [\ \Pi,\>\Theta\ \vdash_s\ W\mathbin:\sigma_i\ ]_i\cr
  \noalign{\hrule}
  \Pi,\>\Theta\ \vdash_s\ KW\mathbin:\dbot\cr
}}

	\vskip2ex
        \noindent\kern5em
\vbox{\offinterlineskip
 \halign{\strut $#$\hfil\cr
  \Pi,\>\Theta\ \vdash_s\ T\mathbin:\neg\underline\kappa\qquad
   [\ \Pi,\>\Theta\ \vdash_s\ K\mathbin:\kappa_i\ ]_i\cr
  \noalign{\hrule}
  \Pi,\>\Theta\ \vdash_s\ TK\mathbin:\dbot\cr
}}

	\vskip2ex
        \noindent\kern5em
\vbox{\offinterlineskip
 \halign{\strut $#$\hfil\cr
  \noalign{\hrule}
  \Pi,\>x\mathbin:\sigma\cap\underline\sigma',\>\Theta\ \vdash_s
   \ x\mathbin:\sigma\cr
}}\kern3em
\vbox{\offinterlineskip
 \halign{\strut $#$\hfil\cr
  \Pi,\>x\mathbin:\underline\sigma,\>\Theta\ \vdash_s
   \ T\mathbin:\tau\cr
  \noalign{\hrule}
  \Pi,\>\Theta\ \vdash_s\ \lambda x.\,T\mathbin:\underline\sigma\rightarrow\tau\cr
}}

	\vskip2ex
        \noindent\kern5em
\vbox{\offinterlineskip
 \halign{\strut $#$\hfil\cr
  \noalign{\hrule}
  \Pi,\>\Theta,\>k\mathbin:\kappa\cap\underline\kappa'\ \vdash_s
   \ k\mathbin:\kappa\cr
}}\kern3em\kern2pt
\vbox{\offinterlineskip
 \halign{\strut $#$\hfil\cr
  \Pi,\>x\mathbin:\underline\sigma,\>\Theta\ \vdash_s
   \ Q\mathbin:\dbot\cr
  \noalign{\hrule}
  \Pi,\>\Theta\ \vdash_s\ \lambda x.\,Q\mathbin:\neg\underline\sigma\cr
}}

	\vskip2ex
\endgroup
        \noindent
 where we assume $\underline\sigma=\bigcap_i\sigma_i$ and
 $\underline\kappa=\bigcap_i\kappa_i$.
The notation $[\cdots]_i$ represents a finite sequence of derivations.
Moreover, we have the following three subsumption rules:

	\vskip2ex
\begingroup\small
        \noindent\kern5em
\vbox{\offinterlineskip
 \halign{\strut $#$\hfil\cr
  \Pi,\>\Theta\ \vdash_s\ T\mathbin:\tau\qquad \tau\leq\tau'\cr
  \noalign{\hrule}
  \Pi,\>\Theta\ \vdash_s\ T\mathbin:\tau'\cr
}}\kern3em
\vbox{\offinterlineskip
 \halign{\strut $#$\hfil\cr
  \Pi,\>\Theta\ \vdash_s\ K\mathbin:\kappa\qquad \kappa\leq\kappa'\cr
  \noalign{\hrule}
  \Pi,\>\Theta\ \vdash_s\ K\mathbin:\kappa'\cr
}}

	\vskip2ex
        \noindent\kern5em
\vbox{\offinterlineskip
 \halign{\strut $#$\hfil\cr
  \Pi,\>\Theta\ \vdash_s\ W\mathbin:\sigma\qquad \sigma\leq\sigma'\cr
  \noalign{\hrule}
  \Pi,\>\Theta\ \vdash_s\ W\mathbin:\sigma'\cr
}}
\endgroup
\end{defi}

	\vskip0ex

\noindent
Only strict types are given to terms in the typing judgments.
The suffix of $\vdash_s$ signifies this.
The inference rules are standard.
We emphasize, however, that the subsumption rules are included.
This makes a contrast to the system by van~Bakel \cite{bkw61}.
The reason is that also $\eta$-reduction is our concern.
Our type system goes in between the system by
 Barendregt et al.~\cite{jix62} and the one by van~Bakel.

	\vskip0ex

\subsection{Soundness and completeness of type systems}

Definition~\ref{seh62} gives the union-intersection type system
 for the CCV $\lambda\mu$-calculus.
Definition~\ref{amg62} gives the intersection type system
 for the target calculus.
In this subsection, we show that the former is sound and
 complete with respect to the latter under the CPS translation.

We start with soundness.
Toward this, we need to extend the CPS translation to types.

	\vskip0ex

\begin{defi}\label{zqv10}
We define $S^*$, $T^+$, and $[\![T]\!]$
 (referring to Def.~\ref{cuq73}).
These three belong to categories $\underline\sigma$, $\underline\kappa$,
 and $\tau$ of Def.~\ref{rle89}, respectively.
The definition is inductively given as follows:

	\vskip2ex

\halign{\kern5em $#$\hfil\cr
 \alpha^*\>=\>\alpha,\qquad (S\rightarrow T)^*\>=\>S^*\rightarrow[\![T]\!],
  \qquad (\bigcap R)^*\>=\>\bigcap R^*\cr
 (\bigcup S)^+\>=\>\bigcap\neg S^*\cr
 [\![T]\!]\>=\>\neg T^+.\cr
}
\end{defi}

\begin{lem}\label{swj63}
If $T_0\leq T_1$ holds, $[\![T_0]\!]\leq[\![T_1]\!]$ holds.
Here the first subtype relation is given in Def.~\ref{mmb04},
 and the second in Def.~\ref{vzy20}.
\end{lem}

\proof\hskip.5em
Simultaneously we verify that $T_0\leq T_1$ implies
 $T_1^+\leq T_0^+$ and that $S_0\leq S_1$ implies $S_0^*\leq
 S_1^*$.
Proof is easy.
\qed

	\vskip0ex

\noindent
We associate typing environments $\Gamma^*$ and $\Delta^+$
 to $\Gamma$ and $\Delta$.
Definition is straightforward.
To each $x\mathbin:S$ is associated
 $x\mathbin:S^*$, and to each $k\mathbin:T$
 is associated $k\mathbin:T^+$.
First, we show soundness.

	\vskip0ex

\begin{prop}\label{qog12}
The following holds:

	\vskip2ex
        \noindent\kern5em
$\Gamma\ \vdash\ M\mathbin:T\ \mathbin|\ \Delta\qquad\Longrightarrow
 \qquad \Gamma^*,\;\Delta^+\ \vdash_s\ [\![M]\!]\mathbin:[\![T]\!]$.
\end{prop}

	\vskip2ex

\proof\hskip.5em
We use the following fact repeatedly.
Suppose that $\Gamma^*,\>\Delta^+,\>k\mathbin:
 (\bigcup S_i)^+\;\vdash_s\;\lc M\rc [k]\mathbin:\dbot$ holds.
Moreover, we suppose $[\;\Gamma^*,\>\Delta^+\;\vdash_s\;K\mathbin:\neg S_i^*\;]_i$.
Then $\Gamma^*,\>\Delta^+\;\vdash_s\;\lc M\rc[K]\mathbin:\dbot$ is derivable.
To prove this fact, we see $\Gamma^*,\>\Delta^+\;\vdash_s\;(\lambda k.\,\lc
 M\rc[k])K\mathbin:\dbot$ is derivable, and use the subject reduction
 \ref{wha02} for $\beta$.

We simultaneously verify two assertions:
(i) $\Gamma^*,\>\Delta^+,\>k\mathbin:
 T^+\;\vdash_s\;\lc M\rc [k]\mathbin:\dbot$ whenever
 $\Gamma\;\vdash\;M\mathbin:T\;\mathbin|\;\Delta$;
(ii) $\Gamma^*,\>\Delta^+\;\vdash_s\;\lc J\rc \mathbin:\dbot$ whenever
 $\Gamma\;\vdash\;J\mathbin:\dbot\;|\;\Delta$.
From the first assertion, the proposition follows immediately.
Proof is by induction on the construction of derivation trees.
We show the case of application $MN$.
Let us see the corresponding inference rule in Def.~\ref{seh62}.
By induction hypothesis, $\Gamma^*,\>\Delta^+,\>k\mathbin:
 \bigcap_i\neg\bigcap_j(S_{ij}^*\rightarrow [\![T]\!])\;\vdash_s\;
 \lc M\rc [k]\mathbin:\dbot$ and
 $\Gamma^*,\>\Delta^+,\>k\mathbin:\bigcap_j\neg S_{ij}^*\;\vdash_s\;
 \lc N\rc [k]\mathbin:\dbot$ for every $i$.
We recall $\lc MN\rc [k]=\lc M\rc [\lambda x.\,\lc N\rc 
 [\lambda y.\,xyk]]$.
From the typing $x\mathbin:\bigcap_j(S_{ij}^*\rightarrow[\![T]\!])$,
 $y\mathbin:S_{ij}^*$, and $k\mathbin:T^+$, we can derive
 $xyk\mathbin:\dbot$.
So $\lambda y.\,xyk\mathbin:\neg S_{ij}^*$ for every $j$.
Thus $\lc N\rc [\lambda y.\,xyk]\mathbin:\dbot$ by the fact
 mentioned in the preceding paragraph.
Consequently
 $\lambda x.\,\lc N\rc [\lambda y.\,xyk]\mathbin:\neg\bigcap_j
 (S_{ij}^*\rightarrow [\![T]\!])$.
This holds for every $i$.
Therefore $\lc MN\rc [k]\mathbin:\dbot$ using the fact again.
If either $M$ or $N$ is a value, $\lc MN\rc [k]$ is obtained
 by $\beta$-reduction of the non-value case.
The subject reduction \ref{wha02} is applied.
For the subsumption rule, we use Lem.~\ref{swj63}.
Other cases are similar.
\qed

	\vskip0ex

\noindent
We turn to completeness.
It is verified in Thm.~\ref{cnp96}.
First, we extend the inverse translation to types.

	\vskip0ex

\begin{defi}\label{tyq65}
We define $\sigma^{-1},\kappa^{-1}$, and $\tau^{-1}$ as well as
 $\underline\sigma^{-1}$ and $\underline\kappa^{-1}$
 in the following way:

	\vskip2ex

\halign{\kern5em $#$\hfil &${}\ =\ #$\hfil\kern3em &$#$\hfil &${}\ =\ #$\hfil\cr
 \alpha^{-1} & \alpha & (\underline\sigma\rightarrow\tau)^{-1}
  &\underline\sigma^{-1}\rightarrow\tau^{-1}\cr
 (\bigcap\sigma)^{-1} & \bigcap\sigma^{-1} & (\neg\underline\sigma)^{-1}
  &\underline\sigma^{-1}\cr
 (\bigcap\kappa)^{-1} & \bigcup\kappa^{-1} & (\neg\underline\kappa)^{-1}
  &\underline\kappa^{-1}\cr
}
	\vskip2ex
        \noindent
The first row defines $\sigma^{-1}$, the second row $\underline\sigma^{-1}$
 and $\kappa^{-1}$, and the third row $\underline\kappa^{-1}$
 and $\tau^{-1}$.
From the first row, in turn, they are raw types, subsidiary types,
 and types in the sense of Def.~\ref{cuq73}.
\end{defi}

\begin{lem}\label{hmf85}\hfill
\shrinktopsep
\begin{enumerate}[(1)]
\item
We have $(S^*)^{-1}=S$, $(T^+)^{-1}=T$, and $[\![T]\!]^{-1}=T$.

\item
The following hold:

	\vskip2ex

\halign{\kern5em $#$\hfil &${}\quad\Longrightarrow\quad #$\hfil\cr
 \tau_0\leq \tau_1 & \tau_0^{-1}\leq\tau_1^{-1}\cr
 \kappa_0\leq \kappa_1 & \kappa_1^{-1}\leq\kappa_0^{-1}\cr
 \sigma_0\leq \sigma_1 & \sigma_0^{-1}\leq\sigma_1^{-1}.\cr
}

	\vskip2ex
	\hangafter0\hangindent0pt
        \noindent
We note that the order is flipped in the case of $\kappa$.
\end{enumerate}
\end{lem}

\proof\hskip.5em
Easy.
\qed

	\vskip0ex

\noindent
We prepare several lemmata needed to prove completeness.
They are the properties of the source-side type systems.

	\vskip0ex

\begin{lem}\label{uws76}
Let $V$ be a value of the CCV $\lambda\mu$-calculus.
If $\Gamma\>\vdash\>V\mathbin:S_i\>|\>\Delta$ holds
 for a finite family of subsidiary types $S_i$, then
 $\Gamma\>\vdash\>V\mathbin:\bigcap_iS_i\>|\>\Delta$.
\end{lem}

\proof\hskip.5em
Easy.
\qed

\begin{lem}\label{oql46}
Let $K$ be a term of sort $K$ in the target calculus.
Moreover, let us take a finite set of subsidiary types $S_i$
 in the source side.

\shrinktopsep
\begin{enumerate}[(1)]
\item
If $\Gamma,\>\square\mathop:\bigcap S_i\;\vdash\;K^{-1}
 \mathbin:\dbot\;|\;\Delta$ holds, and
 if $\Gamma\;\vdash\;V\mathbin:S_i\;|\;\Delta$ holds for every $i$
 where $V$ is a value,
 then $\Gamma\;\vdash\;K^{-1}[V]\mathbin:\dbot\;|\;\Delta$ holds.

\item
If $\Gamma,\>\square\mathop:S_i\;\vdash\:K^{-1}\mathbin:\dbot\;|\;\Delta$
 holds for every $i$, and if $\Gamma\;\vdash\;M\mathbin:\bigcup S_i\;|
 \;\Delta$ holds, then $\Gamma\;\vdash\;K^{-1}[M]\mathbin:
 \dbot\;|\;\Delta$ holds.
\end{enumerate}
\end{lem}

\proof\hskip.5em
(1)
By Lem.~\ref{uws76}, $\Gamma\;\vdash\;V\mathbin:\bigcap S_i\;|\;\Delta$
 holds.
Now the lemma is easy.
\quad
(2)
We split cases by the shape of $K$.
If $K=k$, the typing environment $\Delta$ is of the form
 $\Delta_0,\,k\mathbin:T$.
Since $\Gamma,\,\square\mathbin:S_i\;
\vdash\;[k]\square\mathbin:\dbot\;|\;\Delta_0,\,k\mathbin:T$
 is assumed, we have $S_i\leq T$ for every $i$.
Hence $\bigcup S_i\leq T$.
Thus $\Gamma\;\vdash\;M\mathbin:T\;|\;\Delta_0,\,k\mathbin:T$ by
 the subsumption rule.
From this, the lemma is immediate.
If $K=\lambda x.\,Q$, we have $\Gamma,\,\square\mathbin:S_i\;\vdash
 \;\uplet{Q^{-1}}x\square\mathbin:\dbot\;|\;\Delta$ for every $i$.
Now $\Gamma\;\vdash\;\uplet{Q^{-1}}xM\mathbin:\dbot\;|\;\Delta$
 is a direct application of an inference rule.
\qed

\begin{prop}\label{xjp86}
The following hold:

	\vskip2ex

\halign{\kern5em $#$\hfil &${}\qquad\Longrightarrow\qquad #$\hfil\cr
 \Pi,\;\Theta\ \vdash_s\ T\mathbin:\tau
  & \Pi^{-1}\ \vdash\ T^{-1}\mathbin:\tau^{-1}\ |\ \Theta^{-1}\cr
 \Pi,\;\Theta\ \vdash_s\ Q\mathbin:\dbot
  & \Pi^{-1}\ \vdash\ Q^{-1}\mathbin:\dbot\ |\ \Theta^{-1}\cr
 \Pi,\;\Theta\ \vdash_s\ W\mathbin:\sigma
  & \Pi^{-1}\ \vdash\ W^{-1}\mathbin:\sigma^{-1}\ |\ \Theta^{-1}\cr
 \Pi,\;\Theta\ \vdash_s\ K\mathbin:\kappa
  & \Pi^{-1},\;\square\mathbin:\kappa^{-1}
  \ \vdash\ K^{-1}\mathbin:\dbot\ |\ \Theta^{-1}\cr
}

	\vskip2ex
        \noindent
 where $\Pi^{-1}$ and $\Theta^{-1}$ are naturally defined.
Namely, $\Pi^{-1}$ is the sequence of $x\mathbin:\underline\sigma^{-1}$
 for $x\mathbin:\underline\sigma$ occurring in $\Pi$,
 and $\Theta^{-1}$ is the sequence of $k\mathbin:\underline\kappa^{-1}$
 for $k\mathbin:\underline\kappa$ occurring in $\Theta$.
\end{prop}

\proof\hskip.5em
Proof is by straightforward induction on the derivation.
We pick up several cases.
For readability, let us write $\Gamma,\Delta$ in place of
 $\Pi^{-1},\Theta^{-1}$.
\quad
(i)
Derivation of $W_1W_2$ by the second rule.
By induction hypothesis, $\Gamma\;\vdash\;W_2^{-1}\mathbin:\sigma_i^{-1}
 \;|\;\Delta$ for each $i$.
By Lem.~\ref{uws76}, $\Gamma\;\vdash\;W_2^{-1}\mathbin:\bigcap
 \sigma_i^{-1}\;|\;\Delta$ for $W_2^{-1}$ is a value.
By induction hypothesis, $\Gamma\;\vdash\;W_1^{-1}\mathbin:
 \underline\sigma^{-1}\rightarrow\tau\;|\;\Delta$ where
 $\underline\sigma^{-1}=\bigcap\sigma_i^{-1}$.
Hence $\Gamma\;\vdash\;W_1^{-1}W_2^{-1}\mathbin:\tau^{-1}\;|\;
 \Delta$ is inferred by the application rule of Def.~\ref{seh62}.
\quad
(ii)
Derivations of $KW$ and $TK$ are manipulated
 by (1) and (2) of Lem.~\ref{oql46} respectively.
\quad
(iii)
The subsumption rule for $\kappa$, that is,
 the last rule save one in Def.~\ref{amg62}.
By induction hypothesis, $\Gamma,\>\square\mathbin:\kappa^{-1}\;\vdash
 \;K^{-1}\mathbin:\dbot\;|\;\Delta$.
Now $\kappa\leq\kappa'$ implies $\kappa'{}^{-1}\leq\kappa^{-1}$ by
 Lem.~\ref{hmf85}, (2).
Hence Lem.~\ref{czd76} and Cor.~\ref{iib41} imply
 $\Gamma,\>\square\mathbin:\kappa'{}^{-1}\;\vdash
 \;K^{-1}\mathbin:\dbot\;|\;\Delta$.
Other cases are easy and left to the reader.
\qed

	\vskip0ex

\noindent
Let us see the part
 dealing with $W_1W_2$ in the proof of Prop.~\ref{xjp86}.
To produce the type of $(W_1W_2)^{-1}=W_1^{-1}W_2^{-1}$, we
 use the application rule, that is, the third rule in the table
 of Def.~\ref{seh62}.
We note, however, that the full-fledged potential of the rule
 is not used.
The type of $W_1^{-1}$ is preceded by neither of intersection and union.
The full power is needed in the following lemma, which
 asserts the subject
 reduction property for vertical reductions and the subject expansion
 property for administrative reductions.

	\vskip0ex

\begin{lem}\label{tju43}\hfill
\shrinktopsep
\begin{enumerate}[(1)]
\item
Suppose $L\rightarrow M$ by vertical reduction (Def.~\ref{bme70}).
If $\Gamma\;\vdash\;L\mathbin:T\;|\;\Delta$ holds
$\Gamma\;\vdash\;M\mathbin:T\;|\;\Delta$ holds.

\item
Suppose $L\rightarrow M$ by administrative reduction (Def.~\ref{ofe40}).
If $\Gamma\;\vdash\;M\mathbin:T\;|\;\Delta$ holds
$\Gamma\;\vdash\;L\mathbin:T\;|\;\Delta$ holds.
\end{enumerate}
\end{lem}

\proof\hskip.5em
(1) Immediate.
\quad
(2)
By Lem.~\ref{xsd71}, we can apply the inversion of inference except to values.
This makes our proof slightly simple.
In this proof, the typing context $\Delta$ plays no role.
So let us omit it completely.
There are two cases.
We verify the case of {\small (${\it ad}_2$)}
 $VN\rightarrow \uplet{Vz}zN$ where $z$ is fresh.
Let us suppose $\Gamma\;\vdash\;\uplet{Vz}zN\mathbin:T$.
By inversion, there is a finite family of strict types $S_i$ such that
 $\Gamma,\,z\mathbin:S_i\;\vdash\;Vz\mathbin:T$ holds for every $i$
 and $\Gamma\;\vdash\;N\mathbin:\bigcup S_i$ holds.
The former is, by inversion, obtained from $\Gamma\;\vdash\;V\mathbin:\bigcup_j
 \bigcap_k(\tilde S_{ijk}\rightarrow T)$ and $[\>\Gamma,\,z\mathbin:S_i
 \;\vdash\;z\mathbin:\bigcup_k\smash{\tilde S_{ijk}}\>]_j$.
By Rem.~\ref{qnd21}, (1),
 there is $j$ such that $\Gamma\;\vdash\;V\mathbin:\smash{\bigcap_k(
 \smash{\tilde S_{ijk}}\rightarrow T)}$.
Let us fix such $j=j(i)$ for each $i$.
By the same remark, for each $i$, there is $k$
 such that $S_i\leq \tilde S_{ijk}$.
Thence $\bigcap_k(\tilde S_{ijk}\rightarrow T)\leq S_i\rightarrow T$.
This holds for each $i$.
So Lem~\ref{uws76} yields $\Gamma\;\vdash\;V\mathbin:\bigcap_i(S_i\rightarrow T)$.
By this and $\Gamma\;\vdash\;N\mathbin:\bigcup S_i$, we conclude
 $\Gamma\;\vdash\;VN\mathbin:T$.
The case of {\small (${\it ad}_1$)} is similar.
\qed

	\vskip0ex

\noindent
Now we are in the position to prove completeness for type systems.

	\vskip0ex

\begin{thm}\label{cnp96}
Let $M$ be a term of the CCV $\lambda\mu$-calculus.
We have

	\vskip2ex
        \noindent\kern5em
$\Pi,\>\Theta\ \vdash_s\ [\![M]\!]\mathbin:\tau
 \qquad\Longrightarrow\qquad
 \Pi^{-1}\ \vdash\ M\mathbin:\tau^{-1}\ |\ \Theta^{-1}$.
\end{thm}

\proof\hskip.5em
By Prop.\ref{xjp86}, we have $\Pi^{-1}\;\vdash\;[\![M]\!]^{-1}\mathbin:
 \tau^{-1}\;|\;\Theta^{-1}$.
Let us recall, by Prop.~\ref{ffa70}, that vertical reductions
 from $[\![M]\!]^{-1}$ yield $M^\dagger$, from which we recover
 $M$ by administrative expansions.
Therefore two statements of Lem.~\ref{tju43} yield
 $\Pi^{-1}\;\vdash\;M\mathbin:\tau^{-1}\;|\;\Theta^{-1}$.
\qed

\subsection{Type-theoretic characterization of syntactic properties}

Here is the main part of this section.
We verify that the union-intersection type system of the
 CCV $\lambda\mu$-calculus satisfies the subject reduction as well
 as a weak form of the subject expansion (Thm.~\ref{ehw91}).
Moreover, we give the type-theoretic characterization of the
 termination of call-by-value evaluation as well as
 of normalizability (Thm.~\ref{tno58}).

	\vskip0ex

\begin{thm}\label{ehw91}
Let us suppose $L\mor*M$ in the CCV $\lambda\mu$-calculus.

\shrinktopsep
\begin{enumerate}[(1)]
\item
If $\Gamma\;\vdash\;L\mathbin:T\;|\;\Delta$ holds,
 $\Gamma\;\vdash\;M\mathbin:T\;|\;\Delta$ holds.

\item
If $\Gamma\;\vdash\;M\mathbin:T\;|\;\Delta$ holds,
 there are $\Gamma',\Delta'$ and $T'$ such that
 $\Gamma'\;\vdash\;L\mathbin:T'\;|\;\Delta'$ holds.
\end{enumerate}
\end{thm}

\proof\hskip.5em
(1)
By Prop.~\ref{qog12}, $\Gamma^*,\,\Delta^+\;\vdash_s\;[\![L]\!]
 \mathbin:[\![T]\!]$.
As commented in Rem.~\ref{bqb30}, we need only $\beta\eta$-reduction
 and $\beta$-expansion (but no $\eta$-expansion) to obtain $[\![M]\!]$
 from $[\![L]\!]$.
Hence Prop.~\ref{wha02} and Prop.~\ref{vqu58}, (1), imply
 $\Gamma^*,\,\Delta^+\;\vdash_s\;[\![M]\!]
 \mathbin:[\![T]\!]$.
Thus by Thm.~\ref{cnp96} we have $(\Gamma^*)^{-1}\;\vdash\;
 M\mathbin:[\![T]\!]^{-1}\;|\;(\Delta^+)^{-1}$.
Finally, Lem.~\ref{hmf85}, (1), shows $(\Gamma^*)^{-1}=\Gamma$,
 $(\Delta^+)^{-1}=\Delta$ and $[\![T]\!]^{-1}=T$.
\quad
(2)
To obtain $[\![L]\!]$ from $[\![M]\!]$,
 we need $\eta$-expansion in addition.
We apply Prop.~\ref{vqu58}, (2).
\qed

\begin{rema}\label{dhc26}
We need the change of typing environments and types in
 Thm.~\ref{ehw91}, (2), only in the $\eta_\lambda$-expansion of $V$
 to $\lambda x.\,Vx$.
Recall the proof of soundness~\ref{fgv86}, in which the $\eta$-rule
 of the target calculus is involved only to handle $\eta_\lambda$
 and $\eta_{\it let}$.
However, for the latter, the used $\eta$-rule is
 $\lambda x.\,Kx\rightarrow K$ of the sort $K$.
The subject expansion for $\eta$ of this sort causes no problem
 as mentioned in the proof of Prop.~\ref{vqu58}.
Hence the change is needed only for $\eta_\lambda$.
If $V$ has an atomic type $\alpha$, we replace it by
 an arbitrary raw type $S\rightarrow T$.
\end{rema}

\begin{rema}\label{xgp85}
The rules for application and let-construct in
 Def.~\ref{seh62} are essentially the union-elimination rule.
It is reported in \cite[\S2]{yah42} that, for the call-by-name
 system, the subject reduction/expansion
 property fails if we include the union-elimination
 rule.
The counter-example to the subject expansion is $x(yz)(yz)\mathbin:C$
 where $x$ has type $(A\rightarrow A\rightarrow C)\cap
 (B\rightarrow B\rightarrow C)$, $y$ type $D\rightarrow (A\cup B)$, and
 $z$ type $D$.
The one to the subject reduction is similar.
We comment that the counter-examples are not applicable
 to our call-by-value system.
Indeed, $x(yz)(yz)$ is unable to have type $D$, whereas
 $\uplet{xuu}u{yz}$ has.
Note that the latter term has no reduction to the former in the
 call-by-value calculus.
In exchange, the intersection-introduction rule must be restricted
 in our system (see Example~\ref{bdh35}, (2)), contrary to the
 case of call-by-name.
\end{rema}

\begin{thm}\label{tno58}
Let $M$ be a CCV $\lambda\mu$-term.

\shrinktopsep
\begin{enumerate}[(1)]
\item
The call-by-value evaluation of $M$ terminates if and only if
 typing judgment
 $\Gamma\;\vdash\;M\mathbin:T\;|\;\Delta$ is derivable
 for some $\Gamma$, $\Delta$, and $T$.

\item
$M$ is normalizable if and only if typing judgment
 $\Gamma\;\vdash\;M\mathbin:T\;|\;\Delta$ is derivable
 for some $\Gamma$, $\Delta$, and $T$, all these three containing
 neither $\omega$ nor $\agemo$.
\end{enumerate}
\end{thm}

\proof\hskip.5em
(1)
By Thm.~\ref{pin99}, the call-by-value evaluation of $M$ terminates
 iff $[\![M]\!]$ is solvable.
The latter is equivalent to the existence of typing $\Pi,\,\Theta
 \;\vdash_s\;[\![M]\!]\mathbin:\tau$ by Prop.~\ref{rrx04}.
Now the if-part is a consequence of Prop.~\ref{qog12},
 and the only-if part of Thm.~\ref{cnp96}.
\quad
(2)
Proof is similar.
We use Thm.~\ref{okm14} and Prop.~\ref{njy56} instead.
The condition that $\Gamma$, $\Delta$, and $T$
 contain neither $\omega$ nor $\agemo$ corresponds
 to the condition that $\Pi$, $\Theta$, and $\tau$ do not contain $\omega$,
 by Def.~\ref{zqv10} and \ref{tyq65}.
\qed

	\vskip0ex

\noindent
We remark that, in Thm.~\ref{tno58}, (2), $\omega$ and $\agemo$ are prohibited
 only in the lowermost typing judgment of the derivation tree.
They may appear elsewhere.

	\vskip0ex

\subsection{Conservative extension}\label{qjf05}

Recall the definition of types in Def.~\ref{cuq73}.
The places of union and intersection are constrained.
For example, in the arrow type $S\rightarrow T$, the left-hand $S$
 cannot start with union.
This kind of type system is relatively exotic.
No restriction is imposed, or different constraints are placed, in
 the union-intersection types appearing in the literature.

In this subsection, we give a conservative extension of our
 former type system.
The restriction on types is removed.
We can write union and intersection wherever we want.
The extension itself is not difficult at all.
The motivation we think of it for all that lies in building
 a platform for comparison with related works.

We define the set of types $A$ with no restriction
 forced in their construction:

	\vskip2ex
        \noindent\kern5em
$A\quad\mathrel{::=}\quad \alpha\ \ |\ \ A\rightarrow A
 \ \ |\ \ \bigcap A\ \ |\ \ \bigcup A$

	\vskip2ex
        \noindent
 where $\alpha$ ranges over atomic types.
Moreover $\bigcap A$ denotes finite intersection $A_1\cap A_2
 \cap\cdots \cap A_n$, and likewise for $\bigcup A$.
Case $n=0$ is included.

For the subtype relation, in addition to the standard rules,
 we take the following supplementary axioms:

	\vskip2ex

\halign{{\footnotesize $(#)$}\kern5em &$#$\hfil &${}\ \leq\ #$\hfil\cr
 {\it sup}_1 & A\cap \bigcup_i B_i & \bigcup_i (A\cap B_i)\cr
 {\it sup}_2 & \bigcap_i (A_i\rightarrow B) & (\bigcup_i A_i)\rightarrow B.\cr
}

	\vskip2ex
        \noindent
Specifically, the standard rules are those of
 Def.~\ref{mmb04} where we discard distinction
 among raw types, subsidiary types, and types, understanding $S$ and $T$ to
 range over all types $A$.
Furthermore, we add the inference rules of reflexivity
 and transitivity.
For distinction, let us write $A\leq_{\it ext}B$ for
 the subtype relation of the extended system.

By the distributivity law (${\it sup}_1$), each type $A$
 is equivalent to a disjunctive normal form $\bigcup_i\bigcap_j
 A_{ij}$.
Namely $A\sim_{\it ext}\bigcup_i\bigcap_jA_{ij}$ where
 $A\sim_{\it ext}B$ means $A\leq_{\it ext}B$ and $B\leq_{\it ext}A$.
Moreover, by the second rule (${\it sup}_2$), each $T\rightarrow T'$ with $T=
 \bigcup_iS_i$ is equivalent to $\bigcap_i(S_i\rightarrow T')$.
Therefore each type $A$ is equivalent to a type $T$ subject
 to the formation rule of Def.~\ref{cuq73}.

	\vskip0ex

\begin{lem}\label{cdk62}
We suppose $A\sim_{\it ext}T$ as well
 as $A'\sim_{\it ext}T'$ where $T$ and $T'$ are subject to the
 rule of Def.~\ref{cuq73}.
Then $A\leq_{\it ext}A'$ implies
 $T\leq T'$ where the latter is by
 the subtype relation defined in \ref{mmb04}.
\end{lem}

\proof\hskip.5em
For each $A$ we can find canonical $T$ satisfying $A\sim_{\it ext}T$
 by the translation from inside out.
First we verify $T\leq T'$ for canonical $T$ and $T'$.
The general case is inferred from this.
\qed

	\vskip0ex

\noindent
Next, we give the extended typing judgment $\Gamma
 \;\vdash_{\it ext}M\mathbin:A\;\mathbin|\;\Delta$.
In the typing environment $\Gamma$, we allow $x\mathbin:A$ where
 $A$ is a general type, contrary to the restriction to subsidiary types in
 the former system~\ref{seh62}.
The inference rules include those
 in Def.~\ref{seh62} where we ignore
 distinction between subsidiary types and types,
 regarding $S$ and $T$ to range over all types $A$.
As an effect of allowing arbitrary $A$, we can simplify
 the rule of application:

	\vskip2ex
\begingroup\small
        \noindent\kern5em
\vbox{\offinterlineskip
 \halign{\strut $#$\hfil\cr
  \Gamma\ \vdash_{\it ext}\ M\mathbin:\bigcup_i(A_i\rightarrow A')\ \mathbin|\ \Delta
   \qquad [\ \Gamma\ \vdash_{\it ext}\ N\mathbin:A_i\ \mathbin|\ \Delta\ ]_i\cr
  \noalign{\hrule}
  \Gamma\ \vdash_{\it ext}\ MN\mathbin:A'\ \mathbin|\ \Delta\cr
}}.

	\vskip2ex
\endgroup
        \noindent
Furthermore, we add the following inference rule introducing
 union in typing environments:

	\vskip2ex
\begingroup\small
        \noindent\kern5em
\vbox{\offinterlineskip
 \halign{\strut $#$\hfil\cr
  \Gamma,\,x\mathbin:A_1\ \vdash_{\it ext}\ M\mathbin:B\ \mathbin|\ \Delta
   \qquad \Gamma,\,x\mathbin:A_2\ \vdash_{\it ext}\ M\mathbin:B\ \mathbin|\ \Delta\cr
  \noalign{\hrule}
  \Gamma,\,x\mathbin:A_1\cup A_2\ \vdash_{\it ext}\ M\mathbin:B\ \mathbin|\ \Delta\cr}}

	\vskip2ex
\endgroup

We want to show that the extended type system is conservative.
Since each type $A$ is equivalent to $T$, we may assume,
 from the beginning,
 that the judgment is $\Gamma\vdash_{\it ext}M\mathbin:T\;\mathbin|\;\Delta$
 where $\Gamma$ consists of $x_k\mathbin:\bigcup_{i_k}S_{k\,i_k}$
 ($k=1,2,\ldots,n$)
 for $S_{k\,i_k}$ and $T$ subject to the formation rule of \ref{cuq73}.
Also the types assigned in $\Delta$ are supposed to be of the form $T$.
Let us suppose that each $i_k$ ranges over an index set $I_k$.
We associate a family of $|I_1|\times|I_2|\times\cdots
 \times|I_n|$ typing judgments:

	\vskip2ex
        \noindent\kern5em
$[\ x_1\mathbin:S_{1\,i_1},
 x_2\mathbin:S_{2\,i_2},\ldots,x_n\mathbin:S_{n\,i_n}\;\vdash\;M\mathbin:T
\;\mathbin|\;\Delta\ ]_{i_1,\cdots,i_n}$

	\vskip2ex
        \noindent
 of the former type system.
Now we have the following conservativity result.

	\vskip0ex

\begin{prop}\label{qdr88}
Typing judgment
 $\Gamma\vdash_{\it ext}M\mathbin:T\;\mathbin|\;\Delta$ is derived
 in the extended system if and only if all of
 $x_1\mathbin:S_{1\,i_1},
 x_2\mathbin:S_{2\,i_2},\ldots,x_n\mathbin:S_{n\,i_n}\;\vdash\;M\mathbin:T
\;\mathbin|\;\Delta$ are derived in the system of Def.~\ref{seh62}.
\end{prop}

\proof\hskip.5em
The only-if part is easy induction on derivations for
 $\vdash_{\it ext}$.
We use Lem.~\ref{cdk62} to deal with the subsumption rule.
For the if-part, we observe that derivations for $\vdash$ are
 literally interpreted as derivations for $\vdash_{\it ext}$.
We employ the added inference rule several times
 to bring them together into a single derivation.
\qed

\begin{cor}\label{gqn38}
If $\Gamma\vdash_{\it ext}M\mathbin:T\;|\;\Delta$ holds where
 $\Gamma$ and $\Delta$ are the typing environments allowed in Def.~\ref{seh62},
 then $\Gamma\;\vdash\;M\mathbin:T\;|\;\Delta$ holds.
\qed
\end{cor}

\begin{rema}\label{bdh35}
Let us compare our type system
 with related systems in the literature.
We take the conservative extension above as a workbench.
We focus on call-by-value calculi having both union and
 intersection.

\shrinktopsep
\begin{enumerate}[(1)]
\advance\parindent by15pt
\item
Dunfield and Pfenning propose a functional language having both union
 and intersection (and more), and verified basic properties
 such as type preservation and progress \cite{mji15}.
No control operators are considered.
The subtype relation coincides with the one defined in Def.~\ref{mmb04}.
The two supplementary axioms are not contained.
Related to the intersection type, the following inference rules
 are considered

	\vskip2ex
\begingroup\small
        \noindent\kern5em
\vbox{\offinterlineskip
 \halign{\strut $#$\hfil\cr
  \Gamma\ \vdash\ V\mathbin:A\qquad\Gamma\ \vdash\ V\mathbin:B\cr
  \noalign{\hrule}
  \Gamma\ \vdash\ V\mathbin:A\cap B\cr
}}\kern4em
\vbox{\offinterlineskip
 \halign{\strut $#$\hfil\cr
  \noalign{\hrule}
  \Gamma\ \vdash\ V\mathbin:\omega\cr
}}

	\vskip2ex
\endgroup
        \noindent
 where symbols are changed for the sake of comparison.
We note that the rules are applied only to value $V$.
In \cite{qno45}, it is mentioned that the subject reduction fails
 if they are extended to non-values, under the existence of mutable stores.
These rules are derivable in our system viewing \ref{uws76} and
 \ref{qdr88} (the latter is tacitly used below repeatedly).
For union types, the following inference rules are considered.

	\vskip2ex
\begingroup\small
        \noindent\kern5em
\vbox{\offinterlineskip
 \halign{\strut $#$\hfil\cr
  \Gamma\ \vdash\ M\mathbin:A\cup B\qquad
   \Gamma,\,x\mathbin:A\ \vdash\ E[x]\mathbin:C\qquad
   \Gamma,\,x\mathbin:B\ \vdash\ E[x]\mathbin:C\cr
  \noalign{\hrule}
  \Gamma\ \vdash\ E[M]\mathbin:C\cr
}}

	\vskip2ex
        \noindent\kern5em
\vbox{\offinterlineskip
 \halign{\strut $#$\hfil\cr
  \Gamma\ \vdash\ M\mathbin:\agemo\cr
  \noalign{\hrule}
  \Gamma\ \vdash\ E[M]\mathbin:C\cr
}}

	\vskip2ex
\endgroup
        \noindent
These rules are derivable in our system.
Let us consider the first.
From two hypotheses for $E[x]$, we have $\lambda x.\,E[x]
 \mathbin:(A\rightarrow C)\cap (B\rightarrow C)$ by the inference
 rule for lambda abstraction in \ref{seh62}.
Then the rule for application yields $(\lambda x.\,E[x])M\mathbin:
 C$.
Now we have $(\lambda x.\,E[x])M=_{\it ccv}E[M]$ as seen in Example~\ref{isn97}.
Thus the subject reduction/expansion~\ref{ehw91} yields
 $\Gamma\;\vdash E[M]\mathbin:C$.
(Since $\eta_\lambda$ is not needed in \ref{isn97},
 we need no change of types or typing environments
 as commented in Rem.~\ref{dhc26}.)
The second rule is similar, noticing $\lambda x.\,E[x]$ has type $\omega$.
So the union-intersection fragment of their system is a subsystem
 of ours.

\item
Ishihara and Kurata propose type system ${\rm TA}$ and slightly weaker
 system ${\rm TA}^-$ \cite{muj96}.
They define a class of call-by-value lambda models called ${\rm TA}$-models,
 and show that ${\rm TA}$ is sound while ${\rm TA}^{-}$ is complete
 with respect to certain denotational models.
No control operator exists.
Their subtype relation contains all the rules of ours, including
 the supplementary axioms introduced in this subsection.
The typing rule related to union is

	\vskip2ex
\begingroup\small
        \noindent\kern5em
\vbox{\offinterlineskip
 \halign{\strut $#$\hfil\cr
  \Gamma\ \vdash\ V\mathbin:A\cup B\qquad
   \Gamma,\,x\mathbin:A\ \vdash\ M\mathbin:C\qquad
   \Gamma,\,x\mathbin:B\ \vdash\ M\mathbin:C\cr
  \noalign{\hrule}
  \Gamma\ \vdash\ M\{V/x\}\mathbin:C\cr
}}.

	\vskip2ex
\endgroup
        \noindent
(In ${\rm TA}^-$, the value $V$ is restricted to a variable.)
This rule is derivable in our system as well.
Indeed, from $\lambda x.\,M\mathbin:(A\rightarrow C)\cap(B\rightarrow C)$,
 we obtain $(\lambda x.\,M)V\mathbin:C$.
Now the subject reduction~\ref{ehw91} yields $\Gamma\;\vdash\;
 M\{V/x\}\mathbin:C$.
Hence the rules related to union are simulated in our system.

However, the manipulation of intersection is different.
Their system contains axiom $(A\rightarrow B)\cap (A\rightarrow C)
 \leq A\rightarrow B\cap C$.
Accordingly, it contains an inference rule

	\vskip2ex
\begingroup\small
        \noindent\kern5em
\vbox{\offinterlineskip
 \halign{\strut $#$\hfil\cr
  \Gamma\ \vdash\ M\mathbin:A\qquad \Gamma\ \vdash\ M\mathbin:B\cr
  \noalign{\hrule}
  \Gamma\ \vdash\ M\mathbin:A\cap B\cr
}}

	\vskip2ex
\endgroup
        \noindent
 for arbitrary $M$, not restricted to values.
Let us see what happens with this rule.
We take types such that $A\not\sim_{\it ext}(A\cap B)\not\sim_{\it ext}B$.
Let us suppose $a\mathbin:A$, $b\mathbin:B$, and $f\mathbin:
 ((X\rightarrow X)\rightarrow \omega\rightarrow X)\cap
 (\omega\rightarrow (X\rightarrow X)\rightarrow X)$ where
 $X$ is arbitrary for it plays no role.
Moreover, let $l$ be of type $X$.
We consider $M=\mu k.\,[l]f(\lambda x\mu\delta.\,[k]a)(\lambda x\mu\delta.\,[k]b)$.
The term $M$ has type $A$.
For this, we let the type of $\lambda x\mu\delta.\,[k]b$ be $\omega$
 (see Rem~\ref{qnd21}, (2)).
Symmetrically $M$ has type $B$.
However, $M$ does not have type $A\cap B$ in our system, since
 $f$ cannot have type $\omega\rightarrow\omega\rightarrow X$.
If we naturally extend their system to the $\lambda\mu$-calculus,
 therefore, it is a non-conservative extension of ours.
(System ${\rm TA}$ has no $\omega$ but has $\nu$ behaving similarly.)

Furthermore, the extension violates the subject reduction property.
Let us set $A=\alpha\rightarrow\beta$ and $B=\beta$.
If we take $\uplet {zz}zM$ for $M$ above, it is typeable.
But it contracts to $\mu k.\,[l]f(\lambda x\mu\delta.\,[k]aa)
 (\lambda x\mu\delta.\,[k]bb)$ that is not typeable at $aa$ as well as $bb$.

\item
Palsberg and Pavlopoulou establish a link between
 polyvariant flow analyses and union-intersection type systems \cite{ehr09}.
They show equivalence between flow-safety and typeability.
Although their types are infinite regular trees for their
 purpose, here we ignore that aspect, and
 formally compare their type systems presented
 in \S3.
The types are restricted to disjunctive normal forms.
So axiom $({\it sup}_1)$ is implicitly needed.
The rules of the acceptable type ordering Def.~3.2, p.~278,
 are derivable from our subtype relation.
Also the axioms of a specific acceptable type
 ordering in \S3.6, p.~284, are derivable.
To derive axiom (21) $\bigcap_i(A_i\rightarrow B_i)\leq
 (\bigcup_iA_i)\rightarrow (\bigcup_iB_i)$ therein, we need
 axiom $({\it sup}_2)$.
The typing rules for the lambda-calculus fragment in p.~279
 are the same as our rules \ref{seh62} except for application.
The rule for application is the restriction of ours where
 the union $\bigcup_i$ is not involved.
Therefore their system is a subsystem of ours.

\item
Van~Bakel gives union-intersection type systems for
 the $\overline\lambda\mu\tilde\mu$-calculus and
 the calculus $\mathcal{X}$
 based on the classical sequent calculus \cite{fuu19}\cite{czm05}.
Both of these calculi include the call-by-name and the call-by-value fragments.
Since the syntax is quite different from the $\lambda\mu$-calculus,
 a relation is not clear at this stage.
We leave comparison to the future work.
\end{enumerate}
\end{rema}

\section{Call-by-value catch/throw calculus: a chain of parasitism}\label{weg04}

We can make a chain of parasitism.
We have defined the
 CCV $\lambda\mu$-calculus, and verified several properties
 using proof by parasitism.
Now we can take the calculus as the new target of
 a translation of another calculus.
Namely, we cascade proof by parasitism for the creation of
 new complete calculi.

In this section, we give a call-by-value version of
 Crolard's catch/throw calculus $\lambda_{\rm ct}$
 \cite{ltb70}.
We verify the completeness with respect to the translation
 into the CCV $\lambda\mu$-calculus as a demonstration of
 a chain of parasitism.

	\vskip3ex

\begin{defi}\label{bqz60}
We separate ordinary variables $x$ and continuation variables $k$.
The terms $M$ of the call-by-value catch/throw calculus are defined by

	\vskip2ex
        \noindent\kern5em
$M\ \mathrel{::=}\ x\ \ \ |\ \ \ \lambda x.\,M\ \ \ |\ \ \ MM\ \ \ |
 \ \ \ \uplet MxM\ \ \ |\ \ \ \varepsilon k.\,M\ \ \ |\ \ \ \exraise kM$.

	\vskip2ex
        \noindent
The construct $\varepsilon k$ binds the continuation variable $k$,
 so that it obeys $\alpha$-convertibility.
The displayed occurrence of $k$ in $\exraise kM$ is free.
We note that continuation variables are used only in this form.
\end{defi}

	\vskip0ex

\noindent
We change notations from \cite{ltb70}.
Binding operator $\varepsilon k$ builds a new exception block tagged
 by identifier $k$.
The construct $\exraise k$ throws an exception that is
 caught by the encapsulating block tagged by $k$.
Since $\varepsilon k$ is subject to $\alpha$-conversion,
 the calculus gives the static catch/throw mechanism.

	\vskip0ex

\begin{defi}\label{ttf80}
We introduce the following equality rules between terms:
	\vskip2ex

\halign{\kern5em $#$\hfil &${}\ =\ #$\qquad\hfil &#\hfil\cr
 \uplet Lx{(\uplet MyN)} & \uplet{(\uplet LxM)}yN &if $y\not\in L$\cr
 \uplet {(\varepsilon k.\,M)}xN & \varepsilon k.\,(\uplet MxN)\ &if $k\not\in N$\cr
 \uplet{(\exraise kM)}xN & \exraise k{(\uplet MxN)}\cr
}

	\vskip2ex
        \noindent
We usually omit brackets as in the CCV $\lambda\mu$ calculus (see
 Rem.~\ref{bmv07}).
For the reduction rules, we replace $\beta_\mu,\beta_{\it jmp}$, and
 $\eta_\mu$ in Def.~\ref{ofe40} by the following seven rules:

	\vskip2ex

\halign{\kern5em $#$\hfil &${}\quad\rightarrow\quad #$\hfil\cr
 \varepsilon\delta.\,M & M\cr
 \varepsilon k.\,\exraise kM & \varepsilon k.\,M\cr
 \uplet Mx{\exraise kN} & \exraise kN\cr
 \exraise l{\exraise kN} & \exraise kN\cr
 \uplet Mx{\varepsilon k.\,N} & \varepsilon k.\,\uplet MxN\{\exraise k\square
  \mapsto \exraise k{\uplet Mx\square}\}\cr
 \exraise l{\varepsilon k.\,M} & \exraise l{M\{l/k\}}\cr
 \varepsilon l\varepsilon k.\,M & \varepsilon l.\,M\{l/k\}\cr
}

	\vskip2ex
        \noindent
 where $\delta$ is a dummy continuation variable that does not
 occur free in $M$.
The notation $\{\exraise k\square\mapsto \exraise k\uplet Mx\square\}$ 
 denotes the structural substitution similar to the one in
 the $\lambda\mu$-calculus.
We notice that the substitution is applied recursively.
Moreover $\{l/k\}$ is the abbreviation of
 structural substitution $\{\exraise k\square\mapsto
 \exraise l\square\}$.
\end{defi}

	\vskip0ex

\noindent
Continuation variables $k$ are, unlike ordinary variables,
 never substituted with values.
They behave only as placeholders to mark the positions to
 insert continuations as in the $\lambda\mu$-calculus.

	\vskip0ex

\begin{rema}\label{svi34}\hfill
\shrinktopsep
\begin{enumerate}[(1)]
\advance\parindent by15pt
\item
The operational semantics of call/cc is different
 from that of catch.
The former creates a continuation object at
 each call, whilst the latter simply waits for an exception
 to be raised \cite{ztk07}.
Though called catch/throw for historical reasons,
 our calculus equally simulates the first-class continuation.
The following encoding is found in \cite[p.~646]{ltb70}:

	\vskip2ex

\halign{\kern5em #\hfil\cr
$\mathop{\sf call/cc}M\ \ \mathrel{:=}\ \ \varepsilon k.\,M(\lambda x.\exraise kx)$.
\cr}

	\vskip2ex
        \noindent
Let $\hat\tau$ be a fixed continuation variable representing the topmost
 context.
Let us use the equality symbol for the congruence relation 
generated from the equality and reduction rules in Def.~\ref{ttf80}.
Then we have

	\vskip2ex

\halign{\kern5em $#$\hfil &${}\ =\ #$\hfil\cr
 \exraise{\hat\tau}E[\,\mathop{\sf call/cc}M]
  & \exraise{\hat\tau}E[\varepsilon k.\,M(\lambda x.\exraise kx)]\cr
 & \exraise{\hat\tau}\uplet{E[x]}x{\varepsilon k.\,M(\lambda x.\exraise kx)}\cr
 & \exraise{\hat\tau}E[M(\lambda x.\exraise{\hat\tau}E[x])]\cr
}

	\vskip2ex
        \noindent
 (see \ref{isn97} for the second equality).
In particular, if $E$ has the form $\uplet Lx\square$,
 the first contracts to the last.
Hence, if we regard $\exraise{\hat\tau}$ as the abort operator $\mathcal{A}$,
 we regain the ordinary operational semantics of the call/cc-operator.
Therefore $\varepsilon k$ is much like a control operator than a simple
 catch mechanism.
In fact, the abstract machine model in \cite{ltb70} copies
 the stack environment at each call of catch.

The little twist in the encoding of call/cc
 above is caused by the character
 of the calculus that
 the continuation variables are unable to be substituted.
The first-class continuation is not directly materialized
 in our calculus.
In view of $\mathop{\sf call/cc}(\lambda y.\,L)=\varepsilon k.\,
 \uplet Ly{\lambda x.\,\exraise kx}$, we observe
 that the continuation $k$ is
 passed to the ordinary variable $y$ by making the first-class closure
 $\lambda x.\,\exraise kx$,  the essence of which is $k$ itself.

\item
Sato defines a calculus of catch/throw in \cite{pcd30},
 simplifying Nakano's calculus
 \cite{ogv33}.
The catch operator $\mathop{\sf catch}_kM$ has
 disjunction type $A\vee B$ where $A$ is the type of $M$ representing
 a normal exit while the type $B$ of $k$ is for non-local exits.
(In \cite{pcd30},
 symbol $\mathord?$ is used for catch and $\mathord!$ for throw.)
In the classical propositional logic, $A\vee B$ is equivalent to
 $(B\rightarrow A)\rightarrow A$.
Two injections are represented by $\mathop{\sf inl}M=
 \lambda f.\,M$ and $\mathop{\sf inr}M=
 \lambda f.\,fM$.
Let us adopt the following encoding:

	\vskip2ex

\halign{\kern5em $#$\hfil &${}\ \mathrel{:=}\ #$\hfil\cr
 \mathop{\sf catch}_kM & \lambda f\,\varepsilon \gamma.\,f(\varepsilon k.
  \exraise \gamma M)\cr
 \mathop{\sf throw}_kM & \exraise kM\cr
 \mathop{\sf tapply}(M,k) & M(\lambda x.\exraise kx)\cr
}

	\vskip2ex
        \noindent
 where $\gamma$ is a fresh continuation variable.
If we develop the simply typed system appropriately, $\mathop{\sf catch}_kM$
 has type $(B\rightarrow A)\rightarrow A$.
The six reduction rules in \cite[p.~230]{qwi06} are

	\vskip2ex

\halign{\kern5em $#$\hfil &${}\ \mathrel{\rightarrow}\ #$\kern3em\hfil &(#)\hfil\cr
 \uplet Lx{\mathop{\sf throw}_kM} & \mathop{\sf throw}_kM\cr
 \mathop{\sf catch}_kM & \mathop{\sf inl}M & if $k\not\in M$\cr
 \mathop{\sf catch}_k\mathop{\sf throw}_kM & \mathop{\sf inr}M & if $k\not\in M$\cr
 \mathop{\sf tapply}(\mathop{\sf inl}M,k) & M\cr
 \mathop{\sf tapply}(\mathop{\sf inr}M,k) & \mathop{\sf throw}_kM\cr
 \mathop{\sf tapply}(\mathop{\sf catch}_kM,l) & M\{l/k\}\cr
}

	\vskip2ex
        \noindent
 where the
 first rule is replaced since we are concerned with call-by-value.
In our favor, this replacement serves to get rid of the
 non-confluence example (p.~229, ibid.).
These reduction rules are then realized by one or more steps of contractions
 in our calculus.
Therefore their catch-throw calculus is a subsystem of ours, respecting
 even the direction of reductions.

\item
Next, we consider de~Groote's system \cite{kwx35}.
We encode the exception handling block as

	\vskip2ex

\halign{\kern5em #\hfil\cr
 ${\sf let}\ k\ {\sf in}\ M\ {\sf handle}\ kx\Rightarrow N
 \quad\mathrel{:=}\quad
 \varepsilon \gamma.\,\uplet Nx{\varepsilon k.\,\exraise\gamma M}$
\cr}

	\vskip2ex
        \noindent
 where $\gamma$ is a fresh continuation variable.
We note that it is equal to $(\mathop{\sf catch}_kM)(\lambda x.\,N)$
 using the encoding of the catch above.
For the raise operator, we identify $\mathop{\sf raise}(kM)$ in the original
 with $\exraise kM$ and avoid using tag-less raise.
Then the reduction rules in Tab.~2, p.~206 are simulated
 after appropriate modification.
Unlike the case of Sato's system, the orientation of the reduction is not
 preserved.
We are content to show that the two sides of the reduction rules are equal.
The most essential is the following rule called (handle/raise):

	\vskip2ex

\halign{\kern5em $#$\hfil &\kern3em $#$\kern3em &$#$\hfil \cr
 {\sf let}\ k_1\ {\sf in} && {\sf let}\ k_1\ {\sf in}\cr
 {\sf let}\ k_2\ {\sf in} && {\sf let}\ k_2\ {\sf in}\cr
 \ \ \vdots && \ \ \vdots\cr
 {\sf let}\ k_n\ {\sf in} && {\sf let}\ k_n\ {\sf in}\cr
 \ \ \exraise{(k_1V)}{} & \rightarrow & \ \ N_1\{V/x\}\cr
 {\sf handle}\ k_nx\Rightarrow N_n && {\sf handle}\ k_nx\Rightarrow N_n\cr
 \ \ \vdots && \ \ \vdots\cr
 {\sf handle}\ k_2x\Rightarrow N_2 && {\sf handle}\ k_2x\Rightarrow N_2\cr
 {\sf handle}\ k_1x\Rightarrow N_1 && {\sf handle}\ k_1x\Rightarrow N_1\cr
}

	\vskip2ex
        \noindent
 where $N_i$ may contain $k_1,k_2,\ldots,k_{i-1}$.
The translations of the two sides contract to a common term
 $\varepsilon\gamma.\,N_1\{x\mapsto
 V\theta_n\cdots\theta_2\theta_1\}$ where
 $\theta_i=\{\exraise k_i\square\mapsto\exraise\gamma \uplet{N_i}x
 \square\}$.

\item
The catch/throw-related part of Krebbers' system \cite{djp22}
 is easily simulated by identifying $\mathop{\sf catch}k$ with
 $\varepsilon k$ and $\mathop{\sf throw}k$ with $\exraise k$.
\end{enumerate}
\end{rema}

	\vskip0ex

\noindent
We show that our call-by-value catch/throw calculus is sound and complete.
We take the CCV $\lambda\mu$-calculus as the target language.

	\vskip0ex

\begin{defi}\label{wwh18}
The translation $[\![\hbox{-}]\!]$
 from the call-by-value catch/throw calculus into the CCV $\lambda\mu$-calculus
 and the inverse translation $(\hbox{-})^{-1}$ are defined.
For the control operators, we define as follows:

	\vskip2ex

\halign{\kern5em $#$\hfil &${}\ =\ #$\hfil &\kern5em $#$\hfil &${}\ =\ #$\hfil\cr
 [\![\varepsilon k.\,M]\!] & \mu k.\,[k][\![M]\!]
  & (\mu k.\,J)^{-1} & \varepsilon k.\,J^{-1}\cr
 [\![\exraise kM]\!] & \mu\delta.\,[k][\![M]\!]
  & ([k]M)^{-1} & \exraise k{M^{-1}}\cr
}
	\vskip2ex
        \noindent
 where $\delta$ is a dummy continuation variable.
For the other constructs, the translations are homomorphic except that
 for the application we set

	\vskip2ex

\halign{\kern5em $#$\hfil &${}\ =\ #$\hfil \cr
 [\![N_1N_2]\!] & \uplet {zw}w{\uplet {[\![N_2]\!]}z{[\![N_1]\!]}}\cr
 [\![NV]\!] & \uplet {z[\![V]\!]}z{[\![N]\!]}\cr
 [\![VN]\!] & \uplet {[\![V]\!]z}z{[\![N]\!]}\cr
 [\![V_1V_2]\!] & [\![V_1]\!][\![V_2]\!]\cr
}

	\vskip2ex
        \noindent
 where $V$ is a value and $N$ a non-value.
This definition is chosen to ensure
 that $[\![M]\!]$ becomes an administrative normal form.
Except this technical point, the essentially same translations
 are found in \cite{ltb70} and \cite{qwi06}.
\end{defi}

\begin{lem}\label{ciu82}
If $[\![M]\!]\mor*N$ holds in the CCV $\lambda\mu$-calculus,
 then $M\mor*(N^{-1})^\downarrow$ holds, where
 $(\hbox{-})^\downarrow$ means to take the normal form
 with respect to vertical reductions.
\end{lem}

\begin{thm}\label{ajz96}
The call-by-value catch/throw calculus satisfies the Church-Rosser property.
\end{thm}

\proof\hskip.5em
Proceed along the same line as in Thm.~\ref{pje26}, using
 completeness~\ref{ciu82} and the Church-Rosser property of the
 target~\ref{pje26}.
\qed

	\vskip0ex

\noindent
To avoid that the paper is too long, we stop here.
The results in previous sections are extended in a relatively simple way.
We refer the reader to \cite{jdv67}.

\section{Conclusion}\label{khk00}

We present a new call-by-value $\lambda\mu$-calculus and
 a catch/throw calculus that are complete
 for the contin\-uation-passing style semantics.
A key idea is to deviate from the standard convention that
 terms are syntactic trees.
If we are to accept this deviation, we can have simple but complete
 systems.

The calculi satisfy
 a certain form of completeness with respect to
 reduction.
Cultivating the completeness, we verify a number of anticipated
 properties by pulling back the
 corresponding ones in the ordinary lambda calculus.
This method is called proof by parasitism.
Although the idea itself is not new, we persistently pursue
 its benefits.
We succeed in showing Church-Rosser, and 
 in characterizing normalizability and termination
 of call-by-value evaluation in terms of the CPS translation
 for the call-by-value $\lambda\mu$-calculus.
We also give a union-intersection type system and verifies
 that various syntactic properties are characterized by typeability.

Related works are mentioned in the relevant parts of this paper.
We do not repeat all.
Here we review the previous results on the completeness of lambda calculi
 and their extensions,
 focusing largely on call-by-value systems having control operators.
Almost all previous systems are presented
 as equational theories.
Hence any results comparable to ours do not
 exist in the previous works
 (except Church-Rosser in \cite{mhx11}).

As far as the author knows, the first work establishing
 completeness of the call-by-value lambda calculus
 is done by Sabry et al.~\cite{yrs06}\cite{vyh68}.
The inverse of the CPS translation is first introduced in
 their papers.
The rules of the call-by-value lambda calculus
 in \cite[Fig.~2, p.~311]{yrs06} are
 sound and complete for the CPS translation.
They consider reduction rules, not only equalities.
Reduction sequences are preserved and reflected
 by the CPS translation.
Furthermore, they extend the call-by-value lambda calculus by adding the
 control operators, call/cc and abort.
The rules of Fig.~4 (p.~319) are sound and complete.
For the extended system, only equalities are considered.
If we encode the control operators as $\mathop{\sf call/cc}M
 =\mu k.\,[k]M(\lambda x\mu\delta.\,[k]x)$ and
 $\mathop{\sf abort}M=\mu\delta.\,[\hat\tau]M$ using
 a fixed special continuation variable $\hat\tau$, the rules of
 their calculus are derived from our CCV $\lambda\mu$-calculus
 up to equality.

Hofmann deals with Felleisen's $\mathcal{C}$-operator
 and abort \cite{lss57}.
The equational theory is defined in Def.~1 in p.~465.
It is sound and complete for the categorical model introduced
 in Def.~5 (p.~470).
The proof of completeness is by the construction of the term model.
The calculus has especially the initial type $0$.
The rules involving the abort operator depend on types.
If $M$ is of type $0$, we have rule ${\mathcal{A}}M=M$ called
 ($\mathcal{A}_0$-{\it Id}).
The rules save this one are derived in the CCV $\lambda\mu$-calculus
 if we encode $\mathcal{C}M=\mu k.\,[\hat\tau]M
 (\lambda x\mu\delta.\,[k]x)$.
Later the system is simplified in \cite[\S5.3]{iba58}.

F\"uhrmann and Thielecke simplify Hofmann's calculus \cite{iab45}.
As a new rule, they introduce $\lambda k.\,k(
 \mathcal{C}V)=V$ called rule ($\mathcal{C}$-Delay).
Completeness is verified in Thm.~40 (p.~261),
 Thm.~45 (p.~263), and Cor.~71 (p.274) in various settings.
The type of operator $\mathcal{C}$ is $((A\rightarrow O)
 \rightarrow O)\rightarrow A$ where $O$ is the initial type.
We warn that rule ($\mathcal{C}$-delay) is
 not valid in the CCV $\lambda\mu$-calculus, if we take the encoding of
 $\mathcal{C}$ above, that can have type $((A\rightarrow B)
 \rightarrow B)\rightarrow A$ for arbitrary $B$.
The rule heavily depends on the fact that $O$ is initial.
To make the rule valid in the call-by-value $\lambda\mu$-calculus,
 we must regard $\dbot$ as an atomic type
 and add rules related to the continuation variables of type $\dbot$
 as done in \cite{fck86}.

Hofmann and Streicher discuss both the call-by-name
 and the call-by-value $\lambda\mu$-calculus \cite{iba58}.
Terms and jumps are not separated.
For the call-by-value language, Fig.~5 in p.~347 gives
 the equational theory.
It is sound and complete for the categorical model in
 Def.~5.1 (p.~346).
The calculus contains several rules involving
 the initial type $0$.
Except these, all rules are derivable from ours.

Selinger's theory of the call-by-value $\lambda\mu$-calculus
 is essentially the same
 as ours, though that paper deals with only equalities \cite[Tab.~10]{fck86}.
The theory is sound and complete for the categorical model,
 called the co-control category (\S.4.2).
His system contains also tensor, unit, and finite coproducts.
As minor differences, the let-construct ${\sf let}\ x=M
 \ {\sf in}\ N$ is an abbreviation of $(\lambda x.\,N)M$,
 and terms and jumps are not separated.

Herbelin and Zimmermann give a reduction system, called
 the $\lambda\mu{\sf tp}_{\it CBV}$-calculus, which is complete
 for the CPS semantics \cite{mhx11}.
Their system uses the let-notation as a primitive.
It contains the rule of type ${\sf let}\ x=M\ {\sf in}\ E[x]\rightarrow
 E[M]$ as well as the let-flat rule.
This calculus satisfies confluence (Prop.~9, p.~152).
The idea is to disallow $E[V\square]$ as the evaluation context if $V$ is a
 lambda abstraction, allowing only $E[x\square]$ for variable $x$.
The contexts that are able to be captured by $\mu$ are restricted
 accordingly.

If we extend the range to delimited control operators,
 we can find the axiomatization by Kameyama and Hasegawa \cite{gnj37}.
It deals with equational theories of shift/reset and the delimited call/cc.
Completeness for the CPS translation
 is verified by the inverse translation.
Moreover, Ariola et al.~give several calculi
 having delimited control operators \cite{vtp88}.
The calculus in \S7.5 (p.~262) is complete with
 respect to the Kameyama-Hasegawa's theory.
Although the system is given with reduction rules, the completeness
 result is obtained for equational theories.

The above is an overview of the complete call-by-value calculi.
For the call-by-name, the pioneering paper \cite{pag41} by
 Plotkin already verifies the completeness of the $\lambda$-calculus
 in Thm.~6 (p.~153).
The proof is by simulation relations.
It is extended by de~Groote to the call-by-name
 $\lambda\mu$-calculus \cite{owr18}.
Fujita establishes completeness by
 the inverse translation \cite{cmv29}\cite{eiq93}.
He manages to deal with reduction using an idea similar to ours.
Also Hofmann-Streicher \cite{iba58} and Selinger \cite{fck86}
 show completeness with respect to categorical models.

\section*{Acknowledgement}

We are grateful to Kazunori Tobisawa for letting the author
 notice the importance of Cor.~\ref{lly89}, especially
 in the context of the coding of cooperative multitasking~\ref{sdp73}.

\appendix

\renewcommand{\thesection}{Appendix \Alph{section}}

\section{Properties of the type system for the target calculus}\label{ada92}

\renewcommand{\thesection}{\Alph{section}}

We collect several results of the intersection type system
 for the target calculus defined in \S\ref{ehd63}.
The argument follows the standard one that uses filter domains
\cite{jix62}\cite{ftj70}.
So we omit proof mostly.

We need subject reduction/expansion for the type
 system in Def.~\ref{amg62}.
The results in the literature are not directly applicable,
 though our system is a subsystem of \cite{jix62}
 if we forget about sorts.
The reason is that the types of terms are constrained
 relative to their sorts.
For example, a type of term $K$ must be of the form
 $\neg\underline\sigma$, i.e, $\underline\sigma\rightarrow\dbot$.
The right hand of the arrow is restricted to $\dbot$.
We must ensure that the subject reduction/expansion
 remain to hold under the constraint.
We extend the standard
 arguments to many-sorted languages.

We extend the types in Def.~\ref{rle89} by introducing
$\underline\tau\;\mathrel{::=}\;\bigcap\tau$.

	\vskip0ex

\begin{defi}\label{jbv06}
A {\it $T$-filter} $d$ is a non-empty set of types $\underline\tau$
 satisfying the ordinary filter conditions:
(i) $\underline\tau\cap\underline\tau'\in d$ whenever
 $\underline\tau,\underline\tau'\in d$;\quad
(ii) $\underline\tau'\in d$ whenever $\underline\tau\in d$
 and $\underline\tau\leq\underline\tau'$.
Similarly we define a $K$-filter and a $W$-filter by replacing
 $\underline\tau$ with $\underline\kappa$ and $\underline\sigma$.
We can define also a $Q$-filter as a filter on the singleton
 set $\{\dbot\}$.
\end{defi}

\begin{defi}\label{xcz79}
Filter domain $\mathcal{T}$ is the set of
 all $T$-filters.
It forms an algebraic complete partial order (CPO)
 \cite{xml79} in regard to inclusion.
Similarly filter domains $\mathcal{K},\mathcal{W}$, and
 $\mathcal{Q}$ are defined.
We comment that $\mathcal{Q}$ is the Sierp\'inski space
 consisting of two filters $\{\omega\}$ and $\{\omega,\dbot\}$.
\end{defi}

	\vskip0ex

\noindent
As usual, we let $[\mathcal{D}\rightarrow\mathcal{D}']$ denote
 the CPO of all Scott continuous maps of $\mathcal{D}$ into $\mathcal{D}'$,
 partially-ordered by the pointwise ordering.
We let $\mathop\uparrow X$ denote the smallest filter
 containing the set $X$.

	\vskip0ex

\begin{defi}\label{sto75}
We define six continuous maps:

	\vskip2ex
        \noindent\kern5em
$\mathcal{T}
 \vcenter{\offinterlineskip
  \halign{\hfil $\ \ #\ \ $\hfil\cr
   {\scriptstyle F_T}\cr \longrightarrow\cr \longleftarrow\cr {\scriptstyle G_T}\cr
 }}
 [\mathcal{K}\rightarrow\mathcal{Q}]$,\kern3em
$\mathcal{K}
 \vcenter{\offinterlineskip
  \halign{\hfil $\ \ #\ \ $\hfil\cr
   {\scriptstyle F_K}\cr \longrightarrow\cr \longleftarrow\cr {\scriptstyle G_K}\cr
 }}
 [\mathcal{W}\rightarrow\mathcal{Q}]$,\kern3em
$\mathcal{W}
 \vcenter{\offinterlineskip
  \halign{\hfil $\ \ #\ \ $\hfil\cr
   {\scriptstyle F_W}\cr \longrightarrow\cr \longleftarrow\cr {\scriptstyle G_W}\cr
 }}
 [\mathcal{W}\rightarrow\mathcal{T}]$,

	\vskip2ex
        \noindent
 these defined as follows:

	\vskip2ex

\halign{\kern5em $#$\hfil &${}\ =\ #$\hfil\cr
 F_T(d) & \lambda e.\,\mathop\uparrow\{\dbot\,|\>\exists\underline\kappa\in e.\,
  \neg\underline\kappa\in d\}\cr
 F_K(d) & \lambda e.\,\mathop\uparrow\{\dbot\,|\>\exists\underline\sigma\in e.\,
  \neg\underline\sigma\in d\}\cr
 F_W(d) & \lambda e.\,\mathop\uparrow\{\tau\,|\>\exists\underline\sigma\in e.\,
  (\underline\sigma\rightarrow\tau)\in d\}\cr
	\noalign{\vskip1ex}
 G_T(f) & \mathop\uparrow\{\neg\underline\kappa\,|\>\dbot\in f(
  \mathop\uparrow\underline\kappa)\}\cr
 G_K(f) & \mathop\uparrow\{\neg\underline\sigma\,|\>\dbot\in f(
  \mathop\uparrow\underline\sigma)\}\cr
 G_W(f) & \mathop\uparrow\{\underline\sigma\rightarrow\tau\,|\>\tau\in f(
  \mathop\uparrow\underline\sigma)\}.\cr
}

	\vskip2ex
        \noindent
In the last three, for example, $\mathop\uparrow\underline\sigma$ denotes
 the principal filter generated by $\underline\sigma$ (namely,
 an abbreviation of $\mathop\uparrow\{\underline\sigma\}$).
We remark that the sets occurring here are upward-closed for
 strict types.
Hence $\mathop\uparrow\{\cdots\}$ means taking the closure
 with respect to finite intersection, especially, adding $\omega$ forcibly.
\end{defi}

	\vskip0ex

\noindent
The interpretation of terms of the target calculus, $[\![T]\!]\xi$
 etc., is naturally defined.
An environment $\xi$ assigns a $\mathcal{W}$-filter $e$ to
 each ordinary variable $x$ and a $\mathcal{K}$-filter $d$ to
 each continuation variable $k$.
We have $[\![T]\!]\xi\in \mathcal{T}$, $[\![K]\!]\xi\in \mathcal{K}$,
 $[\![W]\!]\xi\in \mathcal{W}$, and $[\![Q]\!]\xi\in \mathcal{Q}$. 
The definition of the interpretation is standard.
For example, $[\![\lambda k.\,Q]\!]\xi\mathrel{:=}G_T(\lambda d.\,
 [\![Q]\!]\xi\{d/k\})$ and $[\![W_1W_2]\!]\xi
 \mathrel{:=}F_W([\![W_1]\!]\xi)
 ([\![W_2]\!]\xi)$.

To understand relation between the filter domains and the
 intersection types, it is convenient to
 formally introduce the environments associating filters to variables,
 $x\mathbin:d$ and $k\mathbin:e$ where $d\in \mathcal{W}$
 and $e\in \mathcal{K}$.
We let $\Pi_{\it fil}$ denote an environment
 of $x\mathbin:d$ and $\Theta_{\it fil}$
 an environment of $k\mathbin:e$.
As usual, the variables occurring in the environments
 should be distinct from each other.

	\vskip0ex

\begin{defi}\label{qmv74}
We introduce the typing judgment $\Pi_{\it fil},\,
 \Theta_{\it fil}\>\vdash_s\>T\mathbin:\tau$.
We interpret this judgment as 
 there are $\Pi$ and $\Theta$ satisfying the following three conditions:
(i) Supposed that $\Pi_{\it fil}$ is $x_1\mathbin:d_1,x_2\mathbin:d_2,
 \ldots,x_m\mathbin:d_m$, the typing environment $\Pi$ is of the form
 $x_1\mathbin:\underline\sigma_1,x_2\mathbin:\underline\sigma_2,\ldots,
 x_m\mathbin:\underline\sigma_m$ where each $\underline\sigma_i$ is a member
 of filter $d_i$.
(ii) Supposed that $\Theta_{\it fil}$ is $k_1\mathbin:e_1,k_2\mathbin:e_2,
 \ldots,k_n\mathbin:e_n$, the typing environment $\Theta$ is of the form
 $k_1\mathbin:\underline\kappa_1,k_2\mathbin:\underline\kappa_2,\ldots,
 k_n\mathbin:\underline\kappa_n$ where each $\underline\kappa_i$ is a member
 of filter $e_i$.
(iii) $\Pi,\,\Theta\vdash_sT\mathbin:\tau$ holds.
Similar for other sorts.
\end{defi}

\begin{prop}\label{hbr97}
To each environment $\xi$ associate we the filtering
 environments $\Pi_{\it fil}=\Pi_\xi$ and $\Theta_{\it fil}=
 \Theta_\xi$ in a natural manner.
Namely, we take $x\mathbin:d$ if $\xi(x)=d$ and $k\mathbin:e$
 if $\xi(k)=e$.
The following hold:

	\vskip2ex
        \noindent
\halign{\kern5em $#$\hfil &${}\in #$\hfil
  &${}\quad\Longleftrightarrow\quad #$\hfil\cr
 \tau & [\![T]\!]\xi &\Pi_\xi,\,\Theta_\xi\>\vdash_s\>T\mathbin:\tau\cr
 \dbot & [\![Q]\!]\xi &\Pi_\xi,\,\Theta_\xi\>\vdash_s\>Q\mathbin:\dbot\cr
 \sigma & [\![W]\!]\xi &\Pi_\xi,\,\Theta_\xi\>\vdash_s\>W\mathbin:\sigma\cr
 \kappa & [\![K]\!]\xi &\Pi_\xi,\,\Theta_\xi\>\vdash_s\>K\mathbin:\kappa.\cr
}
\end{prop}

	\vskip0ex

\noindent
The next two propositions establish subject reduction/expansion
 of the target calculus.
For $\eta$-reduction, we have only
 a weak form of subject expansion.

	\vskip0ex

\begin{prop}\label{wha02}
We suppose $T\rightarrow T'$ by $\beta$-reduction.
Then we have

	\vskip2ex
        \noindent\kern3em
$\Pi,\,\Theta\>\vdash_s\>T\mathbin:\tau\qquad
 \Longleftrightarrow\qquad\Pi,\,\Theta\>\vdash_s\>T'\mathbin:\tau$.

	\vskip2ex
        \noindent
Similar for other sorts.
\end{prop}

\proof\hskip.5em
The filter model is sound in regard to $\beta$-conversion
 since $F_a\scirc G_a={\it id}$ where $a=T,W,K$.
\qed

\begin{prop}\label{vqu58}
We suppose $T\rightarrow T'$ by $\eta$-reduction.
The following hold:

\shrinktopsep
\begin{enumerate}[(1)]
\item
If $\Pi,\,\Theta\>\vdash_s\>T\mathbin:\tau$ holds,
 $\Pi,\,\Theta\>\vdash_s\>T'\mathbin:\tau$ holds.

\item
If $T'$ is typeable for some typing environments, so is $T$.
\end{enumerate}
Similar for other sorts.
\end{prop}

\proof\hskip.5em
We note $G_W(F_W(d))=\mathop\uparrow\{\underline\sigma\rightarrow
 \tau\,|\>\underline\sigma\rightarrow\tau\in d\}$.
Hence $G_W(F_W(d))\subseteq d$.
So $\eta$-reduction preserves types whereas $\eta$-expansion does not
 in case of $\lambda x.\,Wx$.
Specifically, if $W$ has atomic type $\alpha$,
 then $\lambda x.\,Wx$ is not provided with the same type.
However, we can substitute the atomic type $\alpha$ with an arbitrary
 type of the form $\underline\sigma\rightarrow\tau$.
Then both $W$ and $\lambda x.\,Wx$ have this type.
Since single $W$ may have two or more typing judgments, we
 apply the substitution repeatedly if needed.
Hence $\eta$-expansion preserves typeability.
On the other hand, $G_a(F_a(d))=d$
 for $a=T,K$.
Hence, for the $\eta$-rule of other sorts, subject reduction/expansion
 actually holds.
\qed

	\vskip0ex

\noindent
Now we characterize solvability and normalizability in
 the target calculus by typeability.
We adapt results in \cite{jix62} to our setting.

	\vskip0ex

\begin{prop}\label{rrx04}
Term $T$ is solvable if and only if there are $\Pi,\Theta$, and $\tau$
 such that $\Pi,\,\Theta\>\vdash_s\>T\mathbin:\tau$.
Similar for other sorts.
\end{prop}

	\vskip0ex

\noindent
In the next proposition,
 we should assume that there is at least one atomic type
 for otherwise no derivations satisfy the assumption.

	\vskip0ex

\begin{prop}\label{njy56}
Term $T$ is normalizable if and only if $\Pi,\,\Theta\>\vdash_s\>
T\mathbin:\tau$ holds for some $\Pi,\Theta$,
 and $\tau$, all of these three not containing $\omega$.
Similar for other sorts.
We emphasize that the typing judgments except the lowermost
 in the derivation tree may contain $\omega$.
\end{prop}

\end{document}